\newcommand*{\ATLASLATEXPATH}{./}

\documentclass[cernpreprint,UKenglish,texlive=2012,txfonts=true]{\ATLASLATEXPATH atlasdoc}

\pdfoutput=1
\PreprintIdNumber{CERN-PH-EP-2015-138}

\usepackage[biblatex=false,siunitx=false]{\ATLASLATEXPATH atlaspackage}
\usepackage{tikz}
\usepackage{array}
\usepackage{arydshln}

\usepackage{atlascontribute}

\usepackage{atlasphysics}

\usepackage{cite}
\usepackage{rotating}

\usepackage{multirow}

\usepackage{stopsbottom_summary-defs}
\usepackage{centernot}

\hypersetup{pdftitle={ATLAS draft},pdfauthor={The ATLAS Collaboration}}

\AtlasTitle{ATLAS Run 1 searches for direct pair production of third-generation squarks at the Large Hadron Collider}

\author{The ATLAS Collaboration}

\AtlasRefCode{SUSY-2014-07}

\AtlasJournal{\EPJC}

\AtlasAbstract{This paper reviews and extends searches for the direct pair production of the scalar supersymmetric partners of the top and bottom quarks in proton--proton collisions collected by the ATLAS collaboration during the LHC Run 1. Most of the analyses use 20 \ifb\ of collisions at a centre-of-mass energy of $\sqrt{s} = 8 \TeV$, although in some case an additional $4.7\ \ifb$ of collision data at $\sqrt{s}= 7 \TeV$ are used. New analyses are introduced to improve the sensitivity to specific regions of the model parameter space. Since no evidence of third-generation squarks is found, exclusion limits are derived by combining several analyses and are presented in both a simplified model framework, assuming simple decay chains, as well as within the context of more elaborate phenomenological supersymmetric models. 
}

\AtlasCoverSupportingNote{Combination of 0 and 1 lepton analyses}{https://cds.cern.ch/record/1755210}
\AtlasCoverSupportingNote{Stop in  $c \neut$ and $bff\neut$}{https://cds.cern.ch/record/1977132}
\AtlasCoverSupportingNote{Stop in  two leptons at intermediate mT2}{https://cds.cern.ch/record/1712507}
\AtlasCoverSupportingNote{Sbottom and stop decays to single top + MET}{https://cds.cern.ch/record/1745975}
\AtlasCoverSupportingNote{ttH analysis re-interpretation}{https://cds.cern.ch/record/1968180}
\AtlasCoverSupportingNote{Sneaky-stop}{https://cds.cern.ch/record/1954147}

\AtlasCoverCommentsDeadline{ --- }

\AtlasCoverAnalysisTeam{J.~Abdallah, S.~Amoroso, J.~Anders, J.-F.~Arguin, M.~Barisonzi, M.I.~Besana, M.~Bianco, J.~Boyd, S.~Brazzale, P.~Butti, R.~Caminal, A.~Cortes, D.~Costanzo~(*), D.~Cot\'e, J.~Da~Costa, C.~Deluca, P.~Dondero, M.~D'Onofrio, O.A.~Ducu, T.~Eifert, P.~Favareto, M. Fiascaris, L.~Gauthier, G.~Gaudio, V.~Giangiobbe, T.~Gilliam, M.P.~Giordani,  C.~Giuliani, T.~Golling, E.~Gorini, P.~Grenier, D.~Guest, R.~Hawkings, A.~Henrichs, G.~Herten, M.~Hohlfeld, W.~Hopkins, M.~Hurwitz, A.~Juste, M.~Kruse, J.~K\"uchler, T.~Lari, M.~LeBlanc, G.~Lerner, C.~Macdonald, S.~Majewski, M.~Martinez, A.~Martyniuk, P.~M\"attig, J.~Maurer, J.~McFayden, F.~Meloni, E. Le Menedeu, J.~Montejo, T.~M\"uller, B.~Nachman, Y.~Nakahama, M. Owen, M.~Pagacova, F.~Paige, P.~Pani, A.~Paramonov, S.~Pataraia, J.~Pilcher, G.~Polesello, J.~Poveda, R.~Povey, P.~Pralavorio, M.~Primavera, T.~Rave, G.~Redlinger, K.~Rosbach, F.~Rubbo, J.~Sh\"affer, C. Schwanenberg, J.-E.~Sundermann, K.~Suruliz, C.~Troncon, V.~Tsiskaridze, A.~Tudorache, F.~Ungaro, G.~Usai, A.~Ventura, I.~Vivarelli~(*), C.~Wanatayaroj, X.~Wang, T.~Yamanaka, C.~Zhou}

\AtlasCoverEdBoardMember{Luciano~Mandelli~(*), Bobby~Acharya, Anna~Lipniacka, Chris~Young}

\AtlasCoverEgroupEditors{atlas-susy-2014-07-editors@cern.ch}

\AtlasCoverEgroupEdBoard{atlas-susy-2014-07-editorial-board@cern.ch}

\begin{document}
\MakeAtlasCoverTitle
\tableofcontents
\newpage

\section{Introduction}

In a theory with broken supersymmetry (SUSY)~\cite{Miyazawa:1966,Ramond:1971gb,Golfand:1971iw,Neveu:1971rx,Neveu:1971iv,Gervais:1971ji,Volkov:1973ix,Wess:1973kz,Wess:1974tw}, the mass scale of the supersymmetric particles is undetermined. However, for SUSY to provide a solution to the hierarchy problem~\cite{Weinberg:1975gm,Gildener:1976ai,Weinberg:1979bn,Susskind:1978ms} some of the new SUSY particles masses are typically required to be below about one TeV~\cite{Barbieri:1987fn,deCarlos:1993yy}, hence they could be within the reach of the LHC. 

The scalar partners of the right-handed and left-handed chiral components of the top-quark state ($\tright$ and $\tleft$ respectively) are among these particles. In many supersymmetric models, the large Yukawa coupling of the top quark to the Higgs sector makes the Higgs boson mass sensitive to the masses of the scalar top (referred to as stop in the following) states, such that, to avoid fine tuning, their masses are often required to be light.
 The $\tright$ and $\tleft$ components mix to form the mass eigenstates $\tone$ and $\ttwo$, $\tone$ being defined as the lighter of the two. The scalar superpartner of the left-handed chiral component of the bottom quark ($\bleft$) belongs to the same weak isospin doublet as the $\tleft$, hence they usually share the same supersymmetry-breaking mass parameter: a light stop can therefore imply the existence of a light scalar bottom. The lightest sbottom mass eigenstate is referred to as $\bone$. 

The ATLAS and CMS collaborations have searched for direct production of stops and sbottoms ~\cite{Aad:2014bva,Aad:2014kra,Aad:2014qaa,Aad:2014nra,Aad:2014mha,Aad:2013ija,Aad:2012cz,:2012si,Aad:2012tx,Aad:2012uu,Aad:2012yr,Chatrchyan:2012uea,Chatrchyan:2012wa,Chatrchyan:2013lya,Chatrchyan:2013xna,Chatrchyan:2013mya,Khachatryan:2015wza,Khachatryan:2014doa,CMS:2014dpa,Chatrchyan:2012paa} using about 4.7\ \ifb\ of data from the proton--proton collisions produced by the LHC at $\sqrt{s} = 7$ TeV and 20 \ifb\ at $\sqrt{s} = 8$ TeV. These searches have found no evidence of third-generation squark signals, leading to exclusion limits in many SUSY models. The aim of this paper is to summarise the sensitivity of the ATLAS experiment to R-parity-conserving\footnote{It is also assumed that the decay of the third-generation squarks is prompt: long-lived and metastable stops/sbottoms are discussed elsewhere~\cite{ATLAS:2014fka, Aad:2013gva}.}~\cite{Fayet:1976et,Fayet:1977yc,Farrar:1978xj,Fayet:1979sa,Dimopoulos:1981zb} models including the direct pair production of stops and sbottoms using the full $\sqrt{s}$=8~TeV proton--proton collision dataset collected
during Run 1 of the LHC.\footnote{The analysis exploiting the measurement of the \ttbar\  cross section discussed in this paper also uses 4.7 \ifb\ of proton-proton collisions at $\sqrt{s} = 7$ TeV.} The third-generation squarks are assumed to decay to the stable lightest supersymmetric particle (LSP) directly or through one or more intermediate stages. The analyses considered are those previously published by the ATLAS collaboration on the topic, together with new ones designed to increase the sensitivity to scenarios not optimally covered so far. A wide range of SUSY scenarios are studied by combining different analyses to improve the global sensitivity. 

The paper is organised as follows: Section~\ref{sec:phenomenology} briefly reviews the expected phenomenology of third-generation squark production and decay; Section~\ref{sec:strategy} reviews the general analysis approach followed by the ATLAS collaboration for SUSY searches; Sections~\ref{sec:interpretations}  and \ref{sec:pMSSM} present the exclusion limits obtained in specific models by combining the results of several analyses. Two different types of models have been considered: simplified models, where the third-generation squarks are assumed to decay into typically one or two different final states, and more complex phenomenological supersymmetric models, where the stop and sbottom have many allowed decay channels. Conclusions are drawn in Section~\ref{sec:conclusions}. 

For the sake of brevity, the body of the paper provides no details of the ATLAS detector and object reconstruction, of the analyses used in the limit derivation, or of how the signal Monte Carlo simulation samples were generated. However, a comprehensive set of appendices is provided to supply additional information to the interested reader. Appendix~\ref{sec:detector} briefly summarises the layout of the ATLAS detector and the general principles used in the reconstruction of electrons, muons, jets, jets containing $b$-hadrons (\bjets), and the missing transverse momentum vector $\bf{p}_{\mathrm{T}}^{\mathrm{miss}}$ (whose magnitude is referred to as $\etmiss$). Appendix~\ref{sec:SRs} discusses the analyses used to derive the exclusion limits presented in Sections~\ref{sec:interpretations} and \ref{sec:pMSSM}. The analyses that have already been published are only briefly reviewed, while those presented for the first time in this paper are discussed in detail. Appendix~\ref{sec:01lepCombine} provides further details of a combination of analyses which is performed for the first time in this paper. Finally, Appendix~\ref{sec:signal_generation} provides details about the generation and simulation of the signal Monte Carlo samples used to derive the limits presented.

\section{Third-generation squark phenomenology}
\label{sec:phenomenology}
 The cross section for direct stop pair production in proton--proton collisions at $\sqrt{s} = 8$ TeV as a function of the stop mass as calculated with \prospino~\cite{prospino,Kramer:2012bx} is shown in Figure~\ref{fig:stopcrosssec}. It is calculated to next-to-leading order accuracy in the strong coupling constant, adding the resummation of soft gluon emission at next-to-leading-logarithmic accuracy (NLO+NLL)~\cite{Beenakker:1997ut,Beenakker:2010nq,Beenakker:2011fu}. In this paper, the nominal cross section and its uncertainty are taken from an envelope of cross-section predictions using different parton distribution function (PDF) sets and factorisation and renormalisation scales described in Ref.~\cite{Kramer:2012bx}. The difference in cross section between the sbottom and stop pair production is known to be small~\cite{Beenakker:2010nq}, hence the values of Figure~\ref{fig:stopcrosssec} are used for both.

\begin{figure}[!htb]
\begin{center}
\subfloat[\label{fig:stopcrosssec}]{
\includegraphics[width=0.45\columnwidth,angle=0]{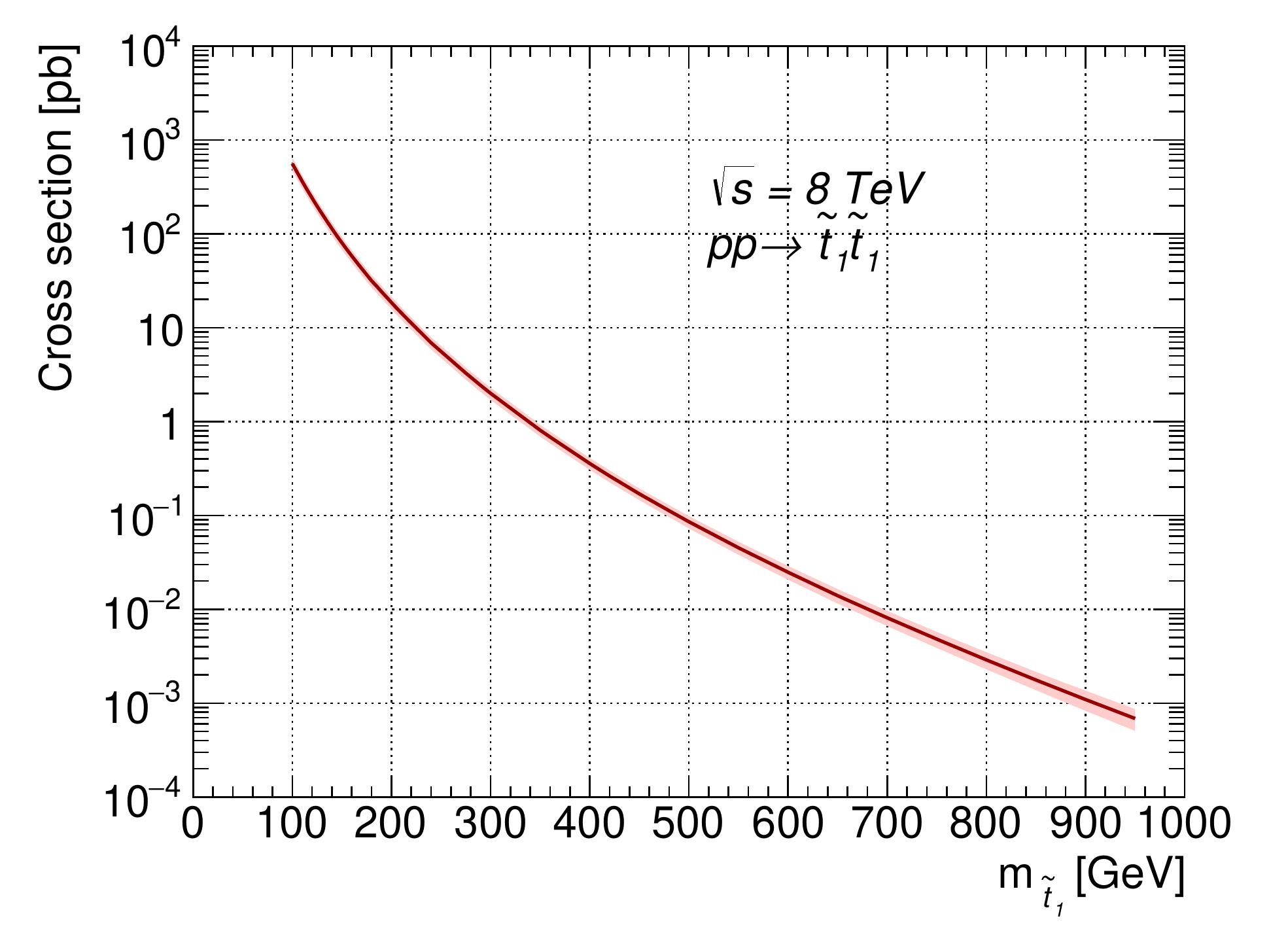}
}
\hfill
\subfloat[\label{fig:stop_pheno}]{
\begin{tikzpicture}
\filldraw[color=blue!5, fill=blue!5, very thick](0,0) -- (0,4.0) -- (3.0,4.0) -- cycle;
\draw [->] (0,0) -- (6.5,0) node [below left]  {\small $\mathrm{m}(\tone)$ [GeV]};
\draw [->] (0,0) -- (0,4) node [above,rotate = 90] {\small $\mathrm{m}(\neut)$ [GeV]};	
\node [above ] at (1.2,3) {\small \DMstopN < 0};
\node [right, rotate=53.13] at (1.,0.8) {\small $\tone \rightarrow c \neut$};
\node [right, rotate=53.13] at (1.7,0.8) {\small $\tone \rightarrow b f f' \neut$};
\node [right, rotate=53.13] at (3.1,0.8) {\small $\tone \rightarrow bW \neut$};
\node [right, rotate=53.13] at (5.0,0.8) {\small $\tone \rightarrow t\neut$};
\node [right, rotate=53.13, color=blue] at (2.5,3) {\tiny $\DMstopN = 0$};
\node [right, rotate=53.13, color=blue] at (3.7,2.4) {\tiny $\DMstopN = m_W + m_b$};
\node [right, rotate=53.13, color=blue] at (5.7,2.5) {\tiny $\DMstopN = m_t$};
\draw [dashed,color=blue] (0,0) -- (3.0,4.0); 
\draw [dashed,color=blue] (1.6,0) -- (4.6,4.0);
\draw [dashed,color=blue] (3.5,0) -- (6.5,4.0); 
\draw (2, -3pt) -- (2, 0) node [below] {\tiny $100$};
\draw (4, -3pt) -- (4, 0) node [below] {\tiny $200$};
\draw (4, -3pt) -- (4, 0) node [below] {\tiny $200$};
\draw (-3pt,2.6666) -- (0, 2.6666) node [left] {\tiny $100$}; 
\draw (-3pt,1.3333) -- (0, 1.3333) node [left] {\tiny $50$}; 
\end{tikzpicture}
\vspace{3cm}
}	
\end{center}
\caption{ (a) Direct stop pair production cross section at $\sqrt{s} = 8$ TeV as a function of the stop mass. The band around the cross section curve illustrates the uncertainty (which is everywhere about 15--20\%) on the cross section due to scale and PDF variations. (b) Illustration of stop decay modes in the plane spanned by the masses of the stop ($\tone$) and the lightest neutralino ($\neut$), where the latter is assumed to be the lightest supersymmetric particle and the only one present among the decay products. The dashed blue lines indicate thresholds separating regions where different processes dominate.
}
\end{figure}

Searches for direct production of stops and sbottoms by the ATLAS collaboration have covered several possible final-state topologies. The experimental signatures used to identify these processes depend on the masses of the stop or sbottom, on the masses of the other supersymmetric particles they can decay into, and on other parameters of the model, such as the stop and sbottom left-right mixing and the mixing between the gaugino and higgsino states in the chargino--neutralino sector. 

Assuming that the lightest supersymmetric particle is a stable neutralino ($\neut$), and that no other supersymmetric particle plays a significant role in the sbottom decay, the decay chain of the sbottom is simply $\bone \rightarrow b \neut$ (Figure~\ref{fig:bN1bN1}). 

\begin{figure}[htb]
  \begin{center}
            \subfloat[\label{fig:bN1bN1}]{
        \includegraphics[width=0.2\textwidth]{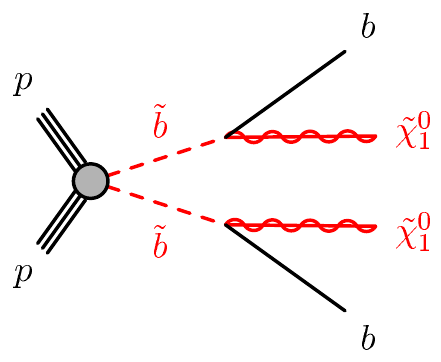}
        }
\hfill
            \subfloat[\label{fig:tN1tN1}]{
        \includegraphics[width=0.2\textwidth]{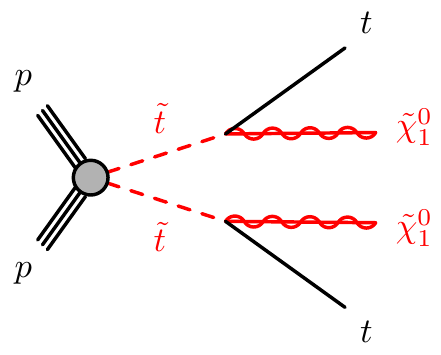}
        }
\hfill
         \subfloat[\label{fig:threebody}]{
        \includegraphics[width=0.2\textwidth]{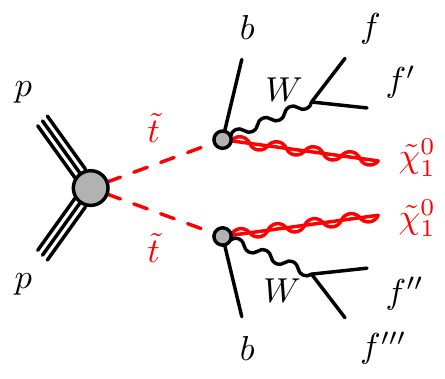}
        }
\hfill
         \subfloat[\label{fig:fourbody}]{
        \includegraphics[width=0.2\textwidth]{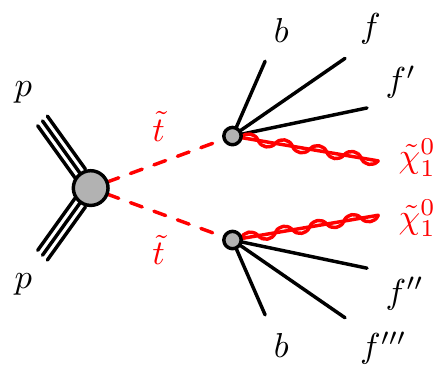}
        }
        \hfill
         \subfloat[\label{fig:cN1cN1}]{
        \includegraphics[width=0.2\textwidth]{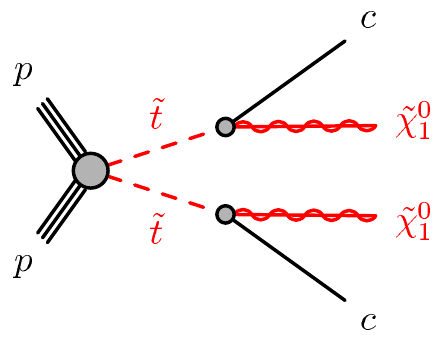}
        }
        \hfill
        \subfloat[\label{fig:feynDiag_bC}]{
        \includegraphics[width=0.2\textwidth]{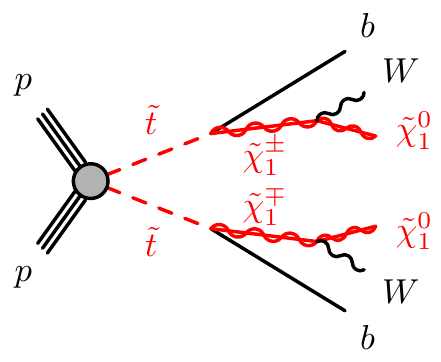}
        }
        \hfill
        \subfloat[\label{fig:feynDiag_bone_tC}]{
        \includegraphics[width=0.2\textwidth]{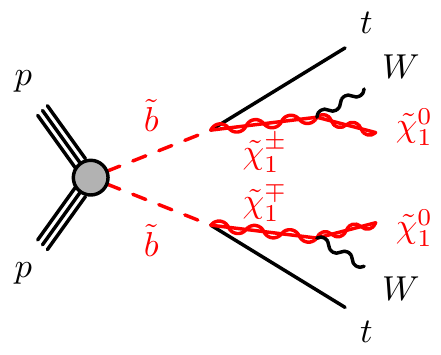}
        }
                \hfill
        \subfloat[\label{fig:feynDiag_bone_bN2}]{
        \includegraphics[width=0.2\textwidth]{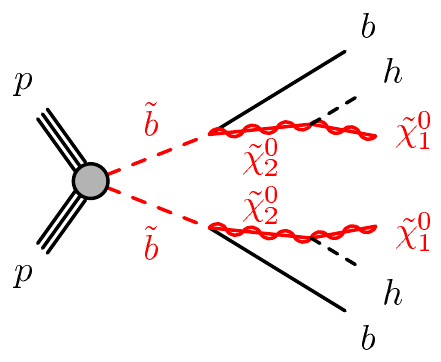}
        }

        	\caption{Diagrams of $\tone$ and $\bone$ pair production and decays considered as simplified models: (a) $\bone\bone\rightarrow b\neut b \neut$; (b) $\tone\tone\rightarrow t\neut t \neut$; (c) three-body decay; (d) four-body decay; (e) $\tone\tone\rightarrow c\neut c\neut$; (f) $\tone\tone\rightarrow b\chipm b\chipm$; (g) $\bone\bone\rightarrow t\chipm t\chipm$; (h) $\bone\bone\rightarrow b\neuttwo b\neuttwo$. The diagrams do not show ``mixed'' decays, in which the two pair-produced third-generation squarks decay to different final states.}
    \label{fig:feynDiag}
  \end{center}
\end{figure}

 A significantly more complex phenomenology has to be considered for the stop, depending on its mass and on the $\neut$ mass. Figure~\ref{fig:stop_pheno} shows the three main regions in the $m_{\tone}$--$m_{\neut}$ plane that are taken into account. They are identified by different values of $\DMstopN = m_{\tone}$-$m_{\neut}$. In the region where $\DMstopN > \mtop$, the favoured decay is $\tone \rightarrow t \neut$ (Figure~\ref{fig:tN1tN1}). The region where $m_W + m_b < \DMstopN <  \mtop$ is characterised by the three-body decay\footnote{In scenarios that depart from the minimal flavour violation assumption, flavour-changing decays like $\tone \rightarrow c \neut$ or $\tone \rightarrow u \neut$ could have a significant branching ratio up to $\DMstopN \sim 100$ GeV \cite{Grober:2015fia}.} ($\tone \rightarrow Wb \neut$ through an off-shell top quark, Figure~\ref{fig:threebody}). The region where the value of $\DMstopN$ drops below $m_W + m_b$, sees the four-body decay $\tone \rightarrow b f f' \neut$, (where $f$ and $f'$ indicate generic fermions coming from the decay of an off-shell $W$ boson, Figure~\ref{fig:fourbody}) competing with the flavour-changing decay\footnote{The decay $\tone \rightarrow u \neut$, in the assumption of minimal flavour violation~\cite{D'Ambrosio:2002ex}, is further suppressed with respect to $\tone \rightarrow c \neut$ by corresponding factors of the CKM matrix.}  $\tone \rightarrow c \neut$ of Figure~\ref{fig:cN1cN1}; the dominant decay depends on the details of the supersymmetric model chosen~\cite{Grober:2014aha}.
 
If the third-generation squark decay involves more SUSY particles (other than the $\neut$), then additional dependencies on SUSY parameters arise. For example, if the lightest chargino ($\chipm$) is the next-to-lightest supersymmetric particle (NLSP), then the stop tends to have a significant branching ratio for $\tone \rightarrow b \chipm$ (Figure~\ref{fig:feynDiag_bC}), or, for the sbottom, $\bone \rightarrow t \chipm$ if kinematically allowed (Figure~\ref{fig:feynDiag_bone_tC}). The presence of additional particles in the decay chain makes the phenomenology depend on their masses. Several possible scenarios have been considered, the most common ones being the gauge-universality inspired $m_{\chipm} = 2 m_{\neut}$, favoured, for example, in mSUGRA/CMSSM models~\cite{Chamseddine:1982,Barbieri:1982eh,Ibanez:1982ee,Hall:1983iz,Ohta:1982wn,Kane:1993td}; other interpretations include the case of a chargino almost degenerate with the neutralino, a chargino almost degenerate with the squark, or a chargino of fixed mass. Another possible decay channel considered for the sbottom is $\bone \rightarrow b \neuttwo \rightarrow b \higgs \neut$ (Figure~\ref{fig:feynDiag_bone_bN2}), which occurs in scenarios with a large higgsino component of the two lightest neutralinos. 

\begin{figure}[htb]
  \begin{center}
            \subfloat[\label{fig:t2Z}]{
        \includegraphics[width=0.2\textwidth]{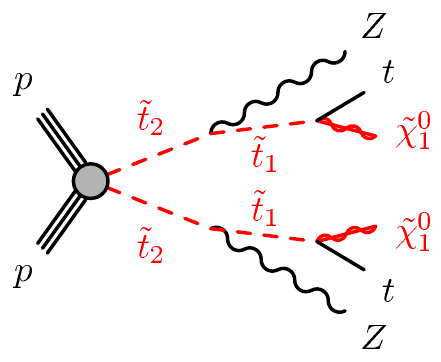}
        }
\hfill
            \subfloat[\label{fig:t2H}]{
        \includegraphics[width=0.2\textwidth]{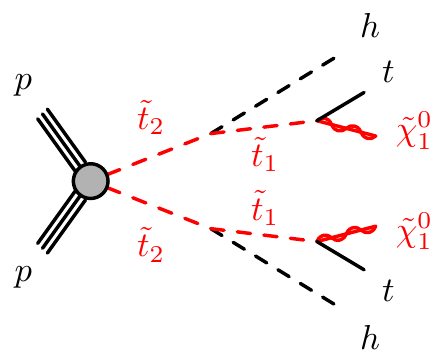}
        }
\hfill
         \subfloat[\label{fig:t2t}]{
        \includegraphics[width=0.2\textwidth]{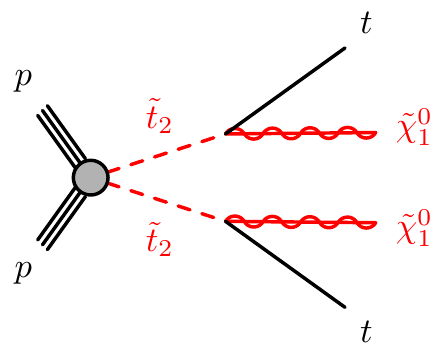}
        }
        	\caption{Diagrams of $\ttwo$ decays considered as simplified models: (a) $\ttwo\ttwo\rightarrow \tone Z \tone Z$; (b) $\ttwo\ttwo\rightarrow \tone \higgs \tone \higgs$; (c) $\ttwo\ttwo \rightarrow t\neut t\neut$. The diagrams do not show ``mixed'' decays, in which the two pair-produced third-generation squarks decay to different final states. The decay $\ttwo \rightarrow \gamma \tone$ is not an allowed process.}
    \label{fig:feynDiag_t2}
  \end{center}
\end{figure}

Despite the lower production cross section and similar final states to $\tone$, the heavier stop state ($\ttwo$) pair production has also been studied: the search for it becomes interesting in scenarios where the detection of $\tone$ pair production becomes difficult (for example if $\DMstopN \sim \mtop$). The diagrams of the investigated processes are shown in Figure~\ref{fig:feynDiag_t2}.

Two types of SUSY models are used to interpret  the results in terms of exclusion limits. The simplified model approach assumes that either a stop or a sbottom pair is produced and that they decay into well-defined final states, involving one or two decay channels. 
Simplified models are used to optimise the analyses for a specific final-state topology, 
rather than the complex (and model-dependent) mixture of different topologies that would arise from a SUSY model 
involving many possible allowed production and decay channels. 
The sensitivity to simplified models is discussed in Section~\ref{sec:interpretations}.

More complete phenomenological minimal supersymmetric extensions of the Standard Model (pMSSM in the following~\cite{Djouadi:1998di}) are also considered, to assess the performance of the analyses in scenarios where the stop and sbottom typically have many allowed decay channels with competing branching ratios. Three different sets of pMSSM models are considered, which take into account experimental constraints from LHC direct searches, satisfying the Higgs boson mass and dark-matter relic density constraints, or additional constraints arising from considerations of naturalness. The sensitivity to these models is discussed in Section~\ref{sec:pMSSM}.

\section{General discussion of the analysis strategy}
\label{sec:strategy}
The rich phenomenology of third-generation supersymmetric particles requires several event selections to target the wide range of possible topologies. A common analysis strategy and common statistical techniques, which are extensively described in Ref.~\cite{Baak:2014wma}, are employed. 

Signal regions (SR) are defined, which target one specific model and SUSY particle mass range. The event selection is optimised by relying on the Monte Carlo simulation of both the Standard Model (SM) background production processes and the signal itself. The optimisation process aims to maximise the expected significance for discovery or exclusion for each of the models considered.

For each SR, multiple control regions (CR) are defined: they are used to constrain the normalisation of the most relevant SM production processes and to validate the MC predictions of the shapes of distributions of the kinematic variables used in the analysis. The event selection of the CRs is mutually exclusive with that of the SRs. It is, however, chosen to be as close as possible to that of the signal region while keeping the signal contamination small, and such that the event yield is dominated by one specific background process. 

A likelihood function is built as the product of Poisson probability functions, describing the observed and expected number of events in the control and signal regions. The observed numbers of events in the various CRs and SRs are used in a combined profile likelihood fit~\cite{Cowan:2010js} to determine the expected SM background yields for each of the SRs. Systematic uncertainties are treated as nuisance parameters in the fit and are constrained with Gaussian functions with standard deviation equal to their value. The fit procedure takes into account correlations in the yield predictions between different regions due to common background normalisation parameters and systematic uncertainties, as well as  contamination from SUSY signal events, when a particular model is considered for exclusion.

 The full procedure is validated by comparing the background predictions and the shapes of the distributions of the key analysis variables from the fit results to those observed in dedicated validation regions (VRs), which are defined to be orthogonal to, and kinematically similar, to the signal regions, with low potential contamination from signal. 
 
 After successful validation, the observed yields in the signal regions are compared to the prediction. The profile likelihood ratio statistic is used first to verify the SM background-only hypothesis, and, if no significant excess is observed, to exclude the signal-plus-background hypothesis in specific signal models.  A signal model is said to be excluded at 95\% confidence level (CL) if the \cls~\cite{Junk:1999kv,Read:2002hq} of the profile likelihood ratio statistics of the signal-plus-background hypothesis is below 0.05.

Several publications, targeting specific stop and sbottom final-state topologies, were published by the ATLAS collaboration at the end of the proton--proton collision run at $\sqrt{s} = 8$ TeV, using a total integrated luminosity of about 20\ \ifb. Each of these papers defined one or more sets of signal regions optimised for different simplified models with different mass hierarchies and decay modes for the stop and/or sbottom. A few additional signal regions, focusing on regions of the parameter space not well covered by existing analyses have been defined since then. All signal regions that are used in this paper are discussed in detail in Appendix~\ref{sec:SRs}, while Table~\ref{tab:SRname} introduces their names and the targeted models. Each analysis is identified by a short acronym defined in the second column of Table~\ref{tab:SRname}. The signal region names of previously published analyses are retained, but, to avoid confusion and to ease the bookkeeping, the analysis acronym is prepended to their names. For example, SRA1 from the \stopZeroLep\ analysis of Ref.~\cite{Aad:2014bva}, which is a search for stop pair production in channels with no leptons in the final state,  is referred to as \stopZeroLep-SRA1.

\begin{sidewaystable}
\begin{center}
\caption{Summary of the ATLAS analyses and signal regions used in this paper. Each signal region is identified by the acronym of the corresponding analysis followed by the original name of the signal region defined either in the published paper or in Appendix~\ref{sec:newSRs}. A dash in the signal region name column indicates that the analysis does not use the concept of signal region. \label{tab:SRname}}

\begin{tabular}{l|c|c|c} 
\hline
\hline
Analysis name and & Analysis & \multirow{2}{*}{Original signal region name} & \multirow{2}{*}{Model targeted}\\
corresponding reference & acronym &  & \\
\hline
\hline
\multirow{3}{*}{Multijet final states~\cite{Aad:2014bva}} & \multirow{3}{*}{\stopZeroLep} & SRA1-4 & \multirow{2}{*}{$\tone \rightarrow t \neut$}\\
& & SRB &\\
\cdashline{3-4}
& & SRC1-3  & $\tone \tone \rightarrow bt\neut\chipm$ with $m_{\chipm} = 2 m_{\neut}$\\
\hline
\multirow{7}{*}{One-lepton final states~\cite{Aad:2014kra}} & \multirow{7}{*}{\stopOneLep} &tN\_diag & $\tone \rightarrow t \neut$ with $m_{\tone} \sim \mtop + m_{\neut}$\\
\cdashline{3-4}
 & & tN\_med, tN\_high, tN\_boost & $\tone \rightarrow t \neut$  \\
 \cdashline{3-4}
 & & bCa\_low, bCa\_med, bCb\_med1, & \multirow{3}{*}{$\tone \rightarrow b \chipm$} \\ 
 & &  bCb\_high, bCb\_med2, bCc\_diag& \\
 & &  bCd\_bulk, bCd\_high1, bCd\_high2 &\\
 \cdashline{3-4}
 & & 3body & $\tone \rightarrow b W \neut$ (three-body decay) \\
 \cdashline{3-4}
 & & tNbC\_mix & $\tone \tone \rightarrow bt\neut\chipm$ with $m_{\chipm} = 2 m_{\neut}$\\
 \hline
 \multirow{2}{*}{Two-lepton final states~\cite{Aad:2014qaa}} & \multirow{2}{*}{\stopTwoLep} & L90, L100, L110, L120, H160  & $\tone \rightarrow b\chipm$, three-body decay \\
 \cdashline{3-4}
 & & M1-4 & $\tone \rightarrow t \neut$\\
 \hline
Final states from compressed stop & \multirow{2}{*}{\stopCharm} & M1-3  & $\tone/\bone \rightarrow \mathrm{anything}$ with $m_{\tone} \sim m_{\neut}$\\
decays~\cite{Aad:2014nra} & & C1-2 & $\tone \rightarrow c \neut$\\
\hline
\multirow{2}{*}{Final states with a $Z$ boson~\cite{Aad:2014mha}} & \multirow{2}{*}{\stopTwo} & SR2A, SR2B, SR2C, & \multirow{2}{*}{$\ttwo \rightarrow \tone Z$ and $\ttwo \rightarrow \tone \higgs$}  \\
 &  & SR3A, SR3B &\\
 \hline
 Final states with two \bjets and \etmiss~\cite{Aad:2013ija} & \sbottom & SRA, SRB & $\bone \rightarrow b \neut$ and $\tone \rightarrow b \chipm$ with $m_{\chipm} \sim m_{\neut}$ \\
 \hline
 Final states with two leptons  & \multirow{2}{*}{\WWlike} & \multirow{2}{*}{SR1-7} & $\tone \rightarrow b \chipm$ with  $m_{\chipm} = m_{\tone} - 10$ GeV and \\
at intermediate \mtTwoll~\ref{sec:WWlike} & &  & $\tone \rightarrow b \ell \nu \neut$ (three- and four-body decays) \\
\hline
Final states containing two top quarks & \multirow{2}{*}{\ttH}&  \multirow{2}{*}{ --- } & \multirow{2}{*}{$\ttwo \rightarrow \tone \higgs$} \\
and a Higgs boson~\ref{sec:ttH} & & & \\
\hline
 \multirow{2}{*}{Final states containing a top and a \bquark~\ref{sec:tbmet}} & \multirow{2}{*}{\TBM} & \multirow{2}{*}{SR1-5} & $\tone \tone \rightarrow b \chipm t\neut$ with \\  
 & & & $m_{\chipm} \sim m_{\neut}$ and pMSSM models\\
\hline
 \multirow{3}{*}{Final states with three \bjets~\cite{Aad:2014lra}} & \multirow{3}{*}{\threeb} & SR-0$\ell$-4j-A, SR-0$\ell$-4j-B, SR-0$\ell$-4j-C, & gluino-mediated $\tone$ and $\bone$ production, \\
 & & SR-0$\ell$-7j-A, SR-0$\ell$-7j-B, SR-0$\ell$-7j-C, & $\bone \rightarrow \neuttwo b \rightarrow \neut \higgs b$ \\
  & & SR-1$\ell$-6j-A, SR-1$\ell$-6j-B, SR-1$\ell$-7j-C & \\
  \hline
   Strongly produced final states & \multirow{2}{*}{\sslll} & SR3b, SR0b, SR1b, & Generic gluino and  \\
      with two same-sign or three leptons~\cite{Aad:2014pda}& & SR3Llow, S3Lhigh & squark production, $\bone \rightarrow t \chipm$ \\	
      \hline
      Spin correlation in & \multirow{2}{*}{\spinCorr} &  \multirow{2}{*}{ --- } & $\tone \rightarrow t \neut$ with \\
       \ttbar\ production events~\cite{Aad:2014mfk} & & & $m_{\tone} \sim \mtop + m_{\neut}$ \\
       \hline
       $\ttbar$ production cross section~\cite{Aad:2014kva} & \xsec & --- & $\tone \rightarrow t \neut$,  three-body decay\\
\hline
\hline
\end{tabular}
\end{center}
\end{sidewaystable}

\newpage

\section{Interpretations in simplified models} 
\label{sec:interpretations}

The use of simplified models for analysis optimisation and result interpretation has become more and more common in the last years. The attractive feature of this approach is that it focuses on a specific final-state topology, rather than on a complex (and often heavily model-dependent) mixture of several  different topologies: only a few SUSY particles are assumed to be produced in the proton--proton collision -- often just one type -- 
and only a few decay channels are assumed to be allowed. 
In the remainder of this section, several exclusion limits derived in different supersymmetric simplified models are presented. Details about how the MC signal samples used for the limit derivations were produced are available in Appendix~\ref{sec:signal_generation}. 

\subsection{Stop decays with no charginos in the decay chain} 
\label{sec:01lep_summary}

A first series of simplified models is considered. It includes direct stop pair production as the only SUSY production process, and assumes that no supersymmetric particle other than the $\tone$ itself and the LSP, taken to be the lightest neutralino $\neut$, is involved in the decay. Under this assumption, there is little model dependence left in the stop phenomenology, as discussed in Section~\ref{sec:phenomenology}. The stop decay modes are defined mainly by the mass separation $\DMstopN$ between the stop and the neutralino, as shown in Figure~\ref{fig:stop_pheno}. The corresponding diagrams are shown in Figure~\ref{fig:feynDiag}.

Figure~\ref{fig:tNsummary} shows the 95\% CL exclusion limits obtained in the $m_{\tone}-m_{\neut}$ plane by the relevant analyses listed in Table~\ref{tab:SRname} and discussed in Appendix~\ref{sec:SRs}, or by their combination. A detailed discussion of which analysis is relevant in each range of $\DMstopN$ follows.

\begin{figure}[htb]
  \begin{center}
    \includegraphics[width=0.7\textwidth]{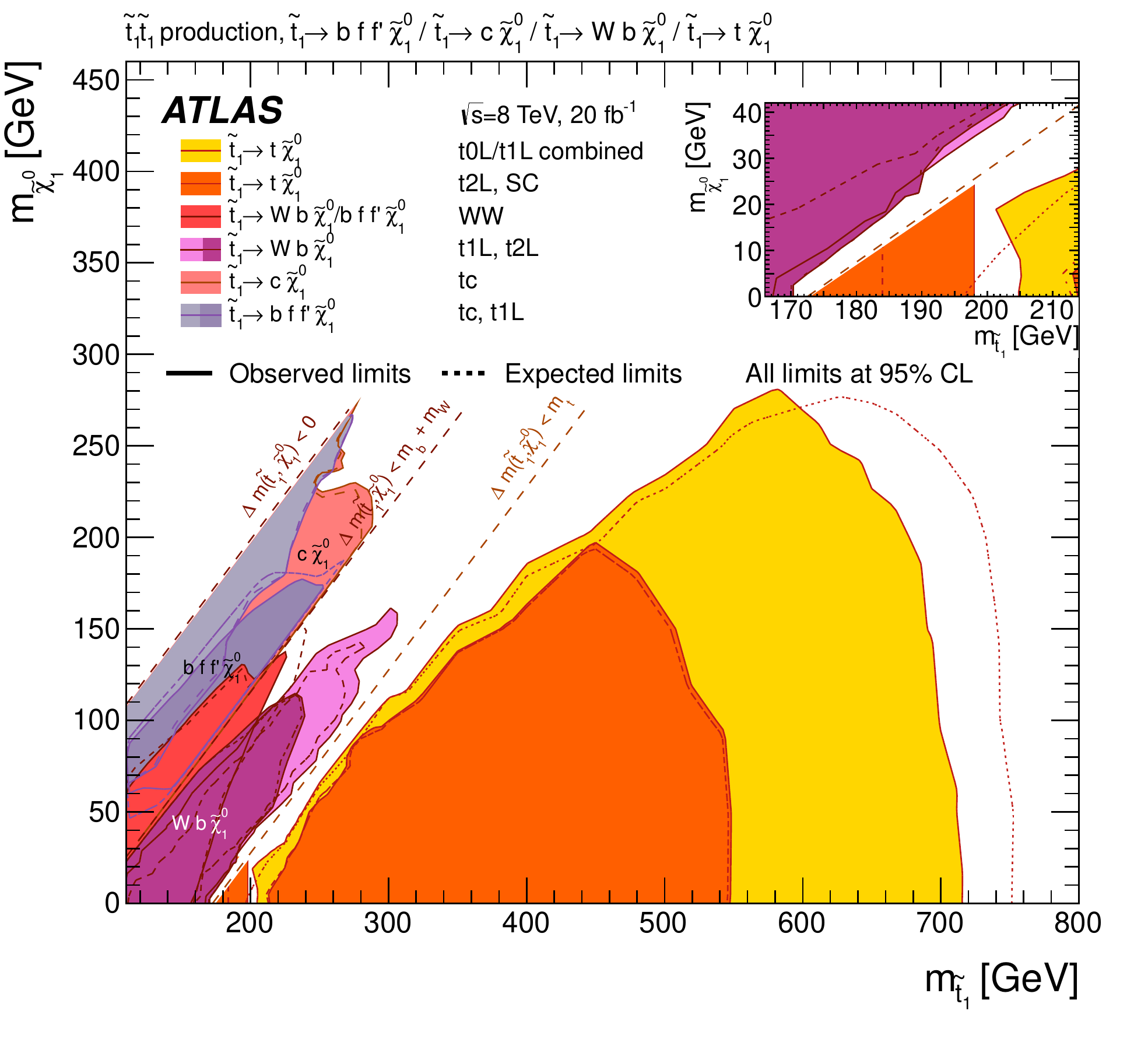}
    \caption{Summary of the ATLAS Run 1 searches for direct stop pair production in models where no supersymmetric particle other than the $\tone$ and the $\neut$ is involved in the $\tone$ decay. The 95\% CL exclusion limits are shown in the $m_{\tone}$--$m_{\neut}$ mass plane. The dashed and solid lines show the expected and observed limits, respectively, including all uncertainties except the theoretical signal cross-section uncertainty (PDF and scale). Four decay modes are considered separately with a branching ratio of 100\%: $\tone \rightarrow t \neut$, where the $\tone$ is mostly $\tright$, for $\DMstopN > \mtop$; $\tone\rightarrow Wb\neut$ (three-body decay) for $m_W + m_b< \DMstopN < \mtop$;  $\tone\rightarrow c \neut$ and $\tone \rightarrow bff'\neut$  (four-body decay) for $\DMstopN < m_W + m_b$. The latter two decay modes are superimposed.}
    \label{fig:tNsummary}
  \end{center}
\end{figure}
\paragraph{$\DMstopN< m_{W}+m_b$.}
This kinematic region is characterised by the presence of two competing decays: the flavour-violating decay $\tone \rightarrow c\neut$ (Figure~\ref{fig:cN1cN1}) and the four-body decay $\tone \rightarrow bff' \neut$ (Figure~\ref{fig:fourbody}). Which one of the two becomes dominant depends on the model details, in particular on the mass separation between the stop and the neutralino, and on the amount of flavour violation allowed in the model~\cite{Grober:2014aha}. 
Several analyses have sensitivity in this region of the $m_{\tone} - m_{\neut}$ plane. The monojet-like signal regions (\stopCharm-M1-3) dominate the sensitivity in the region with $\DMstopN \gtrsim m_b$, regardless of the decay of the stop pair, which goes undetected: their selection is based on the presence of an initial-state radiation (ISR) jet recoiling against the stop-pair system, which is assumed to be invisible.  At larger values of $\DMstopN$, signal regions requiring the presence of a $c$-tagged jet (\stopCharm-C1-2) complement the monojet-like signal regions by targeting the $\tone\rightarrow c \neut$ decay. Limits on four-body decays can be set using signal regions which include low transverse momentum electrons and muons (\stopOneLep-bCa\_low and \WWlike).

The limits reported in Figure~\ref{fig:tNsummary} for these values of $\Delta m$ all assume that the branching ratio of the stop decay into either $\tone \rightarrow c \neut$ or $\tone \rightarrow b f f' \neut$ is 100\%. However, this assumption can be relaxed, and exclusion limits derived as a function of the branching ratio of the $\tone \rightarrow c \neut$ decay,  BR($\tone \rightarrow c \neut$), assuming that BR($\tone \rightarrow c \neut$) + BR($\tone \rightarrow b f f' \neut$) = 1.
Two different scenarios, with $\DMstopN=10, 80$ \GeV, are considered. The first compressed scenario is characterised by low-\pt stop decay products, and the set of signal regions which have sensitivity is  the \stopCharm-M, independently of the decay of the stop. In the second scenario, the phase space available for the  \tone\ decay is larger, and  the full set of \stopCharm-M, \stopCharm-C,  \stopOneLep-bCa\_low, \stopOneLep-bCa\_med and \WWlike-SR selections have different sensitivity, depending on BR($\tone \rightarrow c \neut$). 

The cross-section limit is derived by combining the analyses discussed above. The SR giving the lowest expected exclusion \cls\ for each signal model and for each value of 
BR($\stop \rightarrow c\neut$) is chosen.
\begin{figure}[!htb]
\begin{center}
\subfloat[]{
\includegraphics[width=0.49\textwidth]{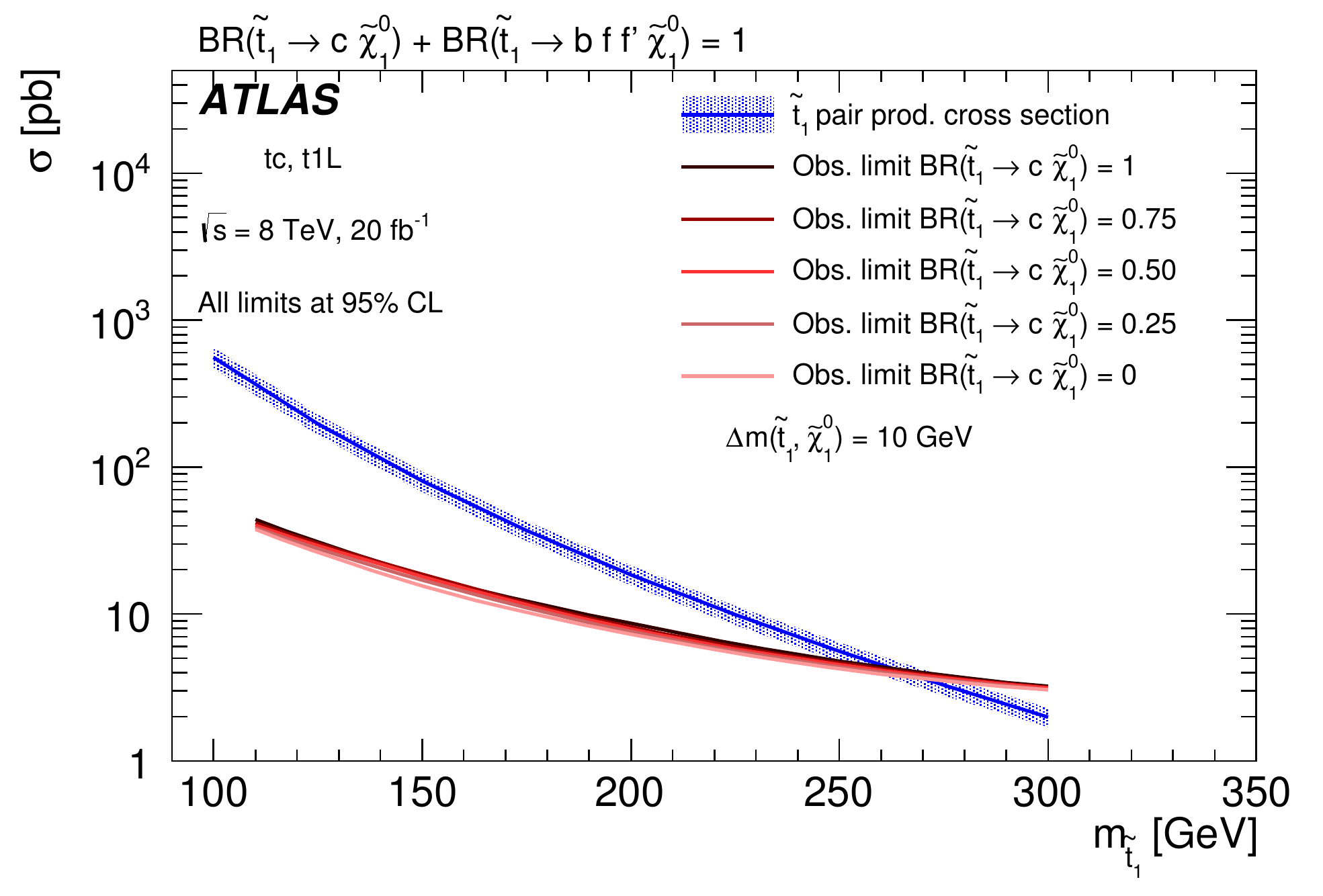}
}
\subfloat[]{
\includegraphics[width=0.49\textwidth]{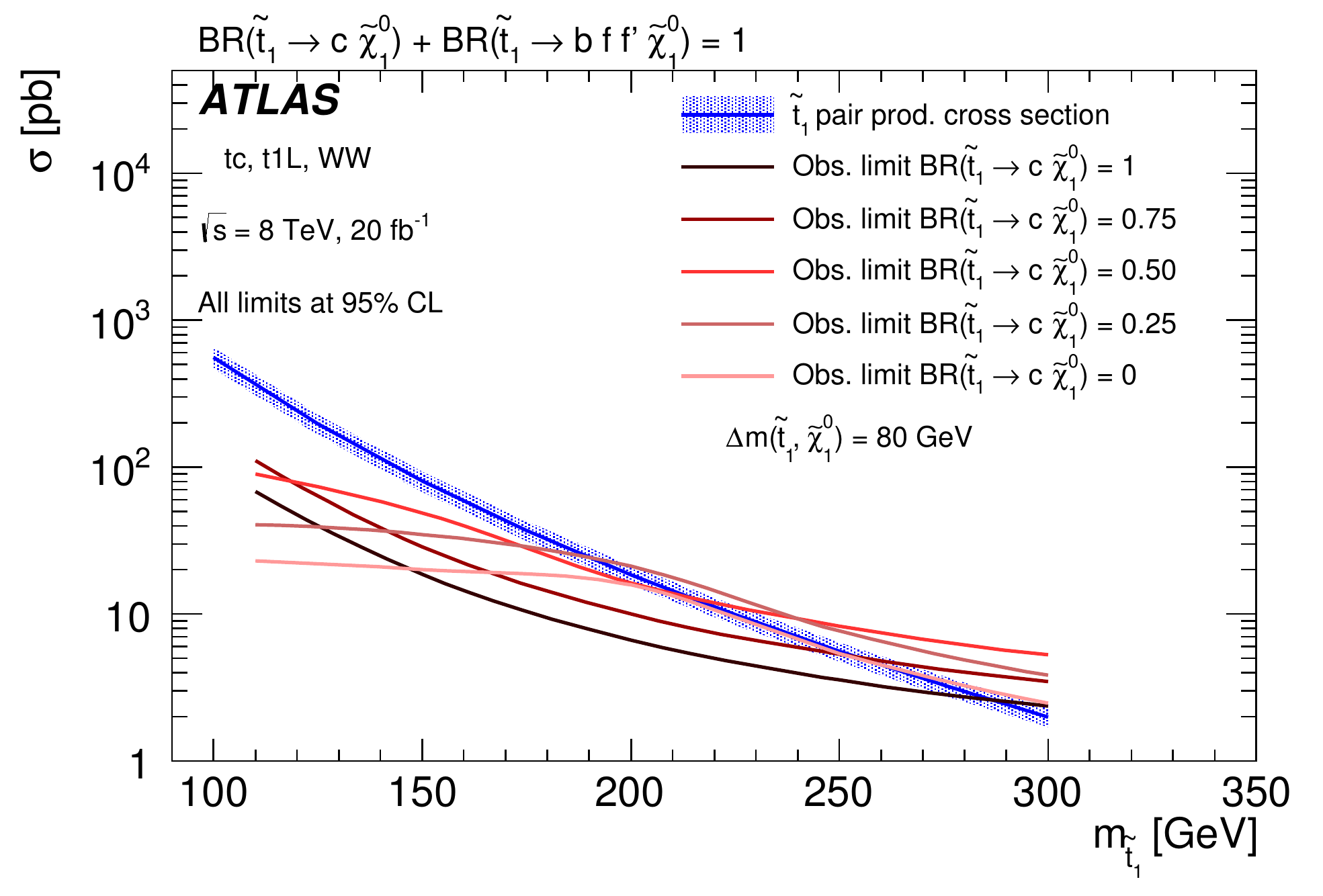}
}
\caption{Upper limits on the stop pair production cross sections for different values of the BRs for the decays $\tone\rightarrow c \neut$ and $\tone \rightarrow ff'b\neut$. 
Signal points with $\DMstopN$ of  10 GeV (a) and  80 GeV (b) are shown.
The limits quoted are taken from the best performing, based on expected exclusion \cls, signal regions from the \stopCharm-M, \stopCharm-C,  \stopOneLep-bCa\_low and \WWlike\ analyses at each mass point. The blue line and corresponding hashed band correspond to the mean value and uncertainty on the production cross section of the stop as a function of its mass. The pink lines, whose darkness indicate the value of BR($\stop \rightarrow c\neut$) according to the legend, indicate the observed limit on the production cross section.}
\label{fig:cNeut_bffneut_combination}
\end{center}
\end{figure}
Figure~\ref{fig:cNeut_bffneut_combination} shows the result of these combinations. For $\DMstopN = 10$ GeV, the sensitivity is completely dominated by the \stopCharm-M 
signal regions, hence no significant dependence on BR($\stop \rightarrow c\neut$) is observed. In this case, stop masses up to about 250 GeV are excluded. For  $\DMstopN = 80$ GeV, 
the sensitivity is dominated by the \stopCharm-C signal regions at high values of BR($\stop \rightarrow c\neut$). For lower values of BR($\stop \rightarrow c\neut$), the ``soft-lepton'' and \WWlike\ signal regions both become competitive, the latter yielding a higher sensitivity at smaller values of the stop mass. The maximum excluded stop mass ranges from about 180 GeV for BR$(\stop \rightarrow c\neut)=25\%$ to about 270 GeV for BR$(\stop \rightarrow c\neut) = 100\%$.

\paragraph{$m_W+m_b < \DMstopN < \mtop$.}
In this case, the three-body decay of Figure~\ref{fig:threebody} is dominant. The signal regions that are sensitive to this decay are the dedicated signal region defined in the analysis selecting one-lepton final states (the \stopOneLep-3body) and the combination of several signal regions from the analysis selecting two-lepton final states, the \stopTwoLep. The exclusion limits shown in Figure~\ref{fig:tNsummary} assume BR($\tone \rightarrow bW\neut) = 1$. The \WWlike\ signal regions are found to be sensitive to the kinematic region  separating the  three-body from the four-body stop decay region. 
\paragraph{$\DMstopN \sim \mtop$.}
In this case, the neutralinos are produced with low $\pt$, and the kinematic properties of the signal are similar to those of SM $\ttbar$ production. Exclusion limits in this region were obtained by two analyses performing precision SM measurements. The first one is the measurement of the \ttbar\ inclusive production cross section $\sigma_{\ttbar}$. Limits on $\tone$ pair production were already set in Ref.~\cite{Aad:2014kva}, which measured $\sigma_{\ttbar}$ in the different-flavour, opposite-sign channel $e\mu$. They were derived assuming a $\tone$ decay into an on-shell top quark, $\tone \rightarrow t \neut$. An extension of the limits into the three-body stop decay is discussed in Appendix~\ref{sec:published_SR}. For a massless neutralino, the analysis excludes stop masses from about 150 GeV to about $\mtop$. The limit deteriorates for higher neutralino masses, mainly because of the softer \bjet\ spectrum and the consequent  loss in acceptance. 
The second analysis considered is that of the top quark spin correlation (\spinCorr) which considers SM $\ttbar$ production with decays to final states containing two leptons (electrons or muons). The shape and normalisation of the distribution of the azimuthal angle between the two leptons is sensitive to the spin of the produced particles, hence it allows the analysis to differentiate between stop pair and $\ttbar$ production. The limit obtained is shown in the bottom middle (dark orange) of the inset of Figure~\ref{fig:tNsummary}.  A small region of $\DMstopN \approx 180$~GeV is excluded with this measurement assuming a small neutralino mass.

\paragraph{$\DMstopN > \mtop$.}
In this kinematic region, the decay $\tone \rightarrow t \neut$ (see Figure~\ref{fig:tN1tN1}) is dominant. 
The best results in this region 
are obtained by a statistical combination of the results of the  multijet (\stopZeroLep) and  one-lepton (\stopOneLep) analyses. They both  have dedicated signal regions targeting this scenario and the expected sensitivity is comparable for the two analyses. The number of required  leptons makes the two signal regions mutually exclusive. 

\begin{figure}[htbp]
  \begin{center}
    \includegraphics[width=0.7\textwidth]{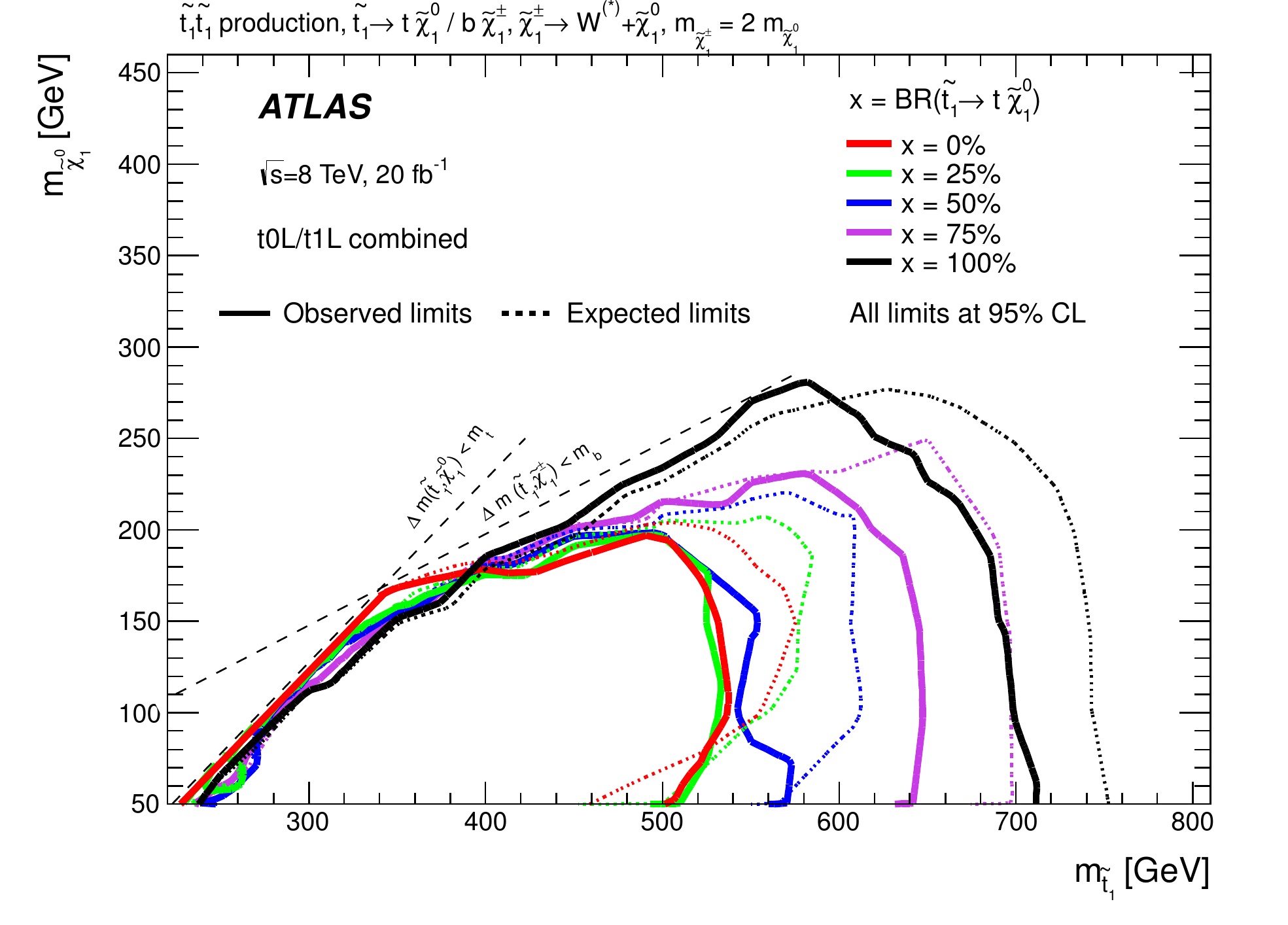}
    \caption{
       Combined exclusion limits assuming that the stop decays through $\tone \rightarrow t \neut$ with different branching ratios $x$ and through $\tone \rightarrow b \chipm$ with branching ratios $1-x$.  The limits assume $m_{\chipm} = 2 m_{\neut}$, and values of $x$ from $0\%$  to $100\%$ 
       are considered. For each branching ratio, the observed
       (with solid lines) and expected (with dashed lines) limits are shown.
    }
    \label{fig:exclusion_zero_one_lepton_combined_asymetric}
  \end{center}
\end{figure}

To maximise the sensitivity to the $\tone \rightarrow t \neut$ decays a statistical combination of the  \stopZeroLep\ and \stopOneLep\ signal regions is performed. The details of the combination are given in Appendix~\ref{sec:01lepCombine} and the final limit is shown in Figure~\ref{fig:tNsummary} by the largest shaded region (yellow). The expected limit on the stop mass is about 50 GeV higher at low $m_{\neut}$ than in the individual analyses. The observed limit is increased by roughly the same amount and stop masses between 200~GeV and 700~GeV are excluded for small neutralino masses.\footnote{This result holds if the top quark produced in the $\tone$ decay has a right-handed chirality. The dependence of the individual limits on the top quark chirality is discussed in Refs.~\cite{Aad:2014bva} and \cite{Aad:2014kra}.}

A similar combination is performed to target a scenario where the stop can decay as $\tone \rightarrow t \neut$ with branching ratio $x$ and as $\tone \rightarrow b \chipm$ with branching ratio $1-x$. Assuming gauge universality, the mass of the chargino is set to be twice that of the neutralino. Neutralino masses below 50 GeV are not considered, to take into account limits on the lightest chargino mass obtained at LEP~\cite{lep2, Heister:2003zk, Abdallah:2003xe, Acciarri:1999km, Abbiendi:2003sc}.  The exclusion limits are derived for $x= 75\%, 50\%, 25\%$ and 0\%.\footnote{A value of $x = 0\%$ is in fact not achievable in a  real supersymmetric model. Nevertheless, this value has been considered as the limiting case of a simplified model.} Regardless of the branching ratio considered, it is always assumed that $m_{\tone} > \mtop + m_{\neut}$ and $m_{\tone} > m_b + m_{\chipm}$, such that the two decays $\stop \rightarrow t \neut$ and $\stop \rightarrow b \chipm$ are both kinematically allowed. A statistical combination, identical to the one described above, is used for $x = 75\%$. For smaller values of $x$, no combined fit is performed, as the sensitivity is dominated by the \stopOneLep\ analysis almost everywhere: rather either the \stopZeroLep\ or the \stopOneLep\ analysis is used, depending which one gives the smaller expected \cls\ value.

Figure~\ref{fig:exclusion_zero_one_lepton_combined_asymetric} shows the result of the combination in the $m_{\tone}-m_{\neut}$ plane. The limit is improved, with respect to the individual analyses, by about 50 GeV for $m_{\neut}= 50$ GeV and $x = 75\%$. For other $x$ values, the \stopOneLep\ analysis is used on the full plane, with the exception of the point at the highest stop mass for  $m_{\neut}= 50$ GeV at $x=50\%$ and 25\%. Stop masses below 500~GeV are excluded for $m_{\neut}<160$~GeV for any value of $x$.

\begin{figure}[h!tbp]
  \begin{center}
 \subfloat[ \label{fig:bCsummary_dmCN}]{
        \includegraphics[width=0.49\textwidth]{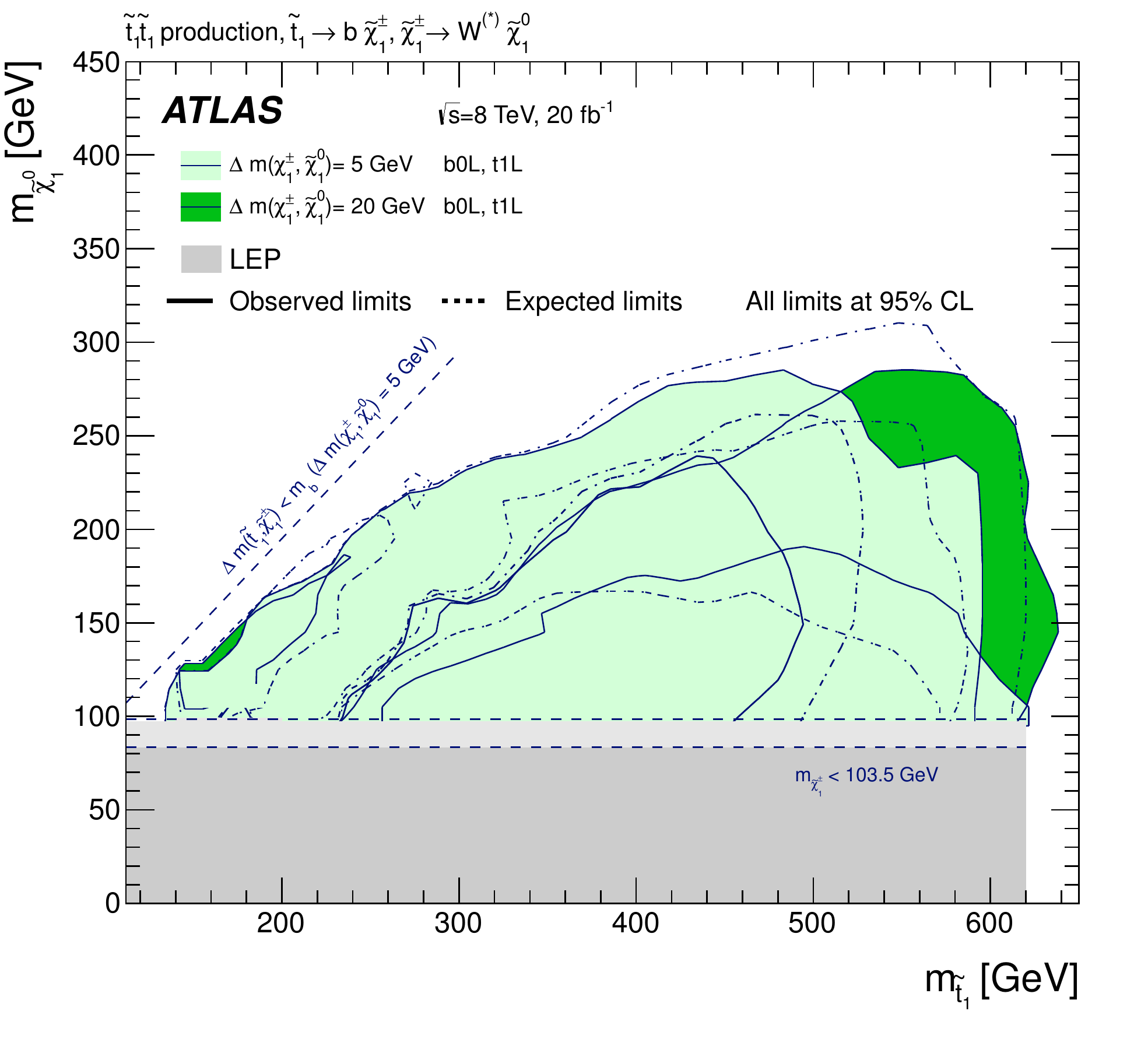}
        }  
         \subfloat[\label{fig:bCsummary_fixedmC}]{
        \includegraphics[width=0.49\textwidth]{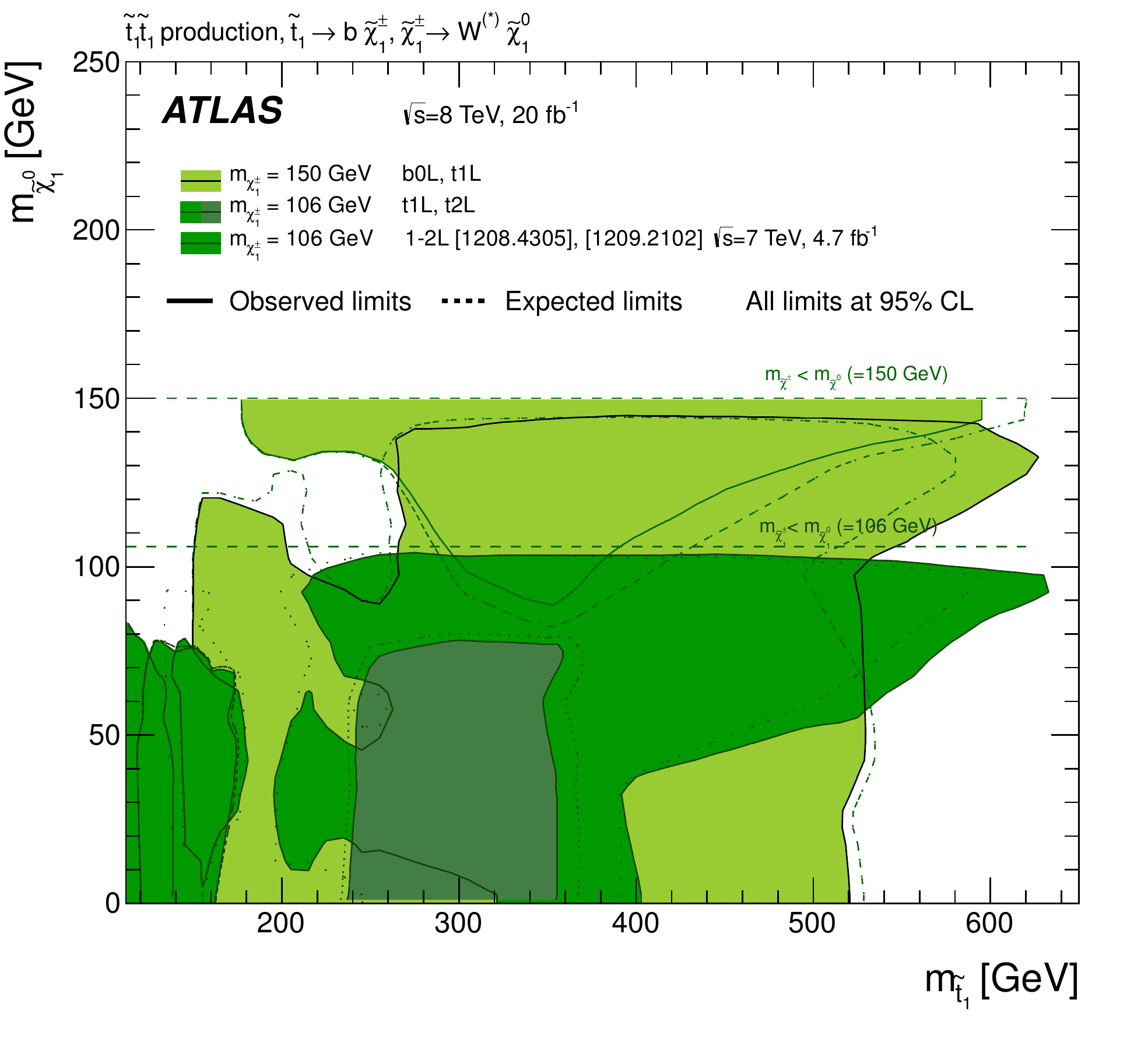}
        }\\
         \subfloat[\label{fig:bCsummary_univ}]{
        \includegraphics[width=0.49\textwidth]{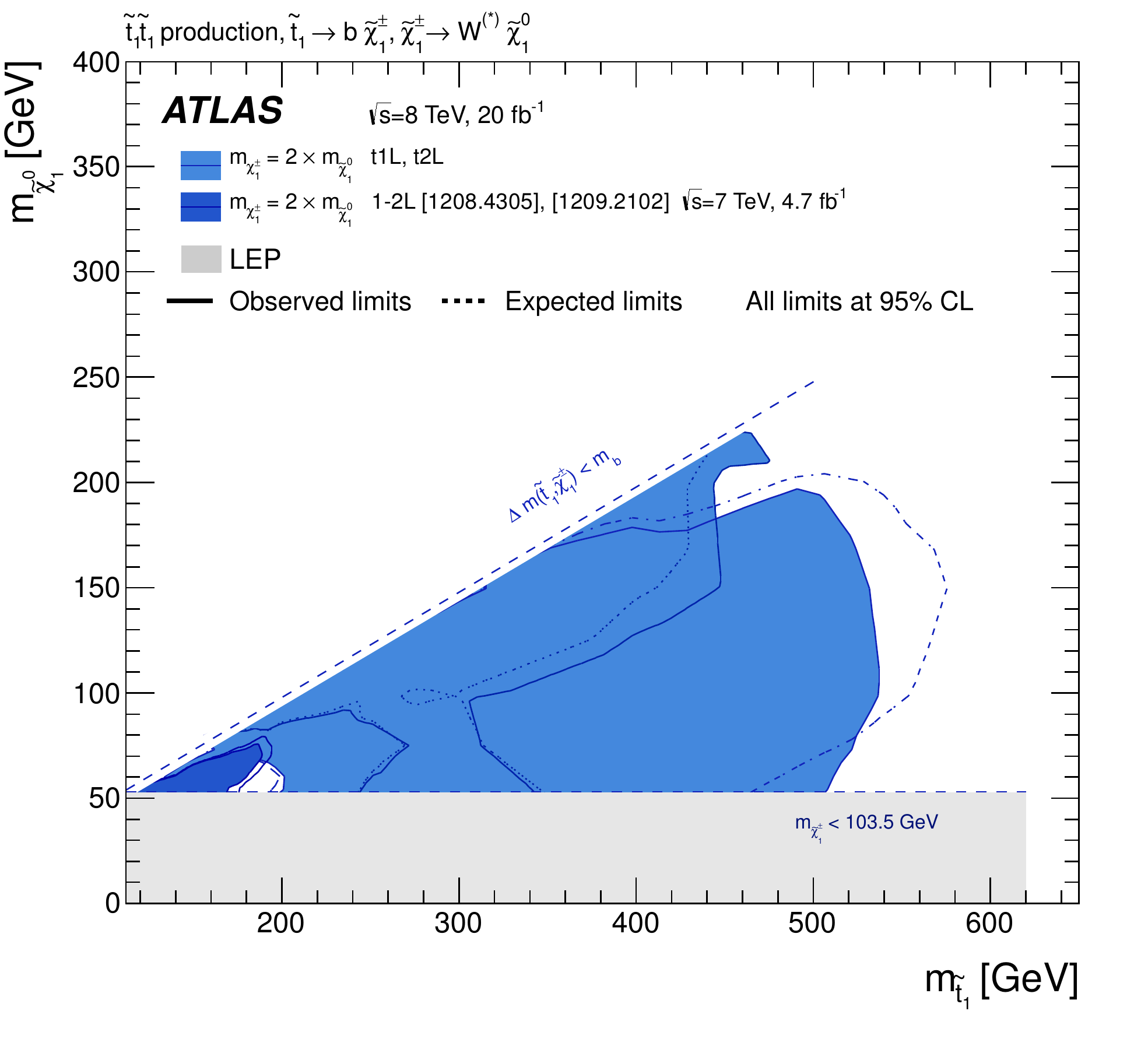}
        } 
         \subfloat[\label{fig:bCsummary_dmtC}]{
        \includegraphics[width=0.49\textwidth]{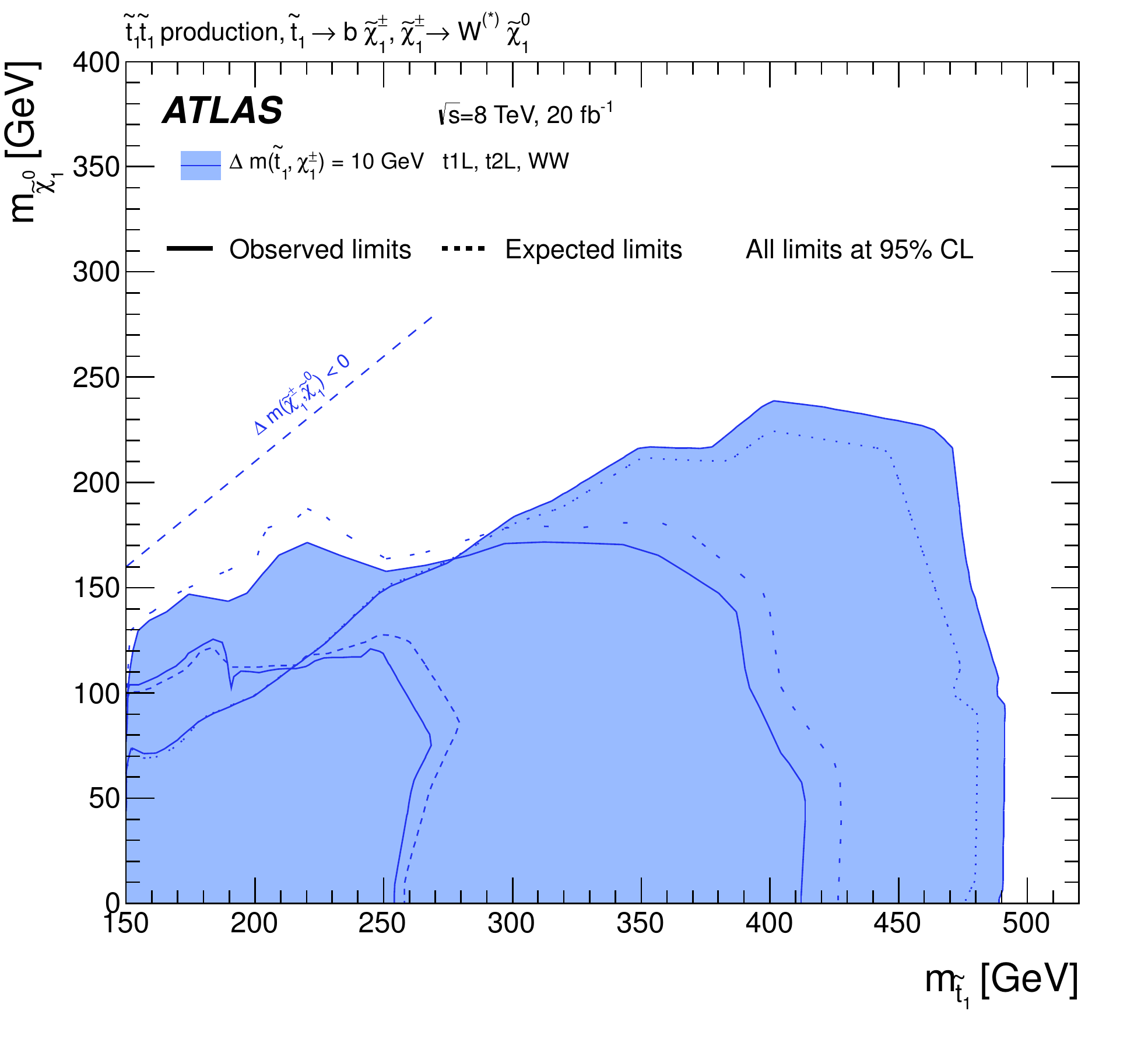}
        }
    \caption{Summary of the ATLAS Run 1 searches for direct stop pair production in models where the decay mode $\tone \rightarrow b \chipm$ with $\chipm\rightarrow W^{*} \neut$ is assumed with a branching ratio of 100\%. Various hypotheses on the $\tone$, $\chipm$, and $\neut$ mass hierarchy are used. Exclusion limits at 95\% CL are shown in the $\tone-\neut$ mass plane. The dashed and solid lines show the expected and observed limits, respectively, including all uncertainties except the theoretical signal cross-section uncertainty (PDF and scale). Wherever not superseded by any $\sqrt{s}=8$ TeV analysis, results obtained by analyses using 4.7 \ifb\ of proton--proton collision data taken at $\sqrt{s} = 7$ TeV are also shown, with the corresponding reference. The four plots correspond to interpretations of (a) the \sbottom\ and \stopOneLep\ soft-lepton analyses in two scenarios ($\dmCN = 5\ \GeV$ in light green and  $\dmCN = 20\ \mathrm{GeV}$ in dark green), for a total of four limits; (b) the \sbottom, \stopOneLep\ and \stopTwoLep\ analyses in scenarios with a fixed chargino mass $m_{\chipm} = 106$~GeV (dark green) and  $m_{\chipm} = 150$ GeV (light green); (c) the \stopOneLep\ and \stopTwoLep\ analyses in scenarios with $m_{\chipm} = 2 m_{\neut}$; (d) interpretations of the \stopOneLep, \stopTwoLep\ and \WWlike\ analyses in senarios with $\Delta m \left(\tone, \chipm \right) = 10\ \mathrm{GeV}$. }
    \label{fig:bCsummary}
  \end{center}
\end{figure}

\subsection{Stop decays with a chargino in the decay chain}

In the pMSSM, unless the higgsino--gaugino mass parameters are related by $M_1 \ll \mu, M_2$, the mass difference between the lightest neutralino  and the lightest chargino cannot be too large. The mass hierarchy $m_{\neut} < m_{\chipm} < m_{\tone}$ is, hence, well motivated, leading to the decay chain shown in Figure~\ref{fig:feynDiag_bC}.  

If additional particles beside the stop and the lightest neutralino take part in the stop decay, the stop phenomenology quickly becomes complex. Even if the chargino is the only other relevant SUSY particle, the stop phenomenology depends on the chargino mass, on the stop left-right mixing, and on the composition of the neutralino and chargino in terms of bino, wino and higgsino states.

Figure~\ref{fig:bCsummary} shows the exclusion limits obtained by the analyses listed in Table~\ref{tab:SRname} and discussed in Appendix~\ref{sec:SRs} if a branching ratio of 100\% for $\stop \rightarrow b \chipm$ is assumed. The exclusion limits are presented in a number of $m_{\tone}$--$m_{\neut}$ planes, each characterised by a different hypothesis on the chargino mass. For all scenarios considered, the chargino is assumed to decay as $\chipm \rightarrow W^{(*)} \neut$, where the $(*)$ indicates a possibly virtual $W$ boson.

\paragraph{$\dmCN =  5, 20$ GeV.} This scenario assumes that the difference in mass between the lightest chargino and the neutralino is small (Figure~\ref{fig:bCsummary_dmCN}), which  is a rather common feature of models where, for example, the LSP has a large wino or higgsino component. Two hypotheses have been considered, with $\dmCN =  5$ GeV and $\dmCN = 20$ GeV. For both, the complete decay chain is $\tone \rightarrow b \chipm \rightarrow b f f' \neut$, where the transverse momenta of the fermions $f$ and $f'$ depend on $\dmCN$ and on the stop mass, given the dependency on the chargino boost. If $\dmCN = 5$ GeV, the fermions have  momenta too low to be efficiently reconstructed. The observed final state then consists of two \bjets and $\met$. This final state is the direct target of the \sbottom\ signal regions. For $\dmCN = 20$ GeV, the signal efficiencies of the \sbottom\ signal regions decrease because of the lepton and jet veto applied. The \stopOneLep\ signal regions with soft leptons, instead, gain in sensitivity, profiting from the higher transverse momentum of the fermions from the off-shell $W$ decay produced in the chargino decay.

\paragraph{$m_{\chipm} = 106, 150$ GeV.} This scenario (Figure~\ref{fig:bCsummary_fixedmC}) assumes a fixed chargino mass. The SR yielding the lowest expected exclusion \cls\ for this scenario depends on the value of $\dmCN$. For $\dmCN<$~20 \GeV, the \sbottom\ signal regions provide the best sensitivity; for larger values of $\dmCN$, the \stopOneLep\ and \stopTwoLep\ signal regions provide better sensitivity because of the
same mechanism as in the $\dmCN =  5, 20$ GeV scenario above.
The exclusion extends up to about 600 GeV  for small values of $\dmCN$. A region of the parameter space with $m_{\tone}$ up to about 260 GeV and  $m_{\neut}$ between 100 GeV and $m_{\chipm}$ is not yet excluded.

\paragraph{$m_{\chipm} = 2 m_{\neut}$.} Inspired by gauge-universality considerations, the third scenario (Figure~\ref{fig:bCsummary_univ}) is characterised by a relatively large $\dmCN$. The \stopTwoLep\ signal regions dominate the sensitivity for $m_{\tone} \sim m_{\chipm}$. The sensitivity of the dedicated \stopOneLep-bC is dominant in a large region of the plane, and determines the exclusion reach for moderate to large values of $\DMstopN$.

\paragraph{$\dmTC = 10$ GeV.} The fourth scenario (Figure~\ref{fig:bCsummary_dmtC}) assumes a rather compressed $\tone-\chipm$ spectrum. The region at low $m_{\tone}$ and large $m_{\neut}$ is characterised by low mass separations between all particles involved, and it is best covered by the \stopOneLep-bCc\_diag, the \stopOneLep\ soft lepton, and the \WWlike\ signal regions. At larger values of the stop mass, the leptons emitted in the $\chipm$ decay have larger $\pt$, and the \stopTwoLep\ signal regions provide the best sensitivity.

\paragraph{$m_{\tone} = 300$ GeV.}

\begin{figure}[htbp]
  \begin{center}
    \includegraphics[width=0.60\textwidth]{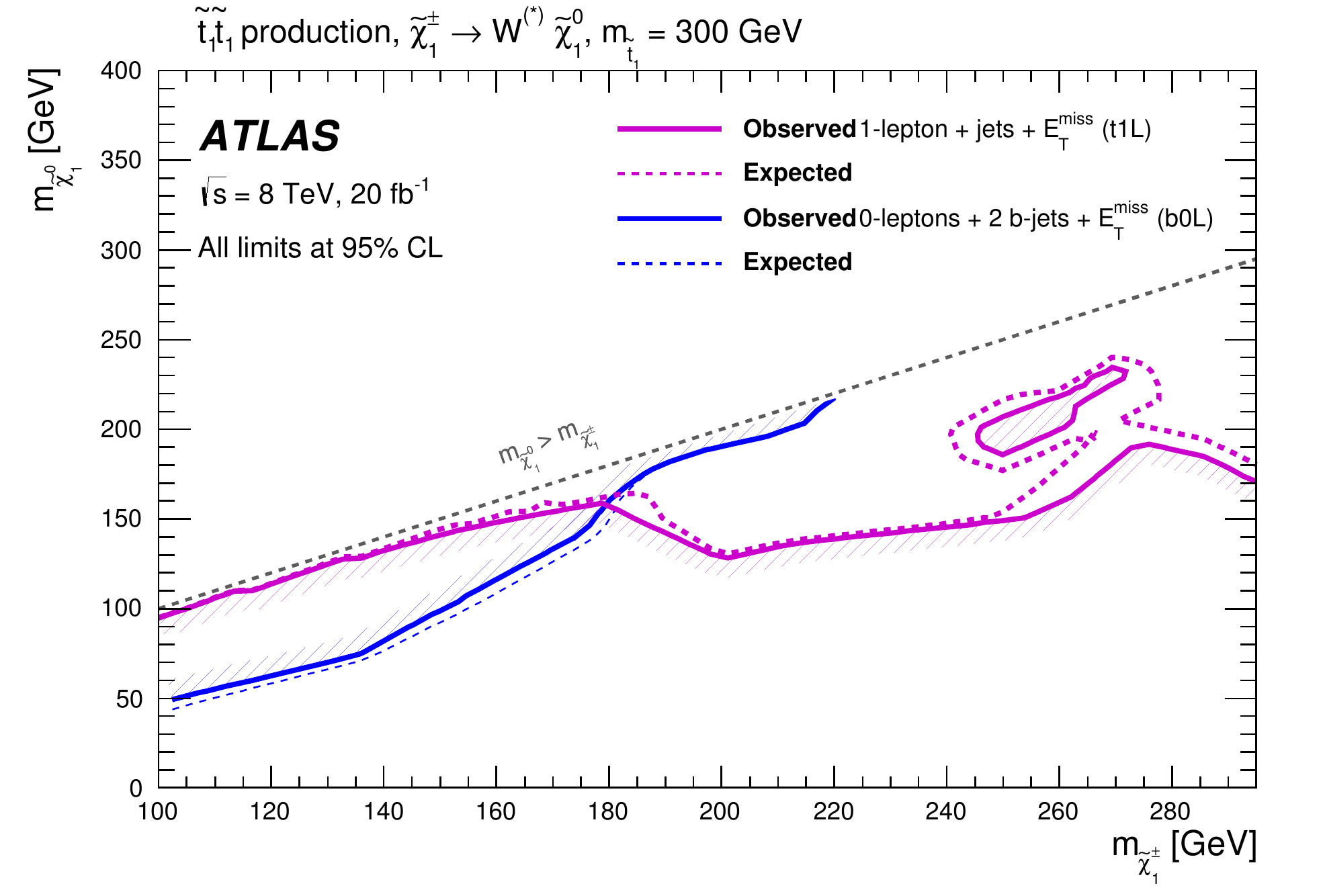}
    \caption{ 
 Exclusion limits assuming that the stop decays through $\tone \rightarrow b + \chipm \rightarrow b + W^{(\ast)} + \neut$ with branching ratio of 100\% assuming a fixed stop mass of $m_{\tone} = 300$ GeV. The region below  the purple line and above the blue line, indicated by a light shading, is excluded.}
       \label{fig:stopBCcombo}
  \end{center}
\end{figure}

The final scenario considered is one where the stop mass is fixed at 300 GeV, and the exclusion limits are expressed in the $m_{\chipm}$--$m_{\neut}$ plane. In the case of the compressed scenario, corresponding to a small mass difference \dmCN, the fermions from the $W^{(\ast)}$ decay can escape detection and only the two \bjets and \etmiss would be identified in the final state. Thus, the \sbottom\ signal regions are expected to have a large sensitivity in this case, while for larger values of \dmCN, the lepton can be observed, yielding a final-state signature investigated by the \stopOneLep\ 
soft-lepton signal region. A combination of the \sbottom\ and \stopOneLep\ signal regions is performed by choosing, for each point of the plane, the SR giving the lowest \cls\ for expected exclusion. The
 result, reported in Figure~\ref{fig:stopBCcombo}, shows that a large portion of the plane is excluded, with the exception of a region where the mass separations between the $\tone$, the $\chipm$ and the $\neut$ are small.

Summarising, in the simplified models with $\tone \rightarrow b \chipm \rightarrow b W^{(*)} \neut$, stop masses
up to 450--600 GeV are generally excluded. Scenarios where
 $\dmTN$ is small are particularly difficult to exclude and in these compressed scenarios, stop masses as low as 200 GeV are
still allowed (Figure~\ref{fig:bCsummary_fixedmC}). A small unexcluded area is also left for a small region around $\left(m_{\tone},m_{\chipm},m_{\neut}\right)= \left(180,100,50\right)$ GeV (Figure~\ref{fig:bCsummary_univ}), where the sensitivity
of the analyses is poor because the signal kinematics are similar to SM $\ttbar$ production.

\subsection{Limits on pair production of $\ttwo$}
\label{sec:stop2}

Although the pair production of $\tone$ has a cross section larger than that of $\ttwo$, and although the decay patterns of the two particles can be similar, it can be convenient to search for the latter in regions where the sensitivity to the former is limited. This is the case, for example, in the region where $\DMstopN \sim \mtop$ of Figure~\ref{fig:tNsummary}, where the separation of $\tone$ pair production from SM top quark pair production is difficult. The \stopTwo\ and \ttH\ analyses are designed to detect $\ttwo$ pair production in this region of the $m_{\tone}-m_{\neut}$ plane, followed by the decays $\ttwo \rightarrow \tone Z$ and $\ttwo \rightarrow \tone h$. The Higgs boson $h$ is assumed to have a mass of 125 GeV and SM branching ratios.

The exclusion limits were first derived in a scenario in which the pair-produced $\ttwo$ decays either through $\ttwo \rightarrow Z \tone$ with a branching ratio of 100\% (Figure~\ref{fig:t2Z}), or through $\ttwo \rightarrow \higgs \tone$ (again with a branching ratio of 100\%; Figure~\ref{fig:t2H}).  In both cases, the $\tone$ is assumed to decay through $\tone \rightarrow t \neut$, and its mass is set to be 180~GeV above that of the neutralino (assumed to be the LSP), which is the region not excluded in Figure~\ref{fig:tNsummary}. The final state contains two top quarks, two neutralinos, and either two $Z$ or two $\higgs$ bosons.

  \begin{figure}[htb]
  \begin{center}
    \includegraphics[width=0.45\textwidth]{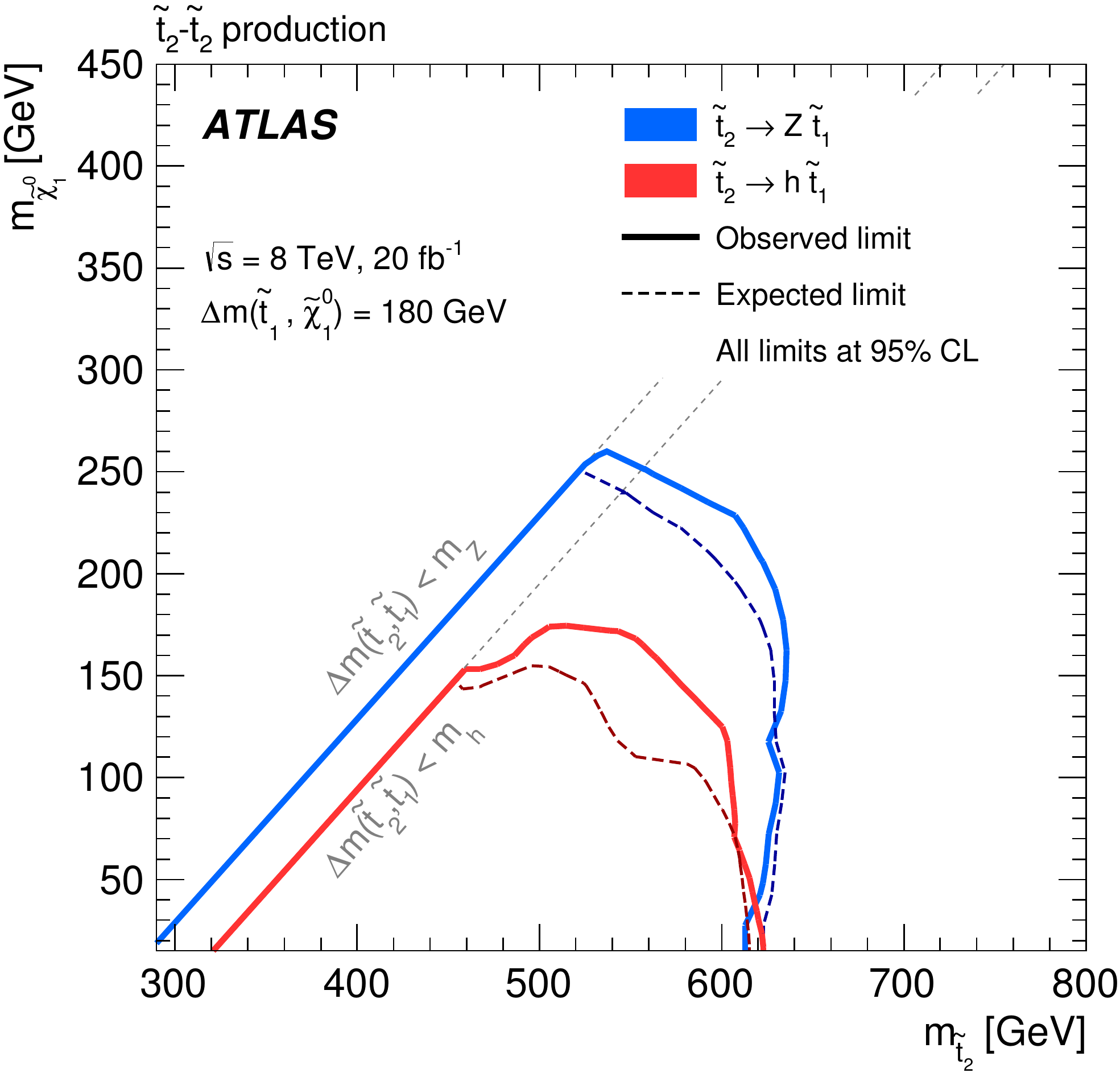}
    \caption{Exclusion limits at 95\% CL in the scenario where $\ttwo$ pair production is assumed, followed by the decay  $\ttwo \rightarrow Z \tone$ (blue) or  $\ttwo \rightarrow \tone \higgs$ (red) and then by $\tone \rightarrow t \neut$ with a branching ratio  of 100\%, as a function of the $\ttwo$ and $\neut$ mass. The $\tone$ mass is determined by the relation $m_{\tone} - m_{\neut} = 180$ GeV. The dashed lines indicate the expected limit and the solid lines indicate the observed limit.}
    \label{fig:stop2100}
  \end{center}
\end{figure}

Figure~\ref{fig:stop2100} shows the exclusion limits for the \ttH\ and the \stopTwo\   analyses. In both cases, a limit on $m_{\ttwo}$ is set at about 600 GeV for a massless neutralino. In the case of a $\ttwo$ decay through a Higgs boson, the limit covers neutralino masses lower than in the case of the decay through a $Z$ boson.

\begin{figure}[htb]
  \begin{center}
    \includegraphics[width=0.7\textwidth]{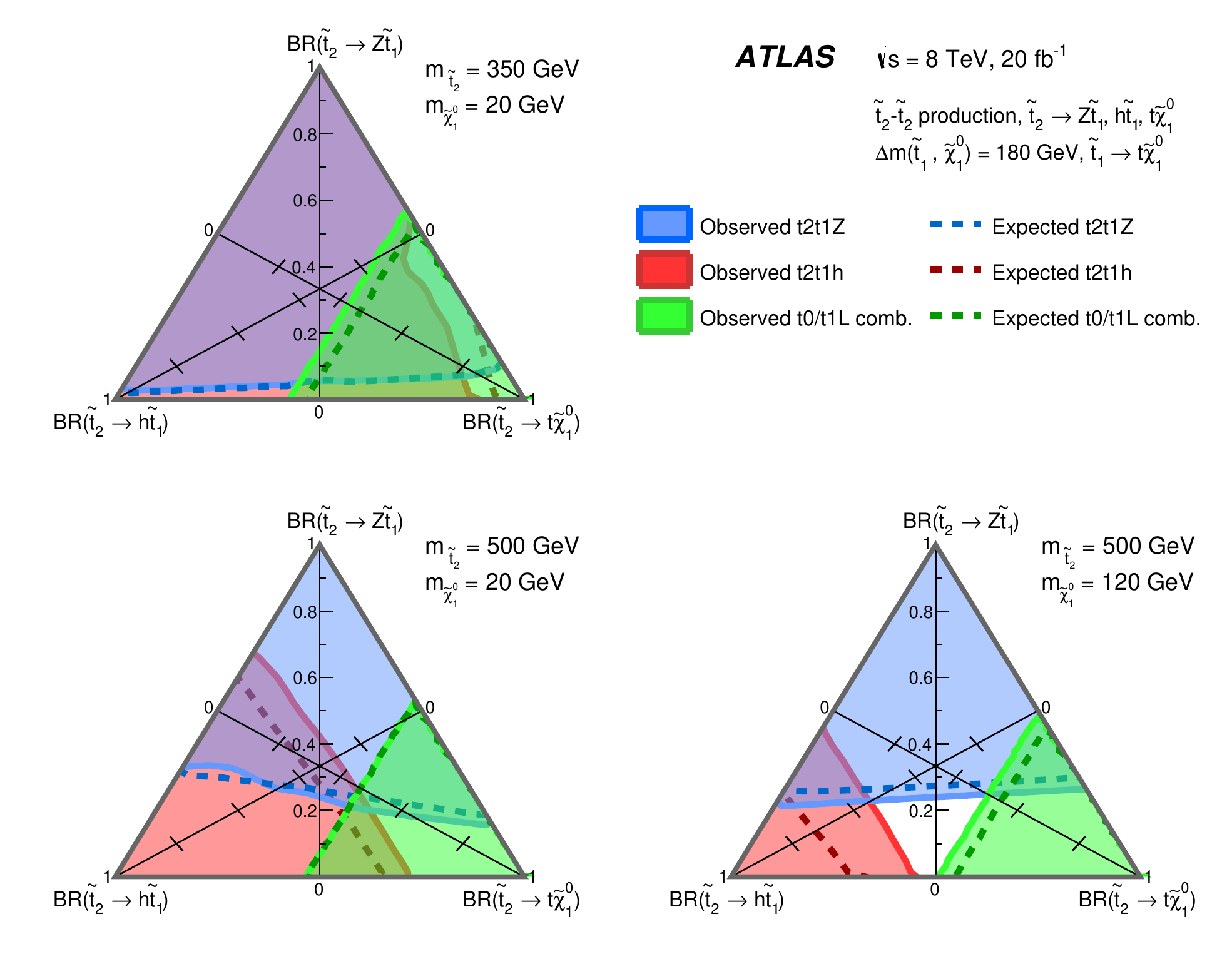}
    \caption{Exclusion limits as a function of the $\ttwo$ branching ratio for $\ttwo \rightarrow \tone \higgs$, $\ttwo \rightarrow \tone Z$ and $\ttwo \rightarrow t \neut$. The blue, red and green limit refers to the \stopTwo, \ttH\ and combination of \stopZeroLep\ and \stopOneLep\ analyses respectively. The limits are given for three different values of the $\ttwo$ and $\neut$ masses.}
    \label{fig:stop2_triangle}
  \end{center}
\end{figure}

The assumption on the branching ratio of the $\ttwo$ has also been relaxed, and limits have been derived assuming that the three decays $\ttwo \rightarrow Z \tone$, $\ttwo \rightarrow \higgs \tone$ and $\ttwo \rightarrow t \neut$ (Figure~\ref{fig:t2t}) are the only possible ones. The limits are shown  in Figure~\ref{fig:stop2_triangle} as a function of the three BRs, for different combinations of the $\ttwo$\ and $\neut$\ masses. Three analyses have been considered: the \stopTwo, \ttH\  and the combination of the \stopZeroLep\ and \stopOneLep\ discussed in Section~\ref{sec:01lep_summary}.\footnote{For the combination of the \stopZeroLep\ and \stopOneLep\ analyses, the limits extracted for the $\tone \rightarrow t \neut$ decay with branching ratio of 100\% have simply been rescaled by appropriate factors depending on the branching ratio of $\ttwo \rightarrow t \neut$ considered here.} The three analyses have complementary sensitivities. Together, they exclude $\ttwo$ pair production with a mass of 350 GeV  and 500 GeV for $m_{\neut} = 20$ GeV. A non-excluded region appears for $m_{\ttwo} = 500$ GeV if larger $\neut$ masses are considered.

\subsection{Sbottom decays}

Under the assumption that no supersymmetric particle takes part in the sbottom decay apart from the lightest neutralino, the sbottom decays as $\bone \rightarrow b \neut$ with a branching ratio of 100\% (Figure~\ref{fig:bN1bN1}). The final state arising from sbottom pair production hence contains two \bjets\ and \etmiss. The \sbottom\ signal regions were explicitly optimised to be sensitive to this scenario. In case of a mass degeneracy between the sbottom and the neutralino, the general consideration that the monojet-like \stopCharm-M selection is almost insensitive to the details of the decay of the produced particles still holds:
the \stopCharm-M signal regions offer the best sensitivity for scenarios where $m_{\bone} \sim m_{\neut}$.  

\begin{figure}[htbp]
  \begin{center}
    \includegraphics[width=0.6\textwidth]{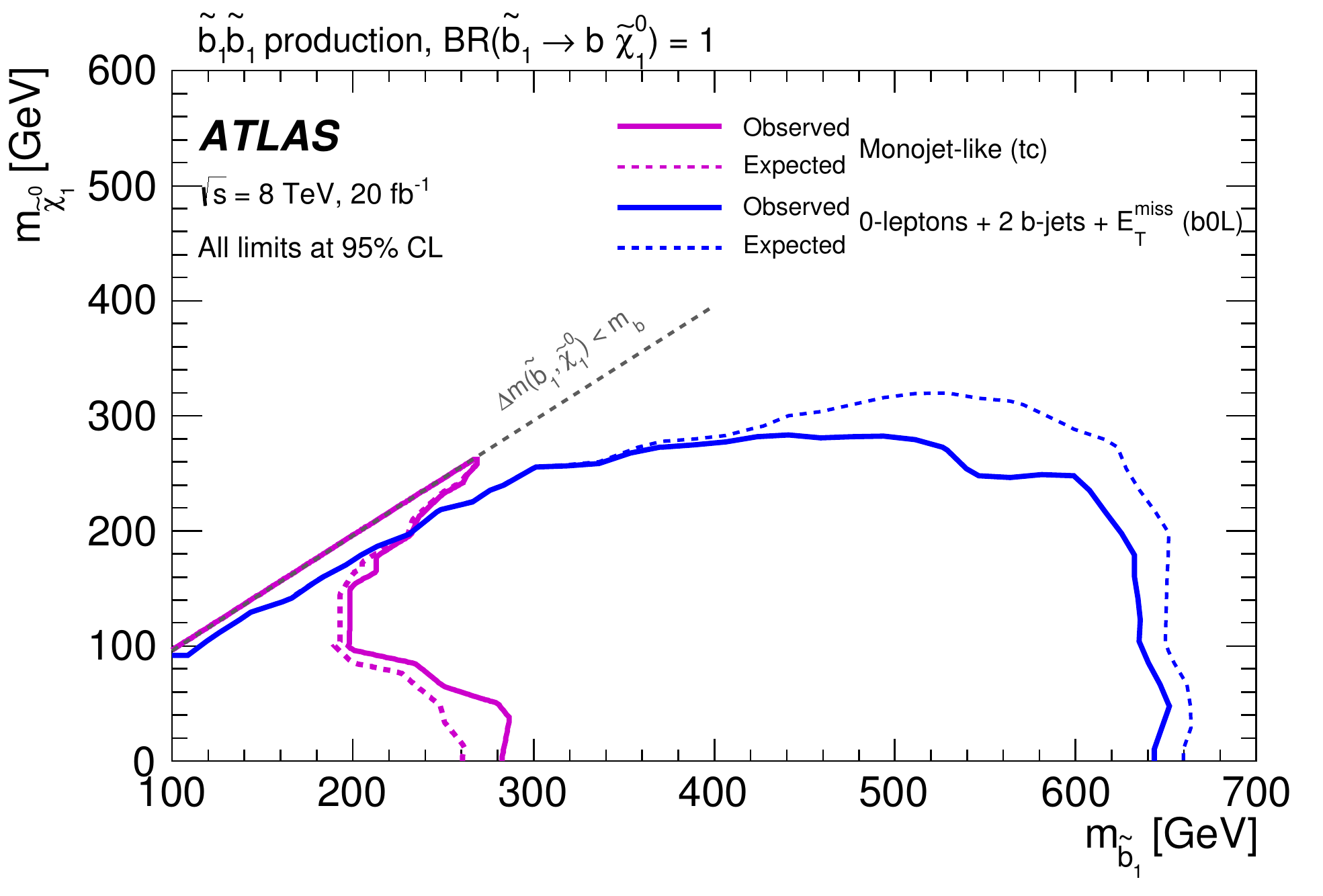}
	\caption{Observed (solid lines) and expected (dashed lines) 95\% CL limits on sbottom pair production where the sbottom is assumed to decay as $\bone \rightarrow b \neut$ with a branching ratio of 100\%. The purple lines refer to the limit of the \stopCharm\ analysis, while the blue lines refer to the \sbottom\ analysis.}

    \label{fig:sbottom_bneut}
  \end{center}
\end{figure}

Figure~\ref{fig:sbottom_bneut} shows the limits of the \stopCharm\ and \sbottom\ analyses on the $m_{\bone} - m_{\neut}$ plane. The monojet-like (\stopCharm-M) SRs  exclude models up to a value of $m_{\bone} \sim m_{\neut} \sim 280\ \GeV$. Sbottom masses are excluded up to about 600 GeV for neutralino masses below about 250 GeV. 

If other supersymmetric particles enter into the decay chain, then multiple decay channels would be allowed. Similarly to the stop, the case in which other neutralinos or charginos have a mass below the sbottom is well motivated. The branching ratios of the sbottom to the different decay channels depend on the supersymmetric particle mass hierarchy, on the mixing of the left-right components of the sbottom, and on the composition of the charginos and neutralinos in terms of bino, wino, and higgsino states.

An exclusion limit is derived under the assumption that the sbottom decays with a branching ratio of 100\% into $\bone \rightarrow t \chipm$ (Figure~\ref{fig:feynDiag_bone_tC}). The chargino is assumed to decay through $\chipm \rightarrow W^{(*)} \neut$ with a branching ratio of 100\%. The final state is a complex one, and offers many handles for background rejection: it potentially contains up to ten jets, two \bjets, and up to four leptons. The limits of Figure~\ref{fig:sbottom_tchipm}, shown in the $m_{\bone}-m_{\neut}$ plane, were obtained by using the three-lepton signal regions \sslll, either fixing the mass of the neutralino to $m_{\neut}=60$~GeV or by making the assumption that $m_{\chipm}= 2 m_{\neut}$. In the two scenarios considered, sbottom masses up to about 440 GeV are excluded, with a mild dependency on the neutralino mass.

\begin{figure}[htb]
  \begin{center}
\subfloat[\label{fig:sbottom_tchipm}]{    
\includegraphics[width=0.49\textwidth]{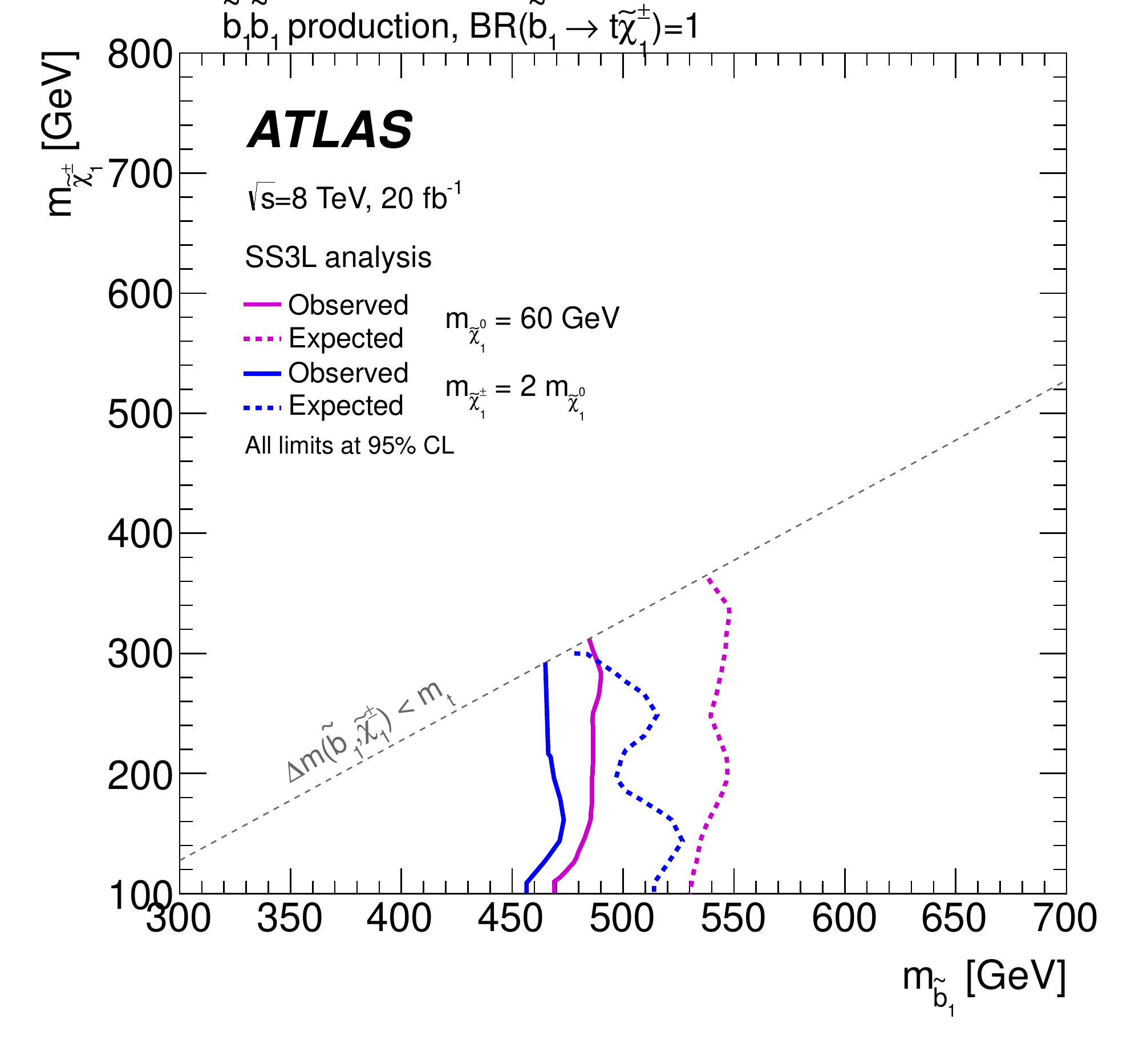}
}
\subfloat[\label{fig:sbottom_neut2}]{
 \includegraphics[width=0.49\textwidth]{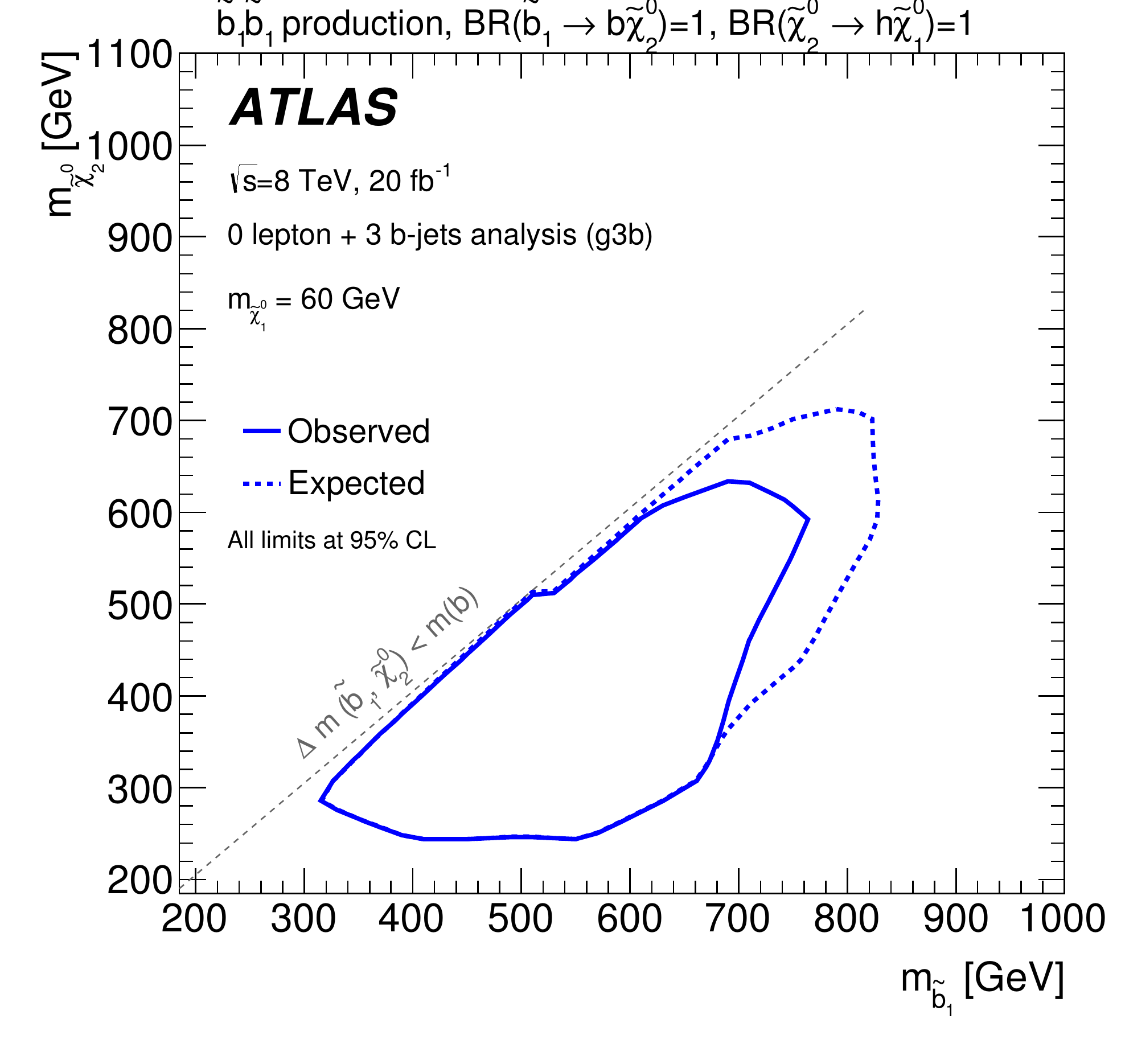}
 }
    \caption{Exclusion limits at 95\% CL for a scenario where sbottoms are pair produced and decay as (a) $\bone \rightarrow t \chipm$ with a BR of 100\% or (b) $\bone \rightarrow b\neuttwo$ with a BR of 100\%. The signal regions used in (a) are the \sslll, and two different models are considered: a fixed neutralino mass of 60 GeV (in purple) or $m_{\chipm} = 2 m_{\neut}$ (in blue). The limits are shown in the $m_{\bone}$--$m_{\chipm}$ plane. The signal regions used in (b) are the \threeb-SR-0j. A fixed neutralino mass of 60 GeV is assumed, and the limit is shown in the $m_{\bone}$--$m_{\neuttwo}$ plane. }
  \end{center}
\end{figure}

The last case considered is one where the pair-produced sbottoms decay through $\bone \rightarrow b \neuttwo$, followed by the decay of $\neuttwo$ into a $\neut$ and a SM-like Higgs boson $\higgs$ (Figure~\ref{fig:feynDiag_bone_bN2}). The final state contains up to six \bjets, four of which are produced by the two Higgs bosons decays. Since multiple \bjets\ are present in the final state, the three-\bjets signal regions (\threeb)  are used to place limits in this model. 

The limit, derived as a function of $m_{\bone}$ and $m_{\neuttwo}$ assuming a fixed neutralino mass of $\neut = 60$ GeV, is shown in Figure~\ref{fig:sbottom_neut2}. Sbottom masses between about 300 and 650 GeV are excluded for $\neuttwo$ masses above 250 GeV.

\section{Interpretations in pMSSM models}
\label{sec:pMSSM}

The interpretation of the results in simplified models is useful to assess the sensitivity of each signal region to a specific topology. However, this approach fails to test signal regions on the complexity of the stop and sbottom phenomenology that appears in a realistic  SUSY model. To this extent, the signal regions are used to derive exclusion limits in the context of specific pMSSM models.

The pMSSM~\cite{Djouadi:1998di} is obtained from the more general MSSM by making assumptions based on experimental results: 

\begin{itemize}
\item[-] No new source of CP violation beyond the Standard Model. New sources of CP violation are constrained by experimental limits on the electron and neutron electric dipole moments.
\item[-] No flavour-changing neutral currents. This is implemented by requiring that the matrices for the sfermion masses and trilinear couplings are diagonal.   
\item[-] First- and second-generation universality. The soft-SUSY-breaking mass parameters  and the trilinear couplings for the first and second generation are assumed to be the same based on experimental data from, e.g., the neutral kaon system~\cite{Ciuchini:1998ix}.
\end{itemize} 

With the above assumptions, and with the choice of a neutralino as the LSP, the pMSSM adds 19 free parameters on top of those of the SM. The complete set of pMSSM parameters is shown in Table~\ref{tab:pMSSM_par}.
\begin{table}[h]
\begin{center}
\caption{Description of the 19 additional parameters of the pMSSM model with a neutralino LSP.  \label{tab:pMSSM_par}}
\vspace{0.5cm}
\begin{tabular}{c|c} \hline\hline
Parameter & Description\\
\hline
$m_{\tilde{u}R}, m_{\tilde{d}R},m_{\tilde{q}L1}, m_{\tilde{e}R},m_{\tilde{\ell}L1}$ & First- and second-generation common mass parameters \\
$m_{\tilde{b}R}, m_{\tilde{t}R},m_{\tilde{q}L3}, m_{\tilde{\tau}R},m_{\tilde{\ell}L3}$ & Third-generation mass parameters \\
$M_1, M_2, M_3$ & Gaugino mass parameters \\
$A_b, A_{\tau}, A_t$ & Trilinear couplings \\
$\mu, M_A$ & Higgs/higgsino mass parameters \\
$\tan \beta$ & Ratio of vacuum expectation values of the two Higgs doublets\\
\hline
\hline
\end{tabular}
\end{center}
\end{table}
A full assessment of the ATLAS sensitivity to a scan of the 19-parameters space has been performed in Ref.~\cite{Aad:2015baa}. Here, a set of additional hypotheses are made, to focus on the sensitivity to a specific, well-motivated set of models with enhanced third generation squark production: 

\begin{itemize}
\item[-] The common masses of the first- and second-generation squarks have been set to a multi-TeV scale, making these quarks irrelevant for the processes studied at the energies investigated in this paper. This choice is motivated by the absence of any signal from squark or gluino production in dedicated SUSY searches performed by the ATLAS~\cite{Aad:2014wea,Aad:2015mia,Aad:2013wta,Aad:2014pda,Aad:2014lra,Aad:2015gna} and CMS~\cite{Khachatryan:2015vra,Chatrchyan:2013fea,Khachatryan:2015pwa,CMS:2014dpa,Chatrchyan:2014lfa,Chatrchyan:2013iqa,Chatrchyan:2013wxa,Chatrchyan:2013lya} collaborations. 
\item[-] All slepton mass parameters have been set to the same scale as the first- and second-generation squarks. This choice has no specific experimental or theoretical motivation, and should be regarded as an assumption. 
\item[-] A decoupling limit with $M_A =3$ TeV and large $\tan \beta$ values ($\tan \beta > 15$) has been assumed. This is partially motivated by results of the LHC searches for higher mass Higgs boson states~\cite{Aad:2014vgg,Khachatryan:2014wca}. 
\item[-] For $\tan \beta \gg 1$, the Higgs boson mass depends heavily on the product of the stop-mass parameters  $M_S = \sqrt{m_{\tone} m_{\ttwo}}$ and the mixing between the left- and right-handed states $X_t = A_t - \mu/\tan \beta$~\cite{Delgado:2012eu}. The stop sector is therefore completely fixed, given the Higgs boson mass, the value of $X_t$ and one of the two stop mass parameters\footnote{In particular, a minimum value of $M_S \sim 800$ GeV is allowed if the maximal mixing condition $X_t/M_S = \sqrt{6}$ is realised.}. 
\item[-] The trilinear couplings $A_b$  in the sbottom sector are found to have limited impact on the phenomenology, and are therefore set to zero. 
\item[-] The gluino mass parameter $M_3$ is set such to evade LHC constraints on gluino-pair production. 
\end{itemize} 

These assumptions reduce the number of additional free parameters of the model to the mass parameters of the electroweak sector ($\mu, M_1, M_2$) and two of the three third-generation squark mass parameters ($\mqlthree, \mtR, \mbR$). All the assumptions made either have a solid experimental basis, or are intended to simplify the interpretation in terms of direct production of stops and sbottoms (as, for example, the assumption on the slepton mass parameters).

Three types of models have been chosen, that, by implementing in different ways  constraints arising from naturalness arguments and the dark-matter relic density measurement, further reduce the number of parameters to be scanned over. They are described below, and summarised in Table~\ref{tab:pMSSM_models} together with additional information on the most relevant production and decay channels.


\begin{sidewaystable}
\begin{center}
\caption{Details of parameters scanned in the three pMSSM models used for interpretations. The settings of additional parameters that are fixed for each model are also given together with the main production and decay channels targeted. \label{tab:pMSSM_models}}
\vspace{0.5cm}
\begin{tabular}{l|c|c|c|c} \hline\hline
\multirow{2}{*}{Model name} & Parameters & Other parameter & Production & Typical \\
 & scanned & settings & channels & decays\\
\hline
\hline
\multirow{4}{*}{\parbox{2.5cm}{Naturalness-inspired pMSSM}} & $350\ \GeV < \mqlthree < 900\ \GeV$   & $M_2 = 3 \mu$   & $pp\rightarrow \tone \tone$  & For $\mu = 110$ GeV, $\mqlthree = 400$ GeV \\
&$100\ \GeV < \mu < \mqlthree - 150$ GeV  & $m_{\tright}$ such that $M_S = 800$ GeV & $pp\rightarrow \bone\bone$ & $\tone \rightarrow t\neut$ (33\%); $\tone \rightarrow t \neuttwo$ (36\%)\\
& & $A_t$ such that $X_t/M_S = \sqrt{6}$ & & $\tone \rightarrow b \chipm$ (26\%); $\bone \rightarrow t \chipm$ (70\%) \\
& &  & & $\bone \rightarrow b \neut$ (16\%); $\bone \rightarrow b \neuttwo$ (13\%)\\
 & &  & & \\
\hline
\multirow{8}{*}{\parbox{2.5cm}{Well-tempered neutralino pMSSM}} & $310\ \GeV < \mqlthree < 810\ \GeV$   & & $pp\rightarrow \tone \tone$  & For $M_1 = 110$ GeV, $\mqlthree = 410$ GeV \\
&$110\ \GeV < M_1 < \mqlthree - 50$ GeV  & & $pp\rightarrow \bone\bone$ & $\tone \rightarrow t\neuttwo$ (35\%); $\tone \rightarrow t \neutthree$ (38\%)\\
& &  $\mu \sim -M_1$   & & $\tone \rightarrow b \chipm$ (20\%); $\bone \rightarrow t \chipm$ (85\%) \\
& &  & & $\bone \rightarrow \tone W$ (6\%); $\bone \rightarrow b \neuttwo$ (4\%)\\
 & & Similar to Naturalness-inspired & & \\
\cline{2-2}\cline{4-5}
& $260\ \GeV < m_{\tright} < 760\ \GeV$& for $A_t$, $m_{\tright}$ or $\mqlthree$, $M_3$ & $pp\rightarrow \tone \tone$  & For $M_1 = 110$ GeV, $m_{\tright} = 410$ GeV \\
&$110\ \GeV < M_1 < \mqlthree - 50$ GeV  & &  & $\tone \rightarrow t\neuttwo$ (17\%); $\tone \rightarrow t \neutthree$ (19\%) \\
& & & &$\tone \rightarrow t \neut$ (6.7\%); $\tone \rightarrow b \chipm$ (57\%) \\
 & &  & & \\
\hline
\multirow{10}{*}{\parbox{2.5cm}{$\higgs/Z$-enriched pMSSM}} & $250\ \GeV < m_{\bright} < 750\ \GeV$   & $M_1 = 100$ GeV;  $M_2 = \mu$  & $pp\rightarrow \bone \bone$  & For $\mu = 300$ GeV, $m_{\bright} = 400$ GeV \\
&$100\ \GeV < \mu < m_{\bright}$  &  $m_{\tright} = 1.6$~TeV;  $\mqlthree = 1.2$~TeV & & $\bone \rightarrow b \neut$ (37\%); $\bone \rightarrow b \neuttwo$ (39\%)\\
 & & $A_t$ fixed by $m_{\higgs} \sim 125$~GeV & & $\bone \rightarrow b \neutthree$ (23\%) \\
 & &  & & $\neuttwo \rightarrow Z \neut$ (29\%); $\neuttwo \rightarrow \higgs \neut$ (71\%)\\
 & & & & $\neutthree \rightarrow Z \neut$ (85\%); $\neutthree \rightarrow \higgs \neut$ (15\%)\\
\cline{2-5}
& $500\ \GeV < \mqlthree < 800\ \GeV$&  $M_1 = 100$ GeV; $M_2 = 1$ TeV & $pp\rightarrow \tone \tone$  & For $\mu = 300$ GeV, $\mqlthree= 600$ GeV \\
&$100\ \GeV < M_1 < \mqlthree$ GeV  & $m_{\bright} = 3$ TeV; $m_{\tright} = 2$~TeV & $pp\rightarrow \bone \bone$ & $\tone \rightarrow t\neuttwo$ (46\%); $\tone \rightarrow t \neutthree$ (39\%) \\
&  & $A_t$ fixed by $m_{\higgs} \sim 125$~GeV  &  & $\tone \rightarrow b\chipm$ (11\%); $\bone \rightarrow t \chipm$ (87\%) \\
 & &  & & $\neuttwo \rightarrow Z \neut$ (24\%); $\neuttwo \rightarrow \higgs \neut$ (76\%)\\
 & & & & $\neutthree \rightarrow Z \neut$ (88\%); $\neutthree \rightarrow \higgs \neut$ (12\%)\\
\hline
\hline
\end{tabular}
\end{center}
\end{sidewaystable}

\paragraph{Naturalness-inspired pMSSM:}The model is inspired by naturalness criteria, which require a value of $\mu$ in the range of a few hundred GeV, favour stop masses below one TeV, place weak constraints on the gluino mass and give no constraints on the mass of other SUSY particles~\cite{Papucci:2011wy}. The exclusion limits are determined as a  function of the higgsino mass parameter $\mu$ and the left-handed squark mass parameter $\mqlthree$. The parameter $\mqlthree$ is scanned in the range $350 \ \GeV < \mqlthree < 900\ \GeV$. The parameter $\mu$ is scanned in the range $100 \ \GeV < \mu < \mqlthree -150\ \GeV$, where the lower bound is determined by limits on the chargino mass arising from LEP~\cite{lep2, Heister:2003zk, Abdallah:2003xe, Acciarri:1999km, Abbiendi:2003sc}. The right-handed stop mass parameter $\mtR$ and the stop mixing parameter $X_t$ are determined by choosing the maximal mixing scenario $X_t/M_S = \sqrt{6}$ and by the requirement of having a Higgs boson mass of about 125 GeV. The other squark and slepton masses, as well as the bino mass parameter $M_1$, are set to 3 TeV. The wino mass parameter $M_2$ is set such that $M_2 = 3\mu$. The gluino mass parameter $M_3$ is set to 1.7 TeV.

With this choice of the model parameters, the spectrum is characterised by two light neutralinos $\left(\neut,\neuttwo\right)$ and one chargino $\left( \chipm \right)$, all with masses of the order of $\mu$, a light $\bone$ with a mass of the order of $\mqlthree$, and a light $\tone$ with mass of the order of $\mqlthree$ up to $\mqlthree \sim 700$ GeV (the constraint on $M_S$ does not allow the mass of $\tone$ to increase beyond about 650 GeV). The production processes considered are direct pair production of $\bone$ and $\tone$ with similar masses. Because of the abundance of light higgsino states, many different decays can occur.

 \paragraph{Well-tempered neutralino pMSSM:} The models are designed to loosely satisfy dark-matter thermal-relic density constraints ($0.09 < \Omega_{\mathrm{c}} h^2 < 0.15$, where $h$ is the Hubble constant), while keeping fine tuning (defined as in Ref.~\cite{CahillRowley:2012rv}) to less than 1\%. The exclusion limits are determined as a function of $M_1$ and $\mqlthree$, or $M_1$ and $\mtR$, with $\mu \sim- M_1$ in both cases to satisfy the dark-matter constraints through the presence of well-tempered neutralinos~\cite{ArkaniHamed:2006mb}. The constraints on the Higgs boson mass are satisfied in a way similar to the naturalness-inspired pMSSM model above. All other parameters are the same as in the naturalness-inspired pMSSM model. These models tend to have three neutralinos and two charginos with masses lower than $\tone$ or $\bone$, giving rise to a diverse phenomenology. 
 
 \paragraph{$\higgs/Z$-enriched pMSSM:} These models are defined such that Higgs and $Z$ bosons are produced abundantly in 
the SUSY particles' decay chains. The assumption of $M_1 = 100$ GeV ensures the presence of a bino-like neutralino LSP, while $M_3 = 2.5$~TeV 
ensures that direct gluino production is highly suppressed compared to third-generation squark production. Two sets of models have been defined: in the first one, $\mu$ and the right-handed sbottom mass parameter $\mbR$ are scanned while keeping $M_2 = \mu$, $\mqlthree = 1.2$ TeV, $\mtR = 1.6$ TeV; in the second one, $\mu$ and $\mqlthree$ are scanned while keeping $M_2 = 1$ TeV, $\mbR = 3$ TeV, $\mtR = 2$ TeV. The former is dominated by sbottom pair production, while both sbottom and stop pair production are relevant for the latter. Stop mixing parameters are chosen with maximal mixing to satisfy Higgs boson mass constraints. 
In these models, the decays of the third generation squarks into the heavier neutralino states ($\neuttwo$ and $\neutthree$) are followed by decays to the lightest neutralino with the emission of a $Z$ or a $h$ boson. Typically the $\neuttwo$ ($\neutthree$) decays into a $Z$ boson 30\% (85\%) of the times, and into a Higgs boson 70\% (15\%) of the times. The subsequent decays of the Higgs boson into $b$-quark pairs (happening with the same branching ratio as in the Standard Model) lead to final states rich in $b$-jets.

\vspace {0.5cm}

Exclusion limits for these pMSSM models are determined by combining many of the SRs defined for the searches discussed in this paper (\stopZeroLep, \stopOneLep, \TBM,\footnote{The \TBM\ signal region, discussed in detail in Appendix~\ref{sec:tbmet}, implement a one-lepton selection, designed to be sensitive to final states containing a top quark, a $b$-quark and $\etmiss$. It complements the selections of the $\stopZeroLep$ and $\stopOneLep$ signal regions targeting $tt\etmiss$ final states.} \stopTwo, \threeb, \stopCharm). For each set of parameters the individual 95\% CL expected limit is evaluated. The combined exclusion contour is determined by choosing, for each model point, the signal region having the smallest expected \cls\ value of the test statistic for the signal-plus-background hypothesis.

\begin{figure}[!htb]
\begin{center}
\includegraphics[width=0.45\columnwidth,angle=0]{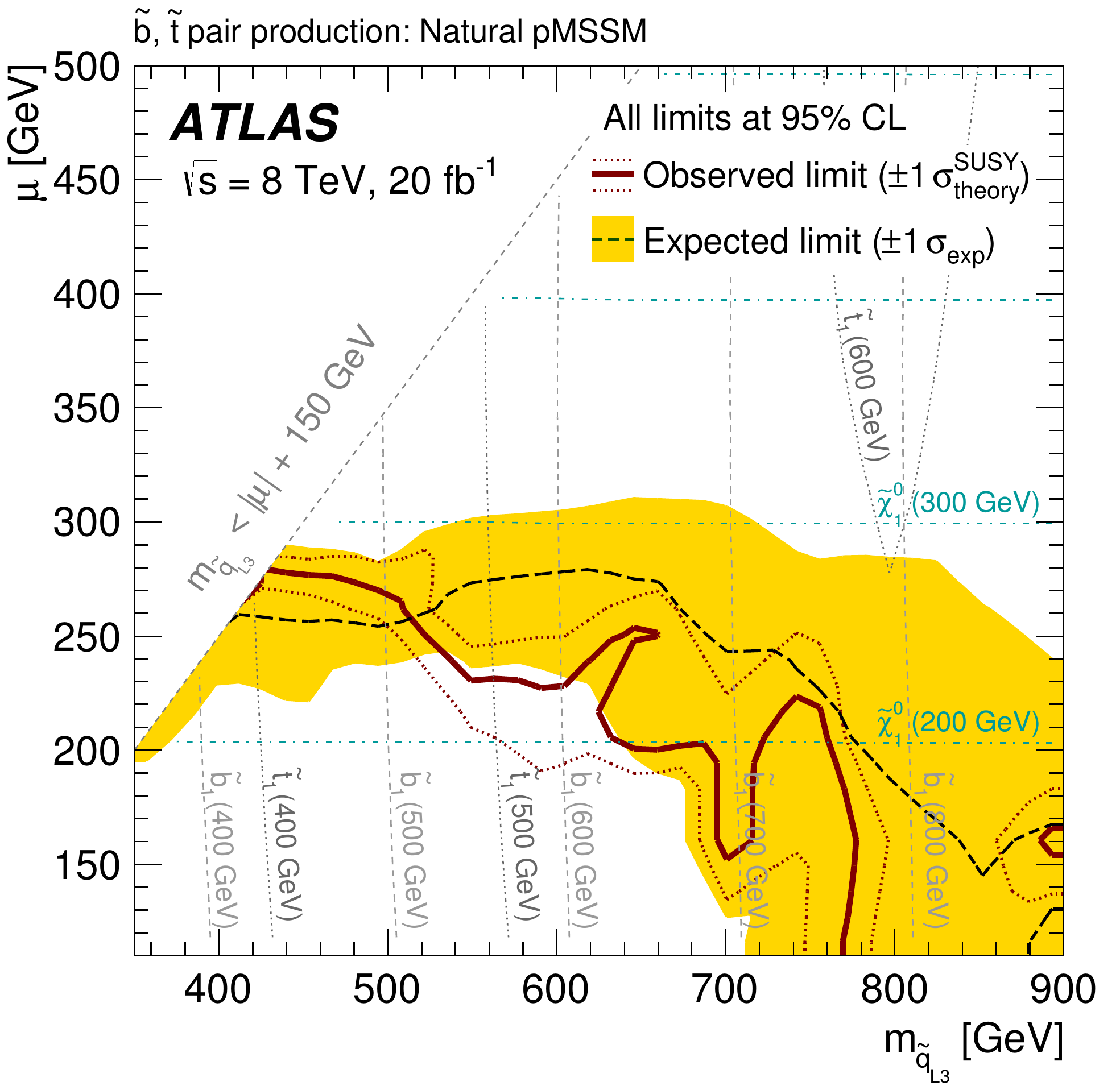}
\end{center}
\caption{Expected and observed 95\% CL exclusion limits for the naturalness-inspired set of pMSSM models from the combination \stopZeroLep, \stopOneLep\ and \TBM\ analyses using the signal region yielding the smallest \cls\ value for the signal-plus-background hypothesis. The dashed black line indicates the expected limit, and the yellow
      band indicates the $\pm 1\sigma$ uncertainties, which include all
      uncertainties except the theoretical uncertainties in the
      signal. The red solid line indicates the observed limit, and
      the red dotted lines indicate the sensitivity to
      $\pm 1\sigma$ variations of the signal theoretical
      uncertainties. The dashed and dotted grey lines indicate a constant value of the stop and sbottom masses, while the dashed light-blue line indicates a constant value of the neutralino mass.}
\label{fig:NatpMSSM}
\end{figure}

Figure~\ref{fig:NatpMSSM} shows the exclusion limit for the naturalness-inspired set of pMSSM models based on  the \stopZeroLep, \stopOneLep\ and \TBM\ analyses. The \stopZeroLep\ and \stopOneLep\ analyses have a similar expected sensitivity. 
These SRs were optimised assuming a 100\% BR for $\tone \rightarrow t \neut$ or $\tone \rightarrow b\chipm$, while for these pMSSM models, the stop decays to $\tone \rightarrow t \neut$, $\tone \rightarrow b \chipm$ and $\tone \rightarrow b \neuttwo$ with similar branching ratios (and the sbottom to both $\bone \rightarrow b \neut$ and $\bone \rightarrow t \chipm$).
The \TBM\ signal regions, discussed in detail in Appendix~\ref{sec:tbmet}, are designed to be sensitive to final states containing a top quark, a $b$-quark and missing transverse momentum and address such mixed-decay scenarios by requiring a lower jet multiplicity. 

 The signal regions that dominate the sensitivity are the \TBM, \stopZeroLep-SRC1 and  \stopOneLep-bCd\_bulk at low values of $\mqlthree$,  and \TBM, \stopZeroLep-SRA1, \stopZeroLep-SRA2 and \stopOneLep-tNbC\_mix at intermediate and high values of $\mqlthree$. The excluded region for models with $\mqlthree \sim 900$ GeV and $\mu \sim 150$ GeV is due to the saturation of $m_{\tone}$ at high $\mqlthree$ values: to satisfy the Higgs boson mass constraint requires $M_S \sim 800$ GeV, hence $m_{\tone}$ at  $\mqlthree \sim 900$ \GeV\ is smaller than that at $\mqlthree \sim 800$ \GeV. The large fluctuations of the observed limit with respect to the expected one are due to transitions between different signal regions providing the best expected exclusion in different regions of the plane.

\begin{figure}[!htb]
\begin{center}
\subfloat[\label{fig:DM_mql3}]{
	\includegraphics[width=0.45\columnwidth,angle=0]{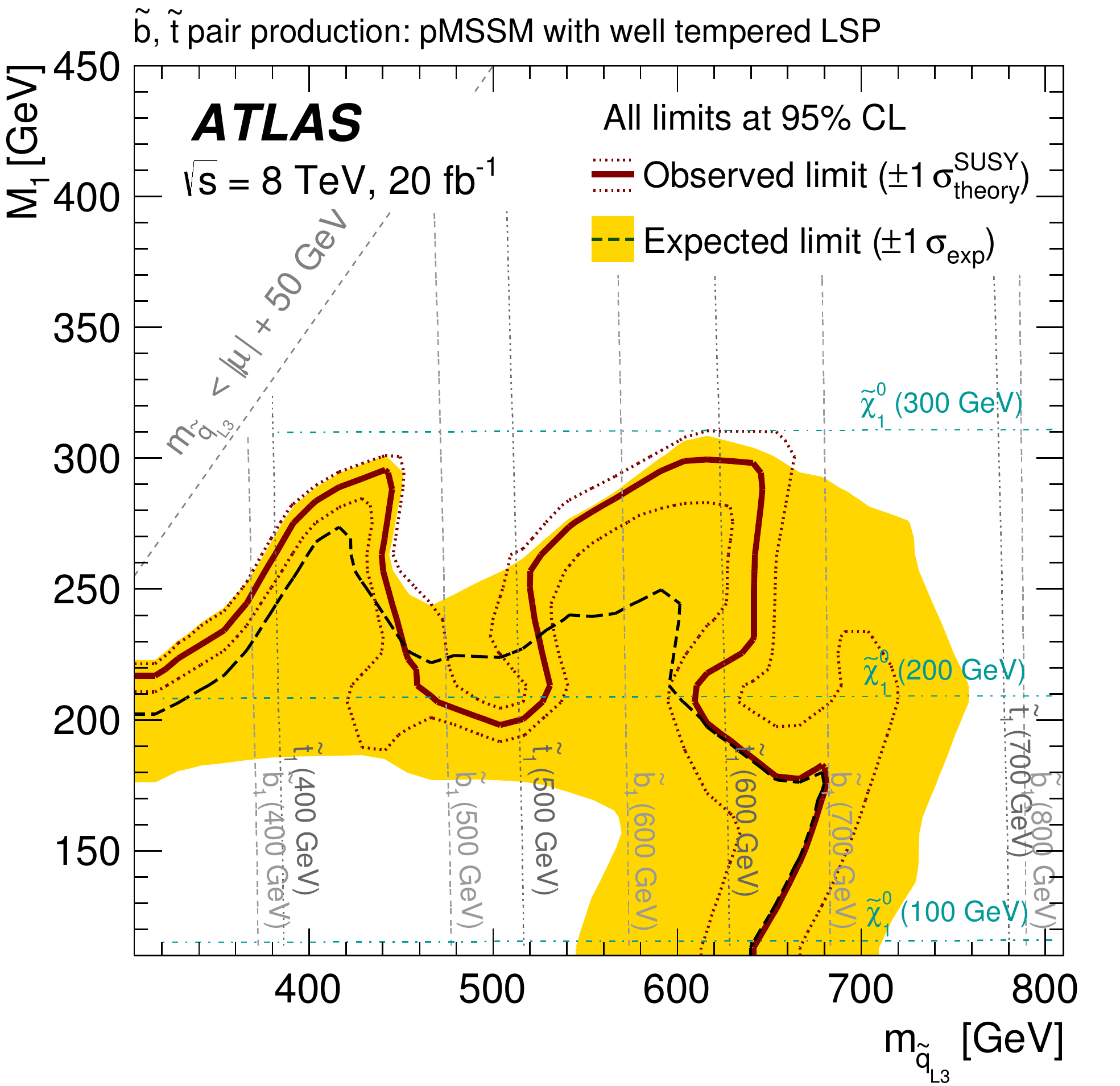}
}
\subfloat[\label{fig:DM_mtR}]{
	\includegraphics[width=0.45\columnwidth,angle=0]{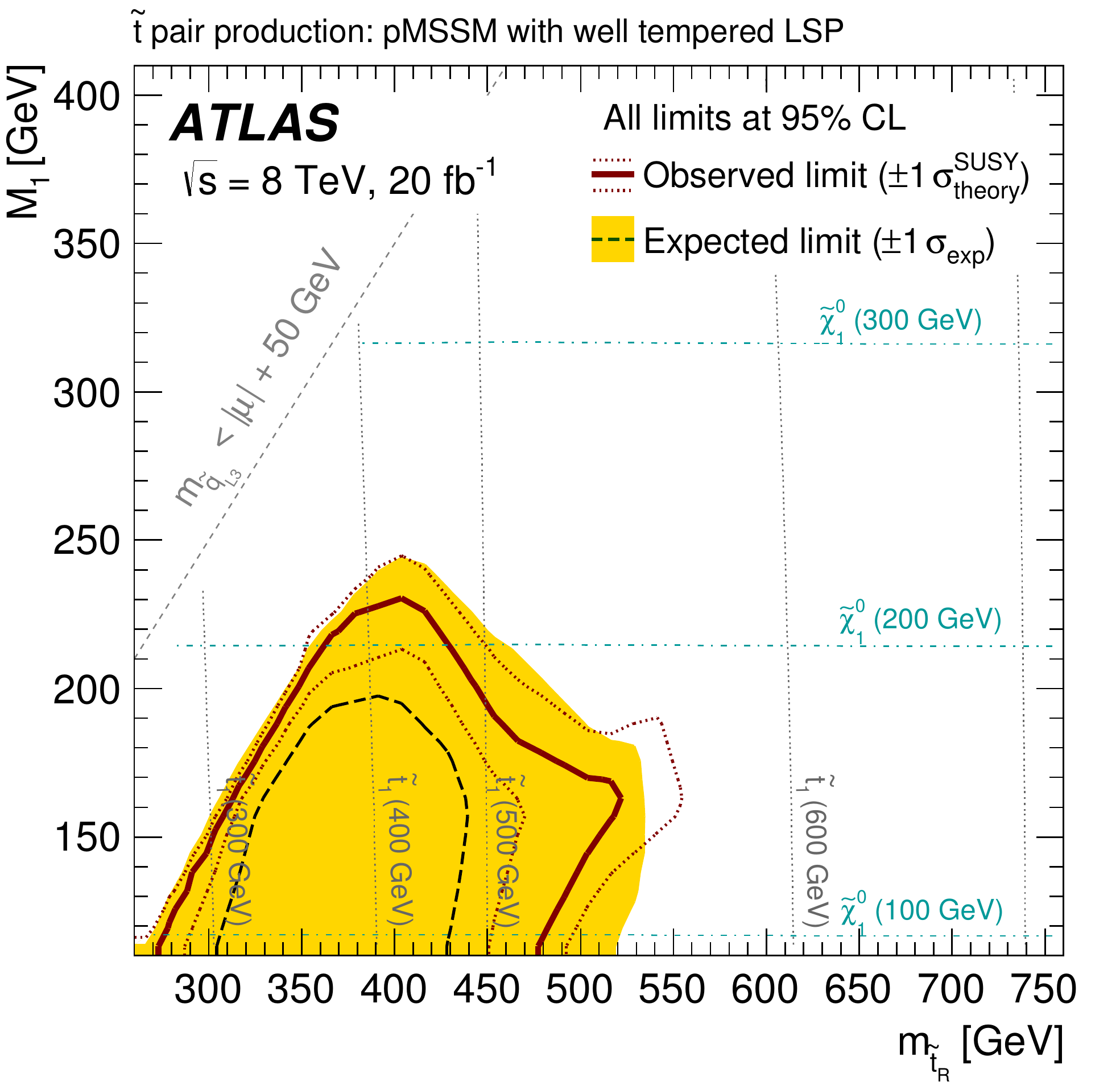}
}
\end{center}
\caption{Expected and observed 95\% CL exclusion limits for the pMSSM model with well-tempered neutralinos as a function of $M_1$ and (a) $\mqlthree$ or (b) $m_{\tright}$. The limit of (a) is obtained as the combination of the \stopZeroLep, \stopOneLep, \TBM\ and \sslll\ analyses, while the \stopZeroLep\ analysis is used for (b). The signal region yielding the smallest \cls\ value for the signal-plus-background hypothesis is used for each point. The dashed black line indicates the expected limit, and the yellow band indicates the $\pm 1\sigma$ uncertainties, which include all
      uncertainties except the theoretical uncertainties in the
      signal. The red solid line indicates the observed limit, and
      the red dotted lines indicate the sensitivity to
      $\pm 1\sigma$ variations of the signal theoretical. The dashed and dotted grey lines indicate a constant value of the stop and sbottom masses, while the dashed light-blue line indicates a constant value of the neutralino mass.}
\end{figure}

Figures~\ref{fig:DM_mql3} and \ref{fig:DM_mtR} show the exclusion limit obtained for the set of pMSSM models with well-tempered neutralinos as a function of $\mqlthree$ and $\mtR$, respectively. In both cases, the exclusion is largely dominated by the \stopZeroLep\ analysis. For Figure~\ref{fig:DM_mql3}, the signal region dominating the sensitivity at low $\mqlthree$ is \stopZeroLep-SRC1, while at higher  $\mqlthree$ values \stopZeroLep-SRA1 and \stopZeroLep-SRA2 dominate the sensitivity. The drop in sensitivity at $\mqlthree = 410\ \GeV$, $M_1 = 260\ \GeV$ is due to the opening of the $\tone \rightarrow t \neuttwo$ and $\tone \rightarrow t \neutthree$ transition, kinematically suppressed for smaller values of  the difference $\mqlthree - M_1$. Such decays introduce more intermediate states in the decay, effectively reducing the transverse momenta of the final state objects. The large fluctuations of the observed limit are again due to transitions between different signal regions. For Figure~\ref{fig:DM_mtR}, the sensitivity is entirely dominated by the various \stopZeroLep-SRC.  The difference in sensitivity between these two scenarios is due to the presence of both a  stop and a sbottom for small $\mqlthree$, while only a stop is present for low values of $\mtR$. 

\begin{figure}[!htb]
\begin{center}
\subfloat[\label{fig:ZH_mql3}]{
	\includegraphics[width=0.45\columnwidth,angle=0]{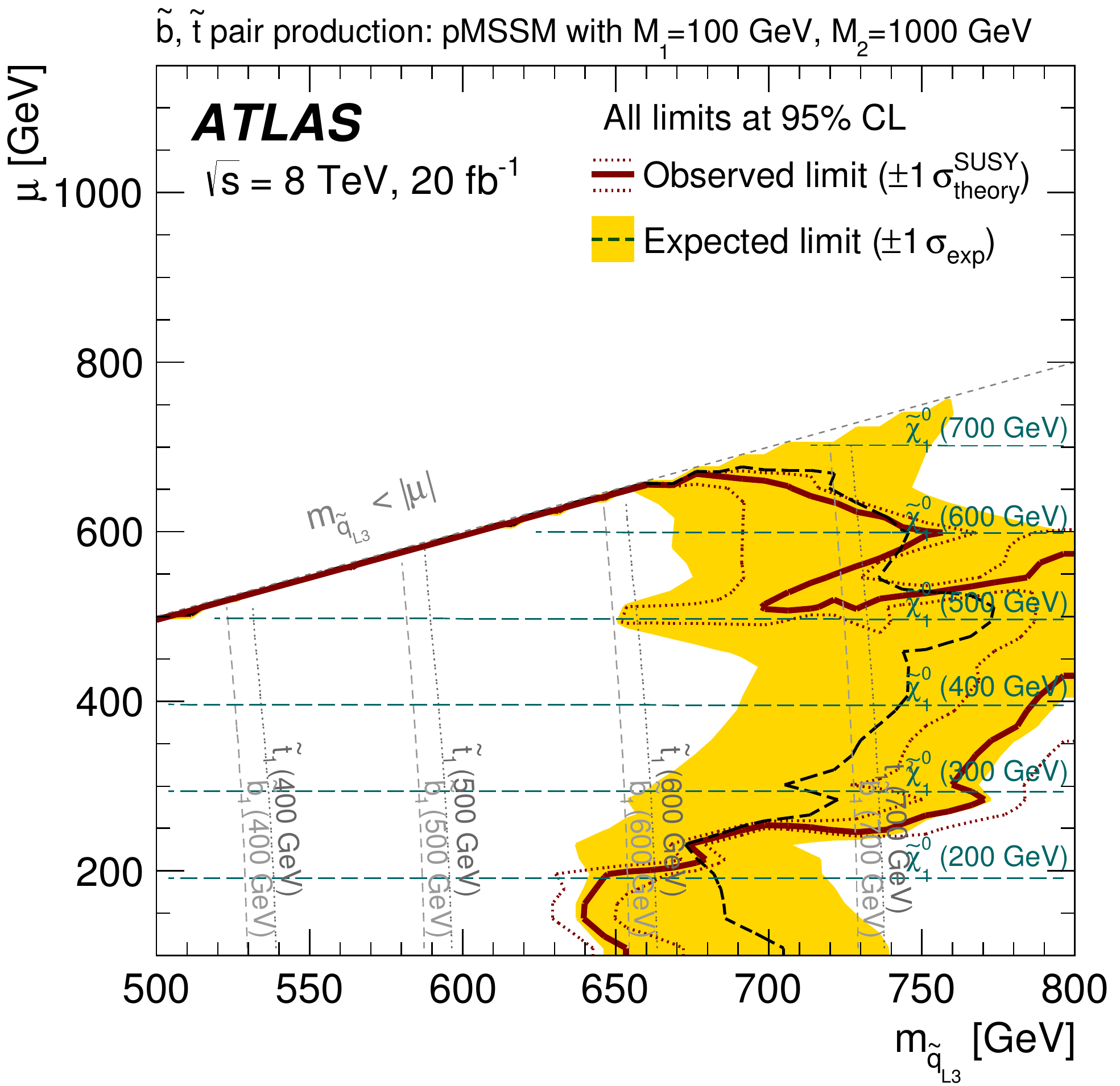}
	}
\subfloat[\label{fig:ZH_mbR}]{
	\includegraphics[width=0.45\columnwidth,angle=0]{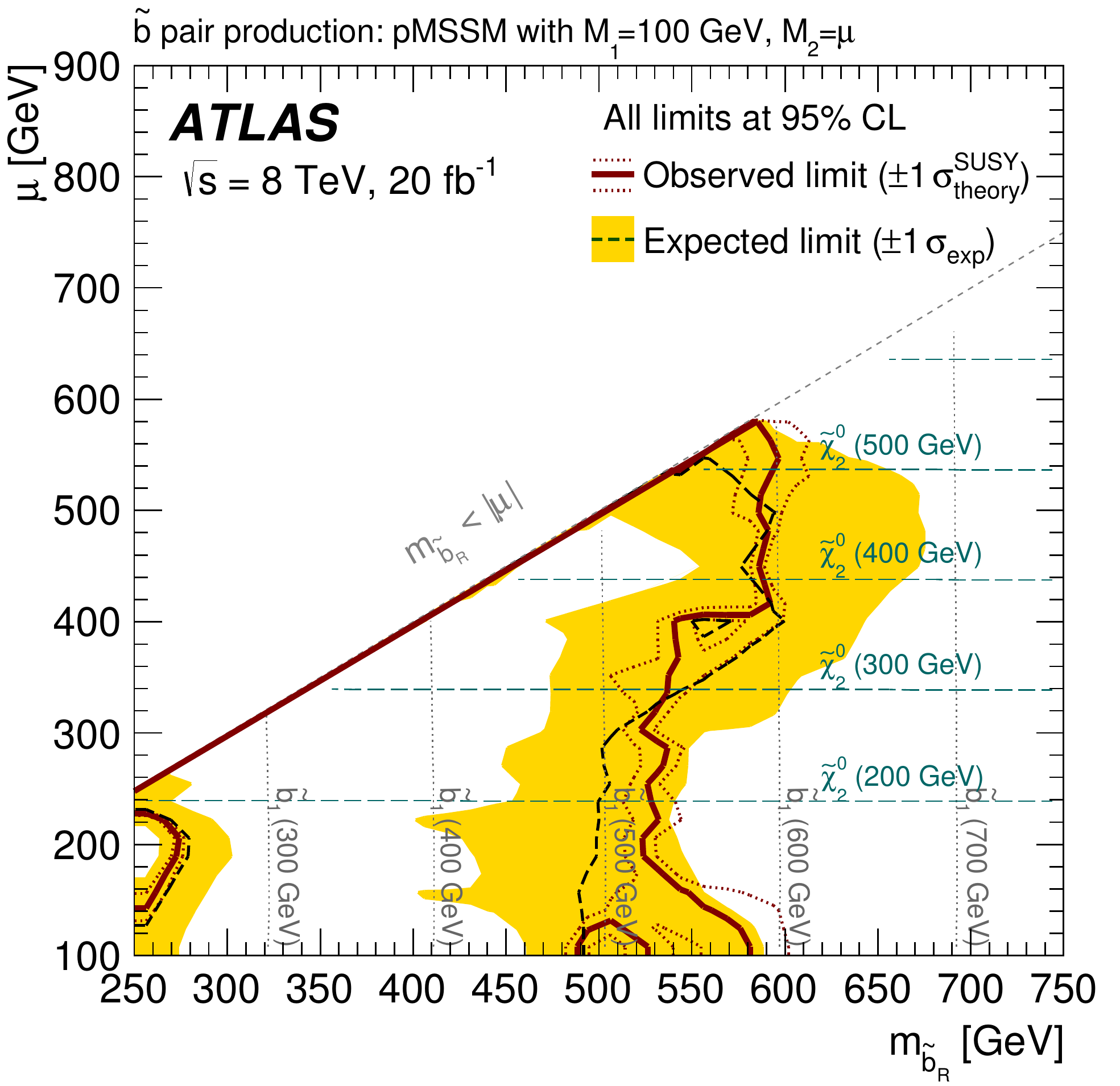}
	}
\end{center}
\caption{Expected and observed 95\% CL exclusion limits for the set of $\higgs/Z$-enriched pMSSM models as a function of $\mu$ and (a) $\mqlthree$ and (b) $m_{\bright}$. The limit of (a) is obtained as the combination of the \stopZeroLep, \threeb, \stopTwo\ and  \sslll\ analyses, while the \stopZeroLep, \stopTwo\ and \TBM\ analysis are used for (b). The signal region yielding the smallest \cls\ value for the signal-plus-background hypothesis is used for each point. The dashed black line indicates the expected limit, and the yellow band indicates the $\pm 1\sigma$ uncertainties, which include all
      uncertainties except the theoretical uncertainties in the
      signal. The red solid line indicates the observed limit, and
      the red dotted lines indicate the sensitivity to
      $\pm 1\sigma$ variations of the signal theoretical. The dashed and dotted grey lines indicate a constant value of the stop and sbottom masses, while the dashed light-blue line indicates a constant value of the neutralino mass.}
\end{figure}

Finally, Figures~\ref{fig:ZH_mql3} and \ref{fig:ZH_mbR} show the exclusion limit obtained for the set of $\higgs/Z$-enriched pMSSM models.  These models  yield large \bjet\ multiplicities to the final state through direct sbottom decays, top-quark decays and $\neuttwo \rightarrow \higgs/Z \neut$. The exclusion is dominated by the \stopZeroLep\ and \threeb\ analyses for Figure~\ref{fig:ZH_mql3} and by and the \stopZeroLep\ analysis for Figure~\ref{fig:ZH_mbR}. 

More informations about the limits obtained, including the SLHA files for the points mentioned in Table~\ref{tab:pMSSM_models}, can be found in Refs.~\cite{hepdata_web} and \cite{hepdata_pub}.

\section{Conclusions}
\label{sec:conclusions}
The search programme of the ATLAS collaboration for the direct pair production of stops and sbottoms is summarised and extended by new analyses targeting scenarios not optimally covered by previously published searches. 
The paper is based on 20 \ifb\ of proton-proton collisions collected at the LHC by ATLAS in 2012 at a centre-of-mass energy $\sqrt{s}$ = 8 TeV.
Exclusion limits in the context of simplified models are presented. In general, stop and sbottom masses up to several hundred GeV are excluded, although the exclusion limits significantly weaken in the presence of compressed SUSY mass spectra or multiple allowed decay chains. Three classes of pMSSM models, based on general arguments of Higgs boson mass naturalness and compatibility with the observed dark-matter relic density have also been studied and exclusion limits have been set. Large regions of the considered parameter space are excluded.   
  
\section*{Acknowledgements}

We thank CERN for the very successful operation of the LHC, as well as the
support staff from our institutions without whom ATLAS could not be
operated efficiently.

We acknowledge the support of ANPCyT, Argentina; YerPhI, Armenia; ARC,
Australia; BMWFW and FWF, Austria; ANAS, Azerbaijan; SSTC, Belarus; CNPq and FAPESP,
Brazil; NSERC, NRC and CFI, Canada; CERN; CONICYT, Chile; CAS, MOST and NSFC,
China; COLCIENCIAS, Colombia; MSMT CR, MPO CR and VSC CR, Czech Republic;
DNRF, DNSRC and Lundbeck Foundation, Denmark; EPLANET, ERC and NSRF, European Union;
IN2P3-CNRS, CEA-DSM/IRFU, France; GNSF, Georgia; BMBF, DFG, HGF, MPG and AvH
Foundation, Germany; GSRT and NSRF, Greece; RGC, Hong Kong SAR, China; ISF, MINERVA, GIF, I-CORE and Benoziyo Center, Israel; INFN, Italy; MEXT and JSPS, Japan; CNRST, Morocco; FOM and NWO, Netherlands; BRF and RCN, Norway; MNiSW and NCN, Poland; GRICES and FCT, Portugal; MNE/IFA, Romania; MES of Russia and NRC KI, Russian Federation; JINR; MSTD,
Serbia; MSSR, Slovakia; ARRS and MIZ\v{S}, Slovenia; DST/NRF, South Africa;
MINECO, Spain; SRC and Wallenberg Foundation, Sweden; SER, SNSF and Cantons of
Bern and Geneva, Switzerland; NSC, Taiwan; TAEK, Turkey; STFC, the Royal
Society and Leverhulme Trust, United Kingdom; DOE and NSF, United States of
America.

The crucial computing support from all WLCG partners is acknowledged
gratefully, in particular from CERN and the ATLAS Tier-1 facilities at
TRIUMF (Canada), NDGF (Denmark, Norway, Sweden), CC-IN2P3 (France),
KIT/GridKA (Germany), INFN-CNAF (Italy), NL-T1 (Netherlands), PIC (Spain),
ASGC (Taiwan), RAL (UK) and BNL (USA) and in the Tier-2 facilities
worldwide.

\newpage  
  
\appendix 
\addcontentsline{toc}{section}{APPENDICES}

\section{The ATLAS detector and object reconstruction}
\label{sec:detector}

The ATLAS detector~\cite{DetectorPaper:2008} consists of inner tracking devices
surrounded by a superconducting solenoid, electromagnetic
and hadronic calorimeters and a muon spectrometer immersed 
in a toroidal magnetic field. 
The inner detector (ID),
in combination with a superconducting solenoid magnet with a central field of 2~T,
provides precision tracking and momentum measurements of charged particles in a pseudorapidity\footnote{ATLAS uses a right-handed
system with its origin at the nominal interaction point (IP) in the centre of the detector and the $z$-axis along
the beam pipe. The $x$-axis points from the IP to the centre of the LHC ring, and the $y$-axis points upward.
Cylindrical coordinates $(r,\phi)$ are used in the transverse plane, $\phi$ being the azimuthal angle around the
beam pipe. The pseudorapidity is defined in terms of the polar angle $\theta$ as $\eta=-\ln\tan(\theta/2)$. The
distance~$\mathrm{\Delta} R$ in the $\eta$--$\phi$ space is defined as
$\mathrm{\Delta} R = \sqrt{(\mathrm{\Delta}\eta)^2+(\mathrm{\Delta}\phi)^2}$.} range $|\eta| < 2.5$.
The ID consists of a silicon pixel detector, a silicon microstrip
detector and a straw tube tracker ($|\eta| < 2.0$) that also provides transition
radiation measurements for electron identification.
A high-granularity electromagnetic calorimeter system, with acceptance covering $|\eta| < 3.2$, uses liquid argon (LAr) as the active medium.  A scintillator-tile calorimeter provides hadronic coverage for $|\eta| < 1.7$. The end-cap and forward regions, spanning $1.5 < |\eta| < 4.9$, are instrumented with LAr electromagnetic and hadronic calorimeters. The muon spectrometer has separate trigger and high-precision tracking chambers which provide trigger coverage for $|\eta| < 2.4$ and muon identification and momentum measurements for $|\eta| < 2.7$.

The data sample used in this analysis was taken during the period from March to December 2012 with the LHC operating at a $pp$ centre-of-mass energy of $\sqrt{s}=8$~TeV.\footnote{The limits derived using a measurement of the \ttbar\ production cross section discussed in Section~\ref{sec:01lep_summary} and Appendix~\ref{sec:published_SR} also uses 4.7 \ifb\ of $pp$ collisions data collected at a centre-of-mass energy of $\sqrt{s} = 7$ TeV.} Following requirements based on beam, detector conditions and data quality, the complete dataset corresponds to an integrated luminosity of 20.3~\ifb, with an associated uncertainty of 2.8\%. The uncertainty is derived following the same methodology as that detailed in Ref.~\cite{Aad:2013ucp}.
Events used in the analyses presented in this paper were selected using the ATLAS three-level trigger following different chains based on the signatures being considered. 
A common set of cleaning cuts, aimed at rejecting events heavily contaminated by non-collision backgrounds, or events containing badly measured or fake jets is applied to all analyses.

\label{sec:objrec}
The experimental signature of third-generation supersymmetric particles includes the production of \bjets in association with missing transverse momentum and possibly additional jets and charged leptons. Different signatures are investigated in this paper to gain sensitivity to a variety of possible topologies arising from the production and decay of stops and sbottoms. Different event selections share common definitions of the final reconstructed objects, which are detailed in the remainder of this Appendix. Analysis-specific departures from those definitions are detailed for each case in Appendix~\ref{sec:SRs} or in the specific analysis paper.    

The reconstructed primary vertex
\cite{PV} is
required to be consistent with the luminous region and to
have at least five associated tracks with $\pt > 400\mev$;
when more than one such vertex is found, the vertex with
the largest summed $\pt^{2}$ of the associated tracks is chosen.

Jets are constructed from three-dimensional clusters of noise-suppressed calorimeter cells~\cite{topoclusters} using the \antikt\ algorithm~\cite{Cacciari:2008gp,Cacciari:2005hq,Cacciari:2011ma} with a distance parameter $R = 0.4$ and calibrated with a local cluster weighting algorithm~\cite{Issever:2004qh}. An area-dependent correction is applied for energy from additional proton--proton collisions based on an estimate of the pileup activity in a given event using the method proposed in Ref.~\cite{Cacciari:2007fd}. Jets are calibrated as discussed in Ref.~\cite{Aad:2011he} and required to have $\pt > 20\,\GeV$ and $|\eta| <4.5$. Events containing jets arising from detector noise, cosmic-ray muons, or other non-collision sources are removed from consideration~\cite{Aad:2011he}. 

Jets arising from a $b$-quark fragmentation and within the acceptance of the inner detector ($|\eta|<2.5$) are identified with an algorithm that
exploits both the track impact parameters and secondary vertex
information~\cite{ATLAS-CONF-2012-043}; this algorithm is based on a
neural network using the output weights of the IP3D, JetFitter+IP3D,
and SV1 algorithms (defined in Refs.~\cite{ATLAS-CONF-2011-089,ATLAS-CONF-2011-102}). A lower cut on the output of the neural network defines the $b$-tagged jets. Three different working points are used, with a nominal efficiency of 60\%, 70\% and 80\% as evaluated on simulated top quark pair production events. The corresponding rejection factors against jets originating from light ($c$) quarks are 25 (3), 135 (5) and 600 (8). 

Electrons are reconstructed from energy clusters in the electromagnetic
calorimeter matched to a track in the inner detector~\cite{Aad:2011mk} and are required to have $|\eta| < 2.47$. Several criteria, including calorimeter shower shape, quality of the match between the track and the cluster, and the amount of transition radiation emitted in the TRT detector, are used to define three selections with decreasing efficiency and increasing purity, named respectively 'loose', 'medium' and 'tight'~\cite{Aad:2011mk}. These three electron selections are used throughout this paper in the definitions of various signal and control regions. Muons, which are identified either as a combined track in the muon spectrometer
and inner detector systems, or as an inner detector
track matched with a  muon spectrometer track segment
\cite{ATLAS-CONF-2011-021,ATLAS-CONF-2011-063}, are required to have
$|\eta| < 2.4$. 

Electrons and muons (generically referred to by the symbol $\ell$) are usually required to have transverse momentum $\pt > 10$~GeV. For specific scenarios with compressed mass spectra, low-\pt leptons are expected and the \pt threshold is lowered to 6~GeV for muons and to 7~GeV for electrons.   

The missing transverse momentum \ptmiss (with magnitude \etmiss) is the negative vector sum
of the \pT\ measured in the clusters of calorimeter cells, which are calibrated
according to their associated reconstructed object (e.g. 
jets and electrons), and the \pT\ of the muons. Calorimeter cells not associated with any reconstructed object are also used
in the calculation of \ptmiss.
The missing transverse momentum from the tracking system (denoted by \ptmisstrk, with magnitude \mettrk)
is computed from the vector sum of the reconstructed inner detector tracks with $\pt > 500\MeV$ and $|\eta|<2.5$, associated with the primary vertex in the event.

\section{Analyses used in the paper}
\label{sec:SRs}

Several signal regions are used in this paper, either standalone or in combination with others, to derive exclusion limits in the many models considered. This Appendix provides a review of the already published analyses and a more extended documentation of the signal regions not previously published.

\subsection{Review of already published signal regions}
\label{sec:published_SR}
The discussion of analyses that have already been published is reduced to a summary for the sake of brevity. Table~\ref{tab:SRname} provides a reference to the papers where full details of the signal, control and validation region selections, together with the strategies adopted for the estimation of the background processes are found.

\paragraph{Multijet final states (\stopZeroLep)}

The analysis is designed to be sensitive to final states arising from all-hadronic decays of directly pair-produced stops~\cite{Aad:2014bva}. Two sets of signal regions were optimised to maximise the sensitivity to topologies arising from $\tone \rightarrow t \neut$ decays, assumed to happen with a branching ratio of one. The first set of signal regions, named \stopZeroLep-SRA, assumes that both top quark hadronic decays can be fully resolved by indentifying the six final-state jets. The SM background (dominated by $\ttbar$ and $Z+$ heavy flavour (HF) jets production) is rejected based on the presence of two hadronic systems consistent with top quarks and large \etmiss. The second set of signal regions, named \stopZeroLep-SRB targets a similar scenario, but aims at topologies where the top quarks have a large boost, and some of the decay products are merged into a single jet. The event selection is designed to select final states with a maximum of five $R= 0.4$ \antikt\  jets, to be mutually exclusive with \stopZeroLep-SRA, and  relies on the presence of  $R=0.8$ and $R=1.2$ \antikt\ jets containing the hadronic decay products of the two top quarks.
 The jet masses, the transverse mass of the \etmiss\ and the nearest $b$-jet, and other variables are used
to discriminate against the dominant SM $\ttbar$, \Zbjets\ and \Wbjets\  production background processes.

Finally, a third set of signal regions, named \stopZeroLep-SRC, is designed to increase the analysis sensitivity to the decay $\tone \rightarrow b \chipm$. The presence of the intermediate chargino state tends to decrease the jet multiplicity: these signal regions require five \antikt\ jets with $R=0.4$, and base the signal selection on a set of transverse mass variables aimed at  rejecting the dominant SM \ttbar\ production process.

\paragraph{One-lepton final states (\stopOneLep)}
\label{sec:1lepton}

The large number of signal regions defined in this analysis stems from the variety and complexity of the possible stop final states considered~\cite{Aad:2014kra}. All signal regions are characterised by the presence of one lepton, a second-lepton veto, a minimum of two jets and large \etmiss. A first set of four signal regions (t1L-tN) were optimised assuming a branching ratio of 100\% for the decay $\stop \rightarrow t \neut$. These signal regions aim at having sensitivity to different $\Delta m(\stop, \neut)$, in particular \stopOneLep-tN\_diag targets scenarios with small $\dmTN$ and makes use of the shape information of the \etmiss and \mt distributions.\footnote{The transverse mass $\mt$ of the lepton  with transverse momentum $\vec{p}_{\mathrm{T}}$ and the missing transverse momentum vector \ptmiss with magnitude \etmiss is defined as 
\begin{equation}
\mt = \sqrt{2\left(|\vec{p}_{\mathrm{T}}| \etmiss - \vec{p}_{\mathrm{T}} \cdot \ptmiss\right)}
\end{equation}
\noindent and it is extensively used in one-lepton final states to reject SM background processes containing a $W$ boson decaying leptonically.}  The \stopOneLep-tN\_boost SR targets models with the largest  $\Delta m(\stop, \neut)$, where the top quark produced by the stop decay has a large boost and large-$R$ jets are used to reconstruct the top quark decays.

The decay $\stop \rightarrow \chipm b$ introduces additional degrees of freedom in the decay. The final-state kinematics is largely driven by the mass separation between the stop and the chargino $\Delta m(\stop,\chipm)$, and by that between the chargino and the neutralino $\Delta m(\chipm, \neut)$. Several signal regions, identified by the prefix \stopOneLep-bC were designed and optimised depending on the mass hierarchy and, consequently, on the different kinematics of the lepton and \bjets.

The four signal regions \stopOneLep-bCa\_low, \stopOneLep-bCa\_med, \stopOneLep-bCb\_med1 and \stopOneLep-bCb\_high  target small values of $\dmCN$ and have the common feature of making use of a dedicated soft-lepton selection: muons and electrons are identified down to a \pt\ threshold of 6 GeV and 7 GeV, respectively, requiring a special treatment for the estimate of possible background processes arising from lepton misidentification. They are collectively referred to as ``soft-lepton'' signal regions. Both \stopOneLep-bCa signal regions require a hard ISR jet to boost the stop pair system and produce a sizeable \etmiss. The \stopOneLep-bCb targets large values of $\dmTC$ and exploits the presence of two relatively hard $b$-jets in the event.

The signal region  \stopOneLep-bCc\_diag targets a mass hierarchy complementary to that of the \stopOneLep-bCb. The small value of $\Delta m(\stop,\chipm)$ gives rise to soft \bjets that go undetected, hence  $b$-tagged jets are vetoed for this region. 

Topologies arising from scenarios where both $\Delta m(\stop,\chipm)$ and $\Delta m(\chipm, \neut)$ are sizeable are targeted by the three \stopOneLep-bCd regions: they all require four jets in the final state, are characterised by different \bjet multiplicities, and  apply different selections on the \etmiss, \mt\ and \amtTwo\footnote{The asymmetric stransverse mass variable is a variant  of the stransverse mass variable~\cite{Lester:1999tx,Barr:2003rg} defined to efficiently reject dileptonic \ttbar\ decays. It assumes that the undetected particle is the $W$ boson for the branch with the lost lepton and the neutrino is the missing particle for the branch with the observed charged lepton. For the  dileptonic \ttbar events, \amtTwo\ is bounded from above by the top quark mass, whereas new physics can exceed this bound.}    variables. A veto on additional isolated tracks and $\tau$ lepton candidates identified with loose criteria helps to suppress the dominant SM background from dileptonic \ttbar\ decays. 

The last two signal regions listed in Table~\ref{tab:SRname},  \stopOneLep-3body and \stopOneLep-tNbC\_mix, were optimised for two additional possible scenarios. If $\Delta m(\stop, \neut) < \mtop$ and the mass hierarchy or the model parameters suppress the decay through a chargino, then the dominant stop decay is $\stop \rightarrow bW\neut$, through an off-shell top quark (three-body decay). The dedicated signal region relies on the shape information from the \mt\ and \amtTwo\ variable distributions. Finally,  \stopOneLep-tNbC\_mix is designed to recover sensitivity in scenarios where the stop is assumed to decay with similar probabilities to $t\neut$ and $b\chipm$: the selection aims to reject the dominant dileptonic \ttbar\ background by making use of the topness~\cite{Graesser:2012qy} variable.

\paragraph{Two-lepton final states (\stopTwoLep)}

If the SUSY mass hierarchy forbids the presence of sleptons in the stop decay chain, final states containing two leptons ($e$ or $\mu$) and a large amount of \etmiss\  would arise from stop pair production.
The main background is given by SM processes containing two $W$ bosons in the final state (mainly $\ttbar$ and $WW$)~\cite{Aad:2014qaa}.  To discriminate the stop signal from the SM background, the stransverse mass variable $\mtTwoll$~\cite{Lester:1999tx,Barr:2003rg} is used. The stransverse mass, computed using the two leptons as visible particles and the missing transverse momentum vector, exhibits a kinematical end-point at $m_{W}$ for most SM processes. Because of the presence of additional \etmiss\ due to the LSP, the end-point for a SUSY signal can be at larger values, depending on the mass separation between the particles involved in the decay. The analysis is optimised assuming $\tone\rightarrow \chipm b$ with BR=100\% and $\Delta m(\chipm, \neut) > m_W$, but it is also sensitive to the  three-body decay mode of the stop. To derive exclusion limits, five signal regions (\stopTwoLep) have been defined, requiring different jet multiplicities and different \mtTwoll\  thresholds.  A selection requiring two \bjets\ and based on $\mtTwoll$ computed using them as visible particles is sensitive to the chargino decay mode with $\Delta m(\tone,\chipm) > \mtop$. Finally, a multivariate discriminant is built which targets the $\tone \rightarrow t \neut$ decay mode.

\paragraph{Final states from  compressed stop decays (\stopCharm)}

If the difference in mass between the stop and the neutralino is smaller than the $W$ boson mass, then the only possible decay channels are $\stop \rightarrow \neut c$ or $\stop \rightarrow W^* b$, where the decay products of the off-shell $W^*$ would,  in general, be soft. This analysis~\cite{Aad:2014nra} has defined two sets of signal regions, both optimised for the $\stop \rightarrow \neut c$ decay. A common preselection requires the presence of a high-\pt\ jet, large \etmiss and applies a lepton veto. The first set of signal regions named \stopCharm-M, targets scenarios with the stop mass almost degenerate with the neutralino mass, and applies a selection that exploits a monojet-like signature arising from the presence of an ISR jet. Three different signal regions have been designed, characterised by increasing thresholds on the leading jet \pt\ and \etmiss. The second set of signal regions, named \stopCharm-C, targets less compressed scenarios, and exploits the presence of jets originating from the fragmentation of $c$-quarks in the final state. A dedicated $c$-tagging algorithm was  used to reject the dominant SM background processes arising mostly from \ttbar\  and $Z\rightarrow \nu \bar{\nu}$ (produced in association with heavy-flavour jets) production. As in the case of the \stopCharm-M signal regions, different thresholds on the leading jet \pt\ and on \etmiss are used to identify a looser and a tighter \stopCharm-C region.

\paragraph{Final states with a $Z$ boson (\stopTwo)}

A $Z$ boson can be emitted in the decay of $\ttwo \rightarrow \tone Z$, producing final states with large lepton multiplicities. It can be useful to look for $\ttwo$ (rather than $\tone$) production if, for example, the mass of $\tone$ is very close to the sum of the top quark and neutralino masses, which would lead to $\tone$ pair production final states difficult to distinguish from SM $\ttbar$ production. Models are investigated with $\DMstopN = 180$~GeV with the decay $\tone \rightarrow t \neut$. The final state would contain, beyond the Z boson, several jets arising from the $\tone$ decay. Similar final states can  be obtained in GMSB models where the $Z$ boson is emitted in the $\neut \rightarrow \tilde{G} Z$ decay if the gravitino $\tilde{G}$ is the LSP and the neutralino the NLSP. 

This analysis~\cite{Aad:2014mha} defines five different signal regions divided into two sets. The first set, named \stopTwo-SR2, requires two same-flavour leptons whose invariant mass is consistent with that of a $Z$ boson, $m_Z$, and at least one $b$-tagged jet. The three signal regions are characterised by the different selection thresholds applied to the \etmiss , to the transverse momentum of the dilepton system \ptll and to the jet multiplicity. The second set of signal regions, named \stopTwo-SR3, requires three leptons, two of which must form an opposite-sign same-flavour pair whose invariant mass is consistent with $m_Z$. Both signal regions require at least five jets, among which at least one has to be $b$-tagged. The two signal regions are characterised by the different selection thresholds applied to \ptll and to the leading lepton \pt.

\paragraph{Final states with two \bjets and \etmiss (\sbottom)}
\label{sec:sbottom}

This signature arises naturally from the sbottom decay $\bone \rightarrow b \neut$. Moreover, one expects the same final state from $\tone \rightarrow \chipm b$ followed by $\chipm \rightarrow f f' \neut$ in the limit of small $\Delta m(\chipm, \neut)$. This analysis~\cite{Aad:2013ija} defines two sets of signal regions, \sbottom-SRA and \sbottom-SRB, targeting scenarios with large and small squark--neutralino mass separations, respectively.

The event selection of \sbottom-SRA requires large \etmiss, exactly two \bjets and vetoes the presence of additional jets; the rejection of the SM \ttbar production background is carried out by making use of the contransverse mass~\cite{Tovey:2008ui} of the two \bjets. Its distributions shows a kinematical end-point at about 135 \GeV\ for \ttbar\ production, while extending to higher values for the signal. 

A selection relying on the presence of an ISR jet is instead needed if the third-generation squark mass is almost degenerate with that of the neutralino. This is the purpose of \sbottom-SRB, which selects a hard, non-\btagged leading jet recoiling against the squark pair system. The selection includes the requirement of two \btagged jets, a veto on additional hadronic activity, and the presence of large \etmiss.

\paragraph{Final states with three \bjets (\threeb)}

This analysis~\cite{Aad:2014lra} is designed to search for gluino-mediated sbottom and stop production in events with no leptons or one lepton (electron or muon) in the final state. However, it was found to have sensitivity for direct $\bone$ production followed by $\bone \rightarrow \neuttwo b \rightarrow \neut \higgs b$, where $h$ is the SM Higgs boson with mass $m_h=125$~GeV, and also sensitivity to some of the pMSSM models considered in this paper. Such final states are characterised by a large multiplicity of \bjets\ both in $\gluino\gluino \rightarrow \tone\tone t t$ and $\gluino\gluino \rightarrow \bone\bone b b$ where there are up to four \bjets in the final state. 

Three sets of signal regions have been designed to target different mass hierarchies of the gluino-mediated sbottom and stop production models. All signal regions have at least four jets with $\pt > 30$ GeV, three identified \bjets, large \met and a large \meff, defined as the scalar sum of the \pt of the jets and \met.

\paragraph{Strongly produced final states with two same sign or three leptons (\sslll)}

Final states containing many leptons or same-sign (SS) leptons can arise from the pair production of gluinos and squarks, when the produced particles decay to the LSP through multiple intermediate stages, or when several top quarks appear as part of the decay chain. The analysis was developed for the gluino-mediated stop production process $\gluino \gluino \rightarrow \tone \tone t t$ followed by $\tone \rightarrow t \neut$,  which can yield final states containing up to four leptons, including SS pairs. Similar final states arise from the sbottom decay $\bone \rightarrow t \chipm$, which are studied in this paper.

This analysis~\cite{Aad:2014pda} concentrates on final states containing either three leptons or a SS lepton pair produced in association with many jets. Five signal regions (identified by the prefix \sslll) are defined, which are characterised by different light- and heavy-flavour jet multiplicities, high selection thresholds on \met\ and $\meff$, and different thresholds on the transverse mass of the lepton with the highest transverse momentum and the \met.

\paragraph{Spin correlation in \ttbar\ production events (\spinCorr)}

If the mass of the $\tone$ is such that $m_{\tone} \sim m_{\neut} + \mtop$, the final-state kinematics are similar to that of Standard Model \ttbar\ production. One possible approach is to derive exclusion limits on the stop mass by performing SM precision measurements. This analysis has measured the azimuthal angle difference between the two leptons arising from the dileptonic \ttbar\ decay~\cite{Aad:2014mfk}. The events are required to contain, beside the two leptons, at least two additional jets, one of which is required to be \btagged. In events containing two leptons of the same flavour, the $Z$ production background is suppressed by applying a selection on the dilepton invariant mass. The distribution of the azimuthal angle between the two leptons is sensitive to the spin correlations of the $\ttbar$ system: it is hence used to extract limits on possible contaminations from direct scalar top production events.

\paragraph{\ttbar\ production cross section (\xsec)}

The measurement of the $\ttbar$ production cross section using events containing two different-flavour leptons $e\mu$ and \btagged\ jets is used in Ref.~\cite{Aad:2014kva} to extract limits on the direct pair production of $\tone$ with mass close to the top quark. The assumed decay is $\tone \rightarrow t \neut$. 

The $\ttbar$ production cross section $\sigma_{\ttbar}$ is obtained by using the equations 

\begin{eqnarray}
N_1 = L \sigma_{\ttbar} \epsilon_{e\mu} 2\epsilon_b (1 - C_b \epsilon_b) + N_1^{\mathrm{bkg}} \label{eq:N1} \\
N_2 = L \sigma_{\ttbar} \epsilon_{e\mu} C_b \epsilon_b^2 + N_2^{\mathrm{bkg}} \label{eq:N2}
\end{eqnarray}

\noindent where $N_1$ and $N_2$ are the number of events with two different flavour leptons having exactly one or two \btagged\ jets, respectively,  $L$ is the integrated luminosity,  $\epsilon_{e\mu}$ the efficiency for a $\ttbar$ event to pass the lepton selection, $\epsilon_b$ is the probability of having a \bjet within acceptance and for it to be  tagged, $C_b$ is a correlation coefficient which is close to unity, and $N_1^{\mathrm{bkg}}$ and $ N_2^{\mathrm{bkg}}$ are the number of events with one or two \btagged\ jets  from SM events different from $\ttbar$ production. The values of $\sigma_{\ttbar}$ and $\epsilon_b$ are extracted from the data by solving the two simultaneous equations~(\ref{eq:N1}) and (\ref{eq:N2}), avoiding the need to estimate $\epsilon_b$ from simulation.

Stop-pair production events with $m_{\tone}>\mtop+m_{\neut}$ have similar $\epsilon_{e\mu}$ and \bjet\ kinematics to SM $\ttbar$ production events, so the fitted value of $\epsilon_b$ in a combined sample is compatible with that from $\ttbar$ production events alone, and the fitted cross section corresponds closely to the sum of $\ttbar$ and stop-pair production cross sections. Limits on stop pair production are extracted by calculating 95\% CL limits on the stop pair production signal strength $\mu$ (defined as the ratio of the obtained stop cross section to
the theoretical prediction) based on the comparison of the measured cross section with that predicted for SM $\ttbar$ production events alone. A 95\% CL signal strength smaller than unity for a given signal point implies its  exclusion.

This interpretation, which made use of collision data with both $\sqrt{s} = 7$ and 8 TeV, is extended here to the three-body decay $\tone \rightarrow W b \neut$. The main difference with respect to the scenario considered in Ref.~\cite{Aad:2014kva} is that the three-body decay tends to yield $b$-jets with lower \pt, leading to a fitted $\epsilon_b$ for the combined sample which is different from that expected for $\ttbar$ events alone.  The limits obtained are summarised in Figure~\ref{fig:stopxsec} for a neutralino mass of 1 GeV. A 95\% CL limit that excludes stop masses below 175 GeV is obtained. The figure also shows the effect on the limit of a ``sneaky top squark'' scenario~\cite{Eifert:2014kea}: the presence of a $\tone$ with mass similar to that of the top quark could bias the measurement of the top-quark mass itself. The bias in the top-mass measurement introduced by the existence of a $\tone$ with mass $m_{\tone} = 170$ GeV depends on the analysis technique and channel, and was evaluated to be at most 1 GeV for the two- and three-dimensional template techniques used in the ATLAS top mass measurement in the lepton+jets channel~\cite{Aad:2015nba}. The effect of a potential bias of 1 GeV and 2.5 GeV on the top-mass measurement  was studied by recalculating the observed 95\% CL limit on $\mu$  when reducing the predicted SM $\ttbar$ production cross section from the baseline value of $\mtop = 172.5 \pm 1.0$ GeV to those obtained for top mass central values of 173.5 and 175 GeV. The corresponding limit on the stop mass is reduced  by about 5 and 15 GeV, respectively.

\begin{figure}[htb]
  \begin{center}
    \includegraphics[width=0.7\textwidth]{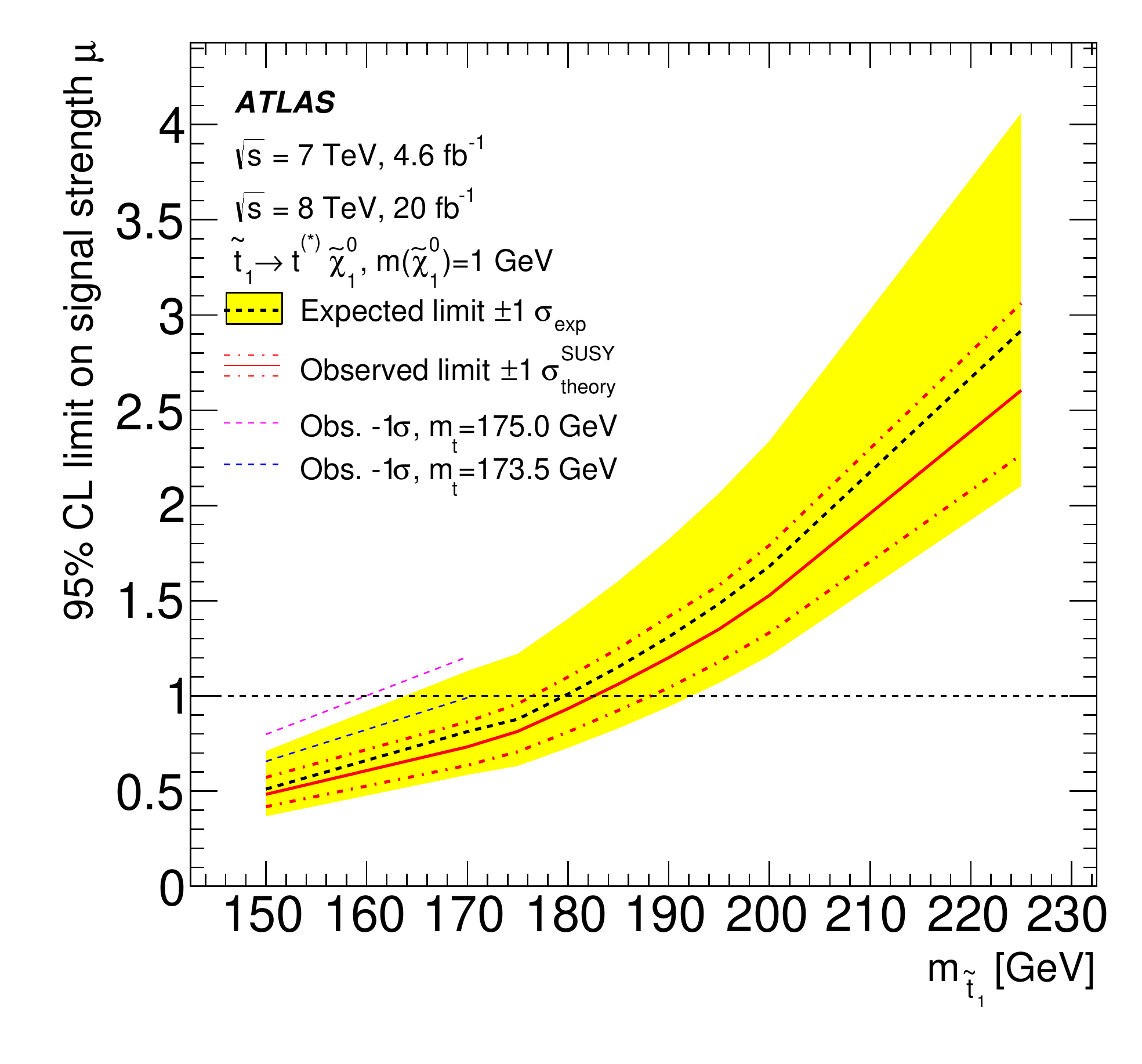}
    \caption{Expected and observed 95\% CL limits on the signal strength $\mu$ (defined as the ratio of the obtained stop cross section to
the theoretical prediction) for the production of $\tone$ pairs as a function of $m_{\tone}$. The stop is assumed to decay as $\tone \rightarrow t \neut$ or through its three-body decay depending on its mass. The neutralino is assumed to have a mass of 1 GeV. The black dotted line shows the expected limit with $\pm 1 \sigma$ uncertainty band shaded in yellow, taking into account all uncertainties except the theoretical cross-section uncertainties on the signal. The red solid line shows the observed limit, with dotted lines indicating the changes as the nominal signal cross section is scaled up and down by its theoretical uncertainty. The short blue and purple dashed lines indicate how the observed limits with the signal cross section reduced by one standard deviation of its theoretical uncertainty for $m_{\tone} < \mtop$ when the top quark mass is assumed instead to be $173.5\pm 1.0$ and $175.0\pm 1.0$~GeV.}
    \label{fig:stopxsec}
  \end{center}
\end{figure}

The dependence of the exclusion limits on the neutralino mass was studied and found to be important: the effect of an increasing neutralino mass is to decrease the \pt of the \bjets, and hence to lower the value of $\epsilon_b$ for the stop pair production signal. For a neutralino mass of 30 GeV, only a small range of stop masses around 150 GeV is excluded. 

The sensitivity of the $\ttbar$ cross-section measurement to $\tone$ pair production assuming  a branching ratio of 100\% into $\tone \rightarrow b \chipm$, followed by $\chipm \rightarrow W^{(*)} \neut$ with $m_{\chipm} = 2 m_{\neut}$ was also investigated. The presence of the intermediate chargino state tends to lower the \pt\ of the leptons and of the \bjets\ significantly, hence decreasing both  $\epsilon_{e\mu}$ and $\epsilon_b$. No exclusion limit can be derived for this scenario.  

Summarising, the limits on stop pair production obtained in Ref.~\cite{Aad:2014kva} have been extended by considering the stop three-body decay. Stop masses between 150 GeV and $\mtop$ can be excluded for a neutralino mass of 1 GeV. The exclusion holds provided that any bias in the top-quark mass measurement by a nearby stop is not significant.  Studies indicate that this potential bias would affect the limit on the stop mass by less than 5 GeV.

\subsection{Description of the new signal regions}
\label{sec:newSRs}

New analyses were developed to target topologies and regions of the SUSY parameter space not well covered by previously published signal regions. They are identified throughout this paper and in Table~\ref{tab:SRname} with the acronym \WWlike, \TBM\ and \ttH. Their contribution to the exclusion limits derived both in simplified and pMSSM models is outlined in Sections~\ref{sec:interpretations} and \ref{sec:pMSSM} respectively. In this Appendix, further details about these analyses are provided for the interested reader. Additional informations about selection efficiencies, sensitivities of the different signal regions and individual limit plots, please refer to Refs.~\cite{hepdata_web} and \cite{hepdata_pub}.
 
\subsubsection{Final states with two leptons at intermediate values of \mtTwoll (\WWlike)}
\label{sec:WWlike}

The measurement of the production cross section of nonresonant $WW$
pairs in the two-lepton channel at the LHC \cite{Aad:2012oea,Chatrchyan:2013yaa,Chatrchyan:2013oev} has given rise to theoretical speculations \cite{Rolbiecki:2013fia,Kim:2014eva, Curtin:2014zua} which interpret the possible excess as due to the production of a light stop. 
The mass hierarchy favoured by these speculations includes a $\tone$ with mass around 200 GeV, a $\chipm$ degenerate with it, and $m_{\chipm} - m_{\neut}$ of a few tens of GeV: possible hadronic decay products of the $\tone \rightarrow b \chipm$ transition would have low \pt\, and would allow the events to survive the tight jet-veto selections applied in the SM cross-section measurement. Dedicated signal regions, defined by requiring two different-flavour opposite-sign leptons in the final states, are designed to have maximum sensitivity to such scenarios. The approach is also sensitive to scenarios where the stop decays predominantly through the three-body $\tone \rightarrow b W\neut $ or four-body  $\tone \rightarrow b \ell \nu \neut $ decay.

MC simulated events are used to model the signal and to describe all backgrounds that produce two prompt leptons from $W$, $Z$ or $h$ decay. For processes whose predicted yield in the signal regions is small, or whose topology resembles very closely that of the signal, making it hard to define a proper control region, the background estimate is fully based on MC simulation. For \ttbar, \Zjets\ and $WW$ production processes, which are the dominant backgrounds, the acceptance of the signal regions selection is estimated with MC simulation, while the normalisation is estimated in dedicated control regions. The MC samples used are the same as in Ref.~\cite{Aad:2014qaa}.

The identification criteria for electrons, muons and jets follow the strategy defined in Appendix~\ref{sec:objrec}: baseline electrons, which are used in the estimation of the fake-lepton background, are selected by applying the ``medium'' identification criteria. Signal electrons are identified using the ``tight'' criteria, and they are further required to be isolated. Signal muons correspond to baseline muons with an additional calorimeter- and track-based isolation requirement applied. Jets that have $|\eta| < 2.5$ and $\pt > 20$ GeV are used for the event selection, although all jets up to $|\eta| < 4.5$ are retained for the computation of the missing transverse momentum. 

Candidate stop production events, preselected by the same trigger and data quality requirements used in Ref.~\cite{Aad:2014qaa}, are further required to contain one electron and one muon of opposite charge, with an invariant mass $m_{\ell \ell} > 20$ GeV. The leading (in $\pt$) and next-to-leading leptons are required to have $\pt > 25$ GeV and $\pt > 20$ GeV, respectively. 

At this stage of the selection, the background is dominated by production of top-quark pairs and $Z\rightarrow \tau \tau$, 
followed by $WW$ and $Wt$ production.  

A requirement of $\mtTwoll > 20$ GeV, where $\mtTwoll$ is the stransverse mass of the two leptons, strongly reduces the $Z\rightarrow \tau\tau$ background, which is expected to have a kinematical end-point at $ \mtTwoll = m_{\tau}$. The ratio $R_1$
of the \met\ and the effective mass, defined as the scalar \pt sum of the \met, the leptons and the jets, is useful in suppressing the \ttbar background, which is typically characterised by a larger hadronic activity than in signal events. The selection chosen is $R_1 > 0.3 + \meff$ (with $\meff$ in TeV).

After the above selections, the SM background is dominated by $WW$ production. Two differences between this process and the stop pair production signal are further exploited: 
firstly, the $WW$ production is dominated by quark-antiquark scattering, while stop pair production is mostly initiated by gluon-gluon processes, and secondly the stop pair production signal has four invisible (two neutralinos and two neutrinos) and two undetected (the two $b$-jets) objects, while the $WW$ process has only two. 
The first difference implies a higher longitudinal boost of the system emerging from the hard scattering in signal events than in background events. The variable 

\begin{equation}
\Delta X = 2\left|\frac{\left( p_z(\ell_1) + p_z(\ell_2)\right)}{\sqrt{s}}\right|
\end{equation}

\noindent was defined in Ref.~\cite{Melia:2011cu}, and it is an estimator of the boost. The second difference implies a higher \met for signal events. This is exploited by making use of 

\begin{equation}
R_2 = \frac{\met}{\met + \pt(\ell_1) + \pt(\ell_2)}.
\end{equation}

Finally, the variable $\cos \theta_\mathrm{b}$, the cosine of the angle between the direction of motion of one of the two leptons and the beam axis in the centre-of-mass frame of the two visible leptons~\cite{Melia:2011cu}, is sensitive to the spin of the produced particles, hence it provides additional rejection power against the $WW$ production process.

A set of seven signal regions were optimised for the discovery of stop pair production, with the stop decaying either as $\tone \rightarrow \chipm b$ with a branching ratio of 100\% (assuming $m_{\tone} - m_{\chipm} < 10$ GeV), or as $\tone \rightarrow bW^{(*)} \neut$. The definitions of the signal regions are shown in Table~\ref{tab:WWSigReg}.

\begin{table}[!h]
\begin{center}
\caption{\label{tab:WWSigReg} Summary of  signal regions used in the analysis. The upper part of the table shows the preselection requirements.}
\begin{tabular}{l|c|c|c|c|c|c|c}
\hline
\hline
SR        & \WWlike-SR1 & \WWlike-SR2 & \WWlike-SR3  & \WWlike-SR4 & \WWlike-SR5 & \WWlike-SR6 & \WWlike-SR7 \\
\hline
$\pT(\ell_1)$                                   & \multicolumn{7}{c}{$> 25 \GeV$} \\
\hline
$\pT(\ell_2)$                                   & \multicolumn{7}{c}{$> 20 \GeV$} \\
\hline
$R_1$ &  \multicolumn{7}{c}{$>0.3+\meff$ (TeV)}\\
\hline
$\mtTwoll$  &   \multicolumn{7}{c}{$>20 \GeV$}\\
\hline
\hline
$\Delta X$&   \multicolumn{7}{c}{$<0.02$}\\
\hline
$R_2$ & \multicolumn{7}{c}{$>0.5$}\\
\hline
$|\cos\theta_\mathrm{b}|$ & $<0.8$ & $<0.8$ & $<0.8$ & - & - & $<0.8$ & - \\
\hline
$\mtTwoll$    & $< 45 \GeV$ & $>25,<55 \GeV$  &  -  & $> 70 \GeV$ & $>90 \GeV$ & $>25,<70 \GeV$ & $>80 \GeV$\\
\hline
\hline
\end{tabular}
\end{center} 
\end{table}
 
 The background from non-prompt leptons originating from heavy-quark decays or from photon conversions in the signal regions, or from hadrons misidentified as leptons (collectively referred to as fake leptons in the following), is estimated as in Ref.~\cite{Aad:2014qaa}.
 
  Specific control regions, whose event yield is expected to be dominated by each of these production processes, are defined and included in the fit to constrain the normalisation parameters. The control region CRT for \ttbar production is defined by changing the following selections with respect to the signal regions: $\mtTwoll > 35$ GeV, $R_1 < 0.3$. Its purity is 92\%. The CR for $WW$ production (CRW) is defined by $\mtTwoll > 35$ GeV, $\Delta X > 0.04$, and has a purity of 72\%. Finally, the CR for \Zjets\  (CRZ) is defined by $\mtTwoll < 20$ GeV, $30\ \mathrm{GeV} < m_{\ell \ell} < 80$ GeV, with a purity of 86\%. The normalisation factors of the $WW$, \ttbar, \Zjets\ production processes ($\mu_{WW}, \mu_{\ttbar}$ and $\mu_Z$ respectively) are determined by a combined profile likelihood fit. When testing the signal-plus-background hypothesis for rejection, the fit takes automatically into account the signal contamination in the control regions. For signal scenarios considering light ($m_{\tone} < 150$ GeV) stops decaying through $\tone \rightarrow bW^{(*)}\neut$, the signal contamination becomes so large that $\mu_{WW}$ becomes unrealistically low. For such cases the fit is performed excluding CRW and taking the normalisation of the $WW$ background from MC simulation.
 
 \begin{figure}[!htb]
\begin{center}
\subfloat[]{
\includegraphics[width=0.48\columnwidth,angle=0]{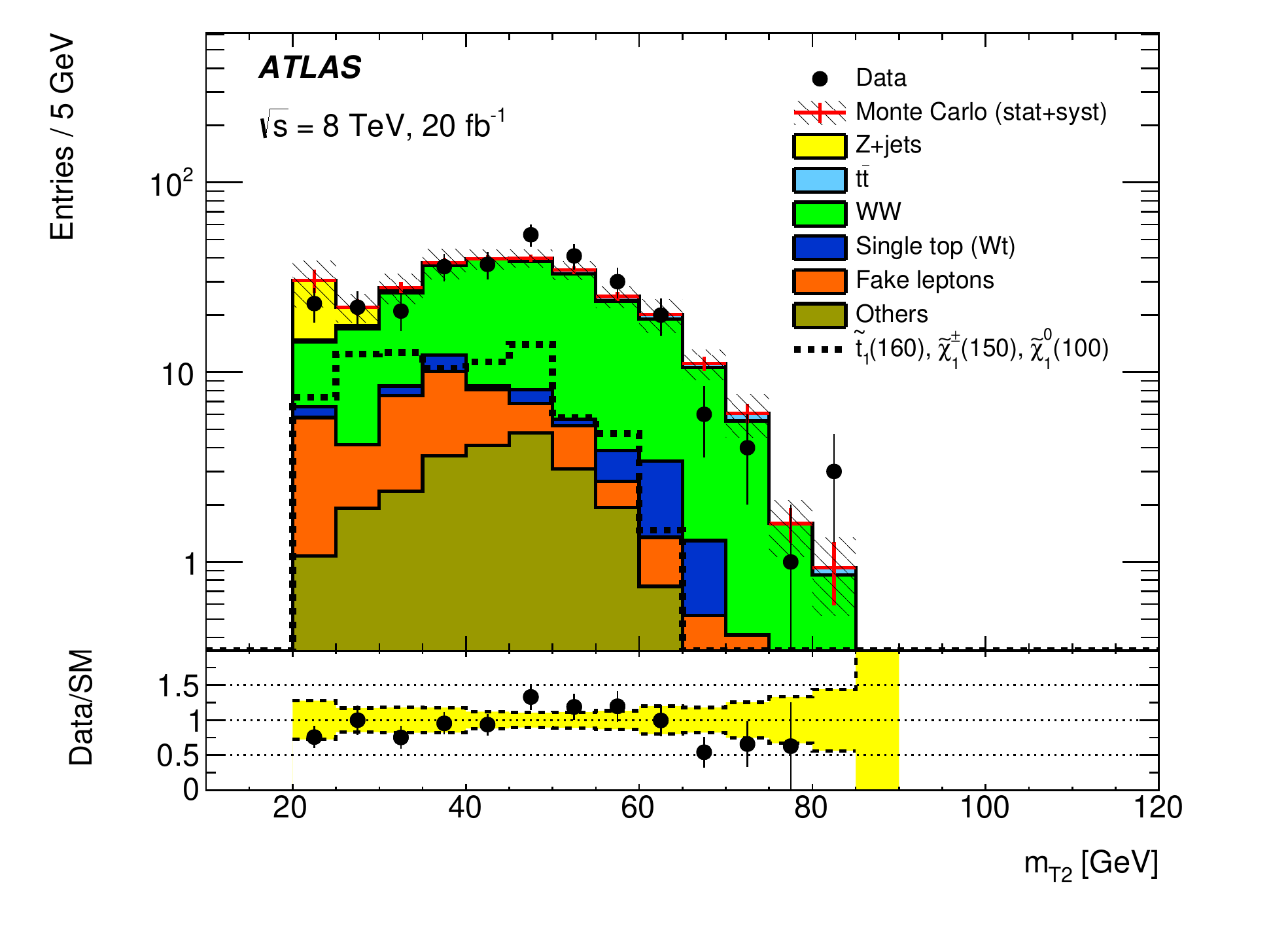}}
\hfill
\subfloat[]{
        \includegraphics[width=0.48\columnwidth,angle=0]{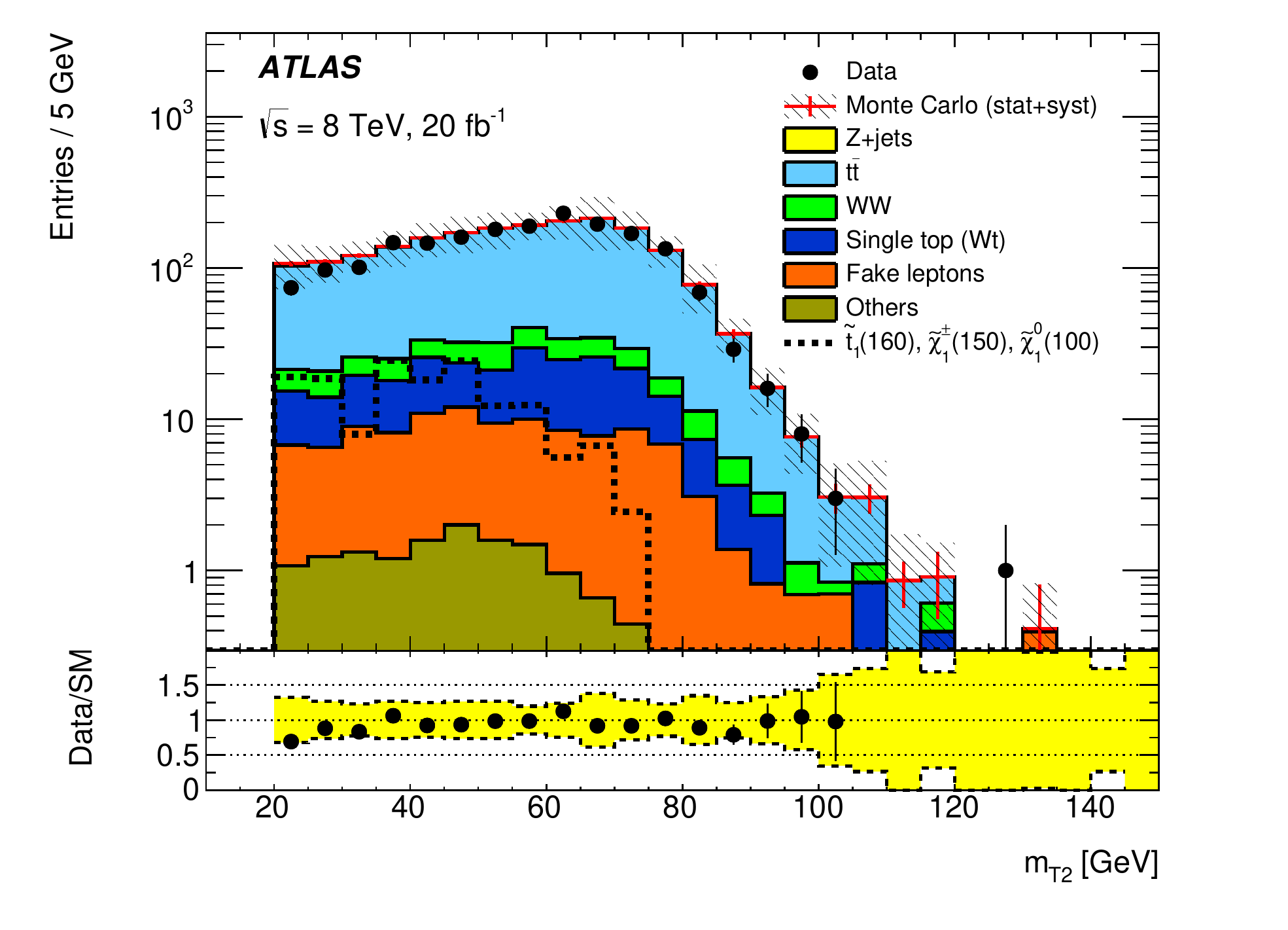}
        }
\end{center}
\caption{Distribution of the stransverse mass $\mtTwoll$ in the (a) \WWlike-VR2 and (b) \WWlike-VRT regions defined in the text. The contributions from all SM processes are shown as a histogram stack. The component labelled as ``Fake leptons'' includes the estimate of the background from non-prompt leptons. The expected signal for a model of stop pair production with the stop decaying as $\tone \rightarrow b\chipm \rightarrow b \ell^{\pm} \nu \neut$ with $m_{\tone} = 160$ GeV, $m_{\chipm} = 150$ GeV and $m_{\neut} = 100$ GeV is also shown. The lower panels show the ratio between the data and the  SM prediction; the yellow band includes statistical and systematic uncertainties on the SM prediction.}
\label{fig:mt2_WW}
\end{figure}

 Systematic uncertainties, affecting both the modelling of the detector response (detector-related systematic uncertainties) and the theoretical prediction of the cross sections and acceptances of the background processes (theory-related systematic uncertainties) affect the predicted rates in the signal regions. Their classification and estimation follows closely those defined in Ref.~\cite{Aad:2014qaa}. A few differences, discussed in the following, exist on the estimation of the theory-related uncertainties. The total uncertainty on the yield of the $WW$ production process is composed of three terms: the  uncertainty on the NLO hard-scattering calculation is taken to be the difference between the prediction of \powheg and aMC@NLO both using \pythia for the parton shower; the uncertainty addressing the  choice of the parton-shower model is estimated as the difference of the aMC@NLO predictions showered either with \herwig or \pythia; the uncertainty due to the choice of the renormalisation and factorisation scale is evaluated by changing the scales independently by a factor of two or one-half and taking the maximum difference. The estimated relative uncertainties on the signal region yields are about 6\% in SR1--SR4 and SR6; 11\% in SR7 and 29\% in SR5. Similar comparisons performed on the $WZ$  and $ZZ$ process yield uncertainties ranging from 30\% to 45\% depending on the signal region considered. Additional systematic uncertainties are assigned to the small expected yields from \Zjets\ production (80\%), $Wt$ (50\% to 100\% depending on the SR considered), and non-prompt lepton background.

The values of the normalisation factors obtained when performing the fit to the control regions only are shown in Table~\ref{tab:WWNorm}. 

\begin{table}[!h]
\begin{center}
\caption{ Normalisation factors for the $\ttbar$, $WW$ and \Zjets\ background processes obtained by the combined fit to the control region yields. The uncertainties include systematic and statistical uncertainties.\label{tab:WWNorm}} 
\begin{tabular}{c|c}
\hline
\hline
Normalisation factor  & Value \\
\hline
$\mu_{\ttbar}$ & $0.94 \pm 0.05$ \\
$\mu_{WW}$ & $1.01 \pm 0.11$ \\
$\mu_{Z}$ & $0.95 \pm 0.62$ \\
\hline
\hline
\end{tabular}
\end{center}
\end{table}

The overall predictions of the fit are compared to the data in dedicated validation region that are kinematically close to the signal region. They are defined by applying the preselection requirements of Table~\ref{tab:WWSigReg} with the additional selections shown in Table~\ref{tab:WWValReg}. The $\mtTwoll$ distribution in \WWlike-VR2 and \WWlike-VRT is shown in Figure~\ref{fig:mt2_WW}.

\begin{table}[!h]
\begin{center}
\caption{\label{tab:WWValReg} Summary of  the validation regions used in the \WWlike\ analysis. The preselection requirements of Table~\ref{tab:WWSigReg} are also applied in all three validation regions.}
\begin{tabular}{c|c|c}
\hline
\hline
 \WWlike-VR1 & \WWlike-VR2 & \WWlike-VRT   \\
\hline
 - & - & $0.3 < R_1 < 0.3 + \meff\ (\TeV)$ \\
$0.02 <|\Delta X| < 0.04$ & $ \Delta X< 0.02$&  $ \Delta X< 0.02$\\
$R_2>0.5$ & $R_2< 0.5$ & $R_2> 0.5$\\
$|\cos\theta_\mathrm{b}|<0.8$ & $|\cos\theta_b|<0.8$ & $|\cos\theta_b|<0.8$  \\
\hline
\hline
\end{tabular}
\end{center} 
\end{table}

 \begin{figure}[!htb]
\begin{center}
\subfloat[]{
\includegraphics[width=0.48\columnwidth,angle=0]{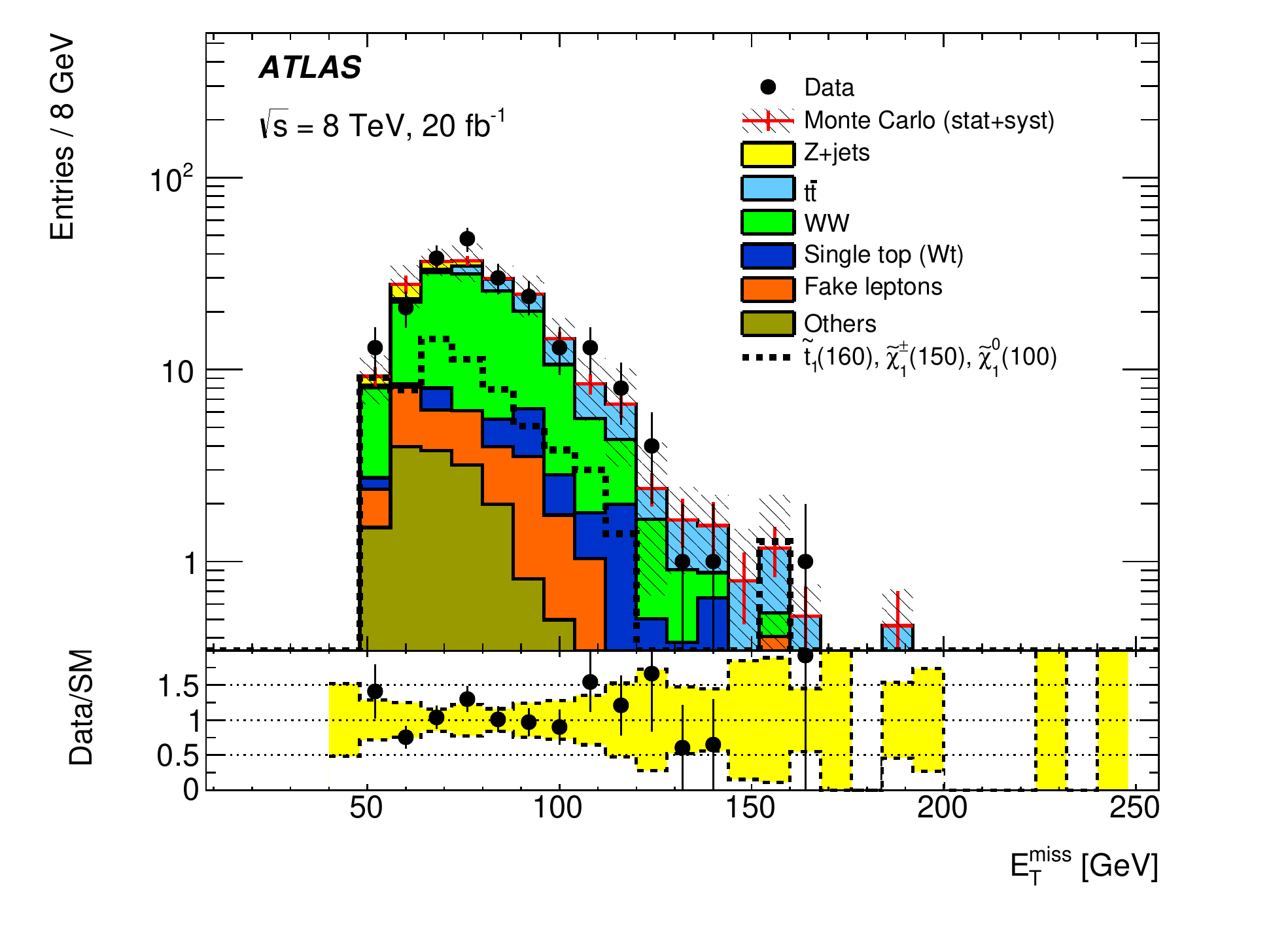}
}
\hfill
\subfloat[]{
        \includegraphics[width=0.48\columnwidth,angle=0]{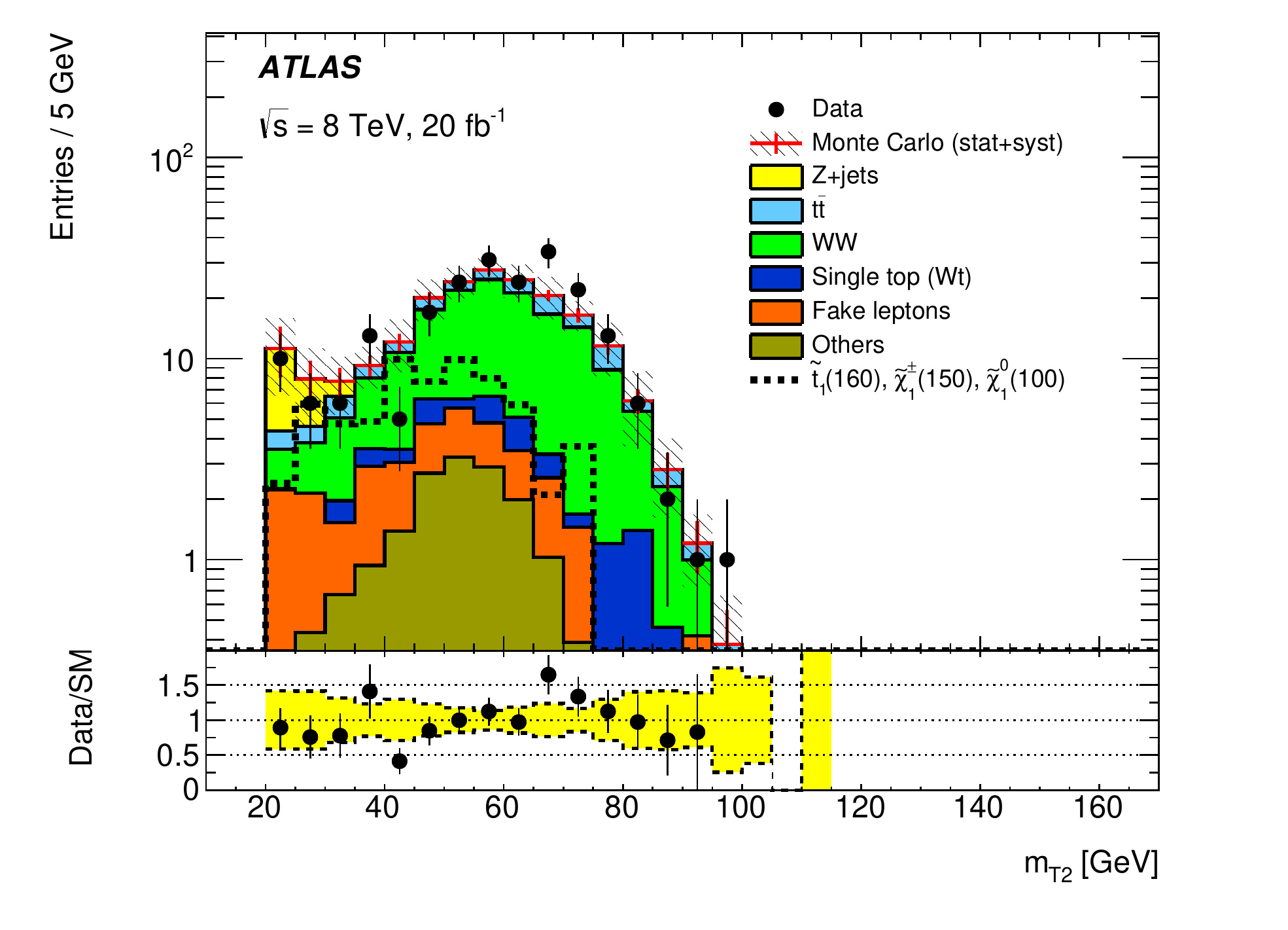}
        }
\end{center}
\caption{Distribution of the (a) magnitude of missing transverse momentum \met and (b) stransverse mass $\mtTwoll$ in \WWlike-SR3. The contributions from all SM processes are shown as a histogram stack. The component labelled as ``Fake leptons'' includes the estimate of the background from non-prompt leptons. The expected signal for a model of stop pair production with the stop decaying into $\tone \rightarrow b\chipm \rightarrow b \ell^{\pm} \nu \neut$ with $m_{\tone} = 160$ GeV, $m_{\chipm} = 150$ GeV and $m_{\neut} = 100$ GeV is also shown. The lower panels show the ratio between the data and the SM prediction; the yellow band includes statistical and systematic uncertainties on the SM prediction. }
\label{fig:SR3_WW}
\end{figure}

 For all signal regions, the expected background yield is dominated by production of $WW$ (35\% in SR1 to 66\% in SR4). Other important background processes are \Zjets\ in SR1 (20\%), non-prompt leptons in SR2 (12\%), \ttbar in all other SR, with contributions of about 10\%. The distributions of \met and \mtTwoll\ in the signal region \WWlike-SR3 are shown in Figure~\ref{fig:SR3_WW}.

\begin{table} [htb!]
\begin{center}
  \caption{Observed (Obs) and predicted (Exp) numbers of events in the signal regions of the \WWlike\ analysis, together with the 95\% CL upper limits on the observed and expected number of signal events ($S_{\mathrm{obs}}^{95}$ and $S_{\mathrm{exp}}^{95}$, respectively), and on the visible cross section ($\langle\epsilon \sigma \rangle_{\mathrm{obs}}^{95}$).
  \label{tab:WWresults}}
\renewcommand\arraystretch{1.2}
\begin{tabular}{lccccc}
\hline
\hline
{\bf Signal channel}                        & Obs & Exp &  $S_{\mathrm{obs}}^{95}$  & $S_{\mathrm{exp}}^{95}$ &  $\langle\epsilon\sigma\rangle_{\mathrm{obs}}^{95}$[fb]  \\
\hline
SR1  &  40 & $47 \pm 14$ &  $22.6$  & ${25.2}^{+9.4}_{-4.3}$ & $1.12$ \\ 
SR2  & 71& $80 \pm 13$ &  $25.3$  & ${27.8}^{+11.5}_{-4.1}$& $1.24$   \\ 
SR3  &  215& $203 \pm 27$ &  $48.4$  & ${46.6}^{+4.9}_{-6.9}$ & $2.38$  \\ 
SR4  &  88& $81\pm 11$&  $35.1$  & ${28.8}^{+11.0}_{-5.4}$ & $1.73$ \\ 
SR5  &  4 & $3.4 \pm 0.9$ &  $6.2$  & ${5.7}^{+2.1}_{-1.4}$ & $0.30$  \\ 
SR6  & 160 & $154 \pm 19 $ &  $45.6$  & ${43.8}^{+19.3}_{-14.4}$  & $2.25$ \\ 
SR7  &  21 &$ 23 \pm 4$ &  $12.4$  & ${13.4}^{+4.8}_{-3.4}$ & $0.61$  \\ 
\hline
\hline
\end{tabular}
\end{center}
\end{table}

\begin{figure}[!htb]
\begin{center}
\subfloat[\label{fig:WW_stop_10}]{
\includegraphics[width=0.48\columnwidth,angle=0]{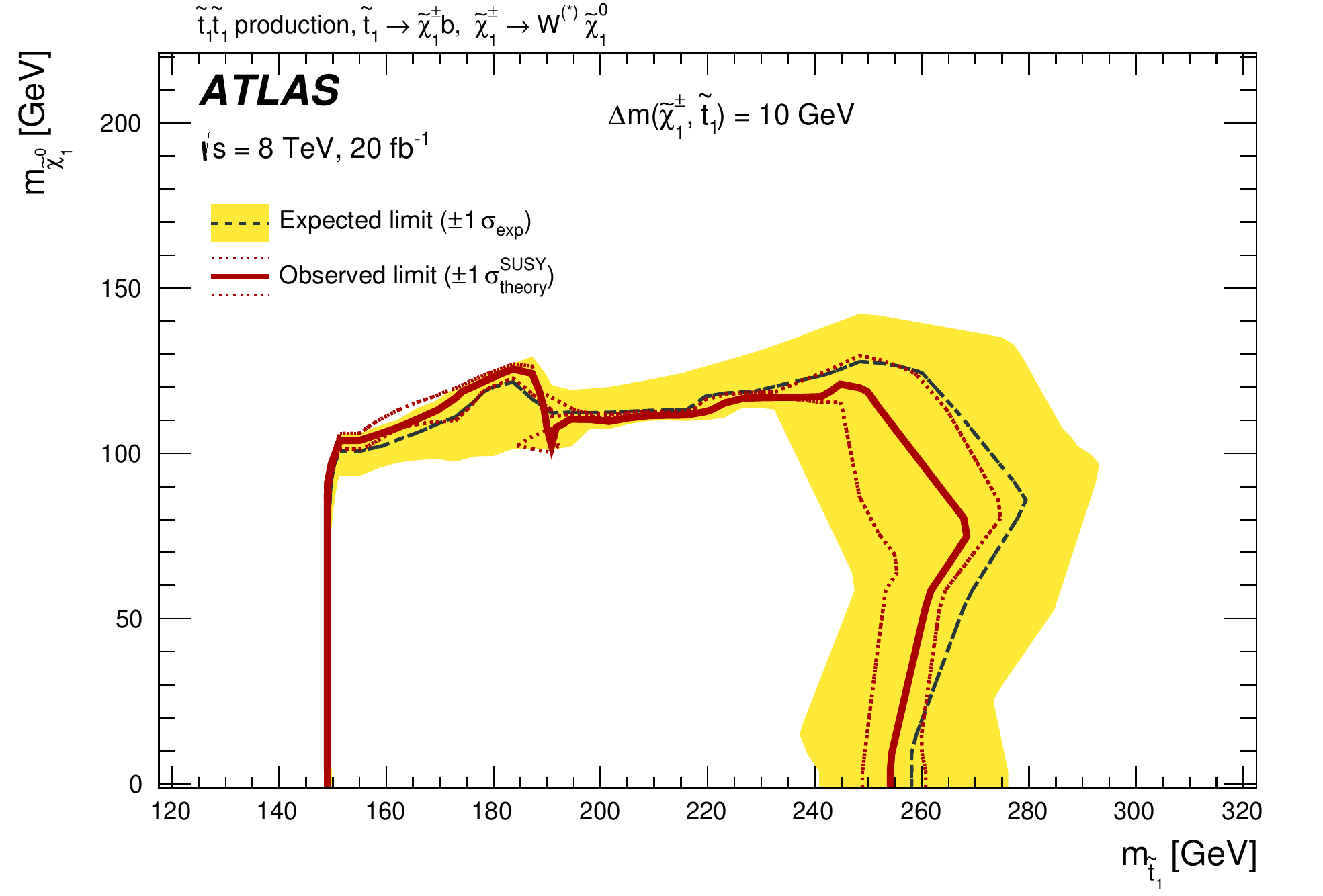}
}
\hfill
\subfloat[\label{fig:WW_threebody}]{
\includegraphics[width=0.48\columnwidth,angle=0]{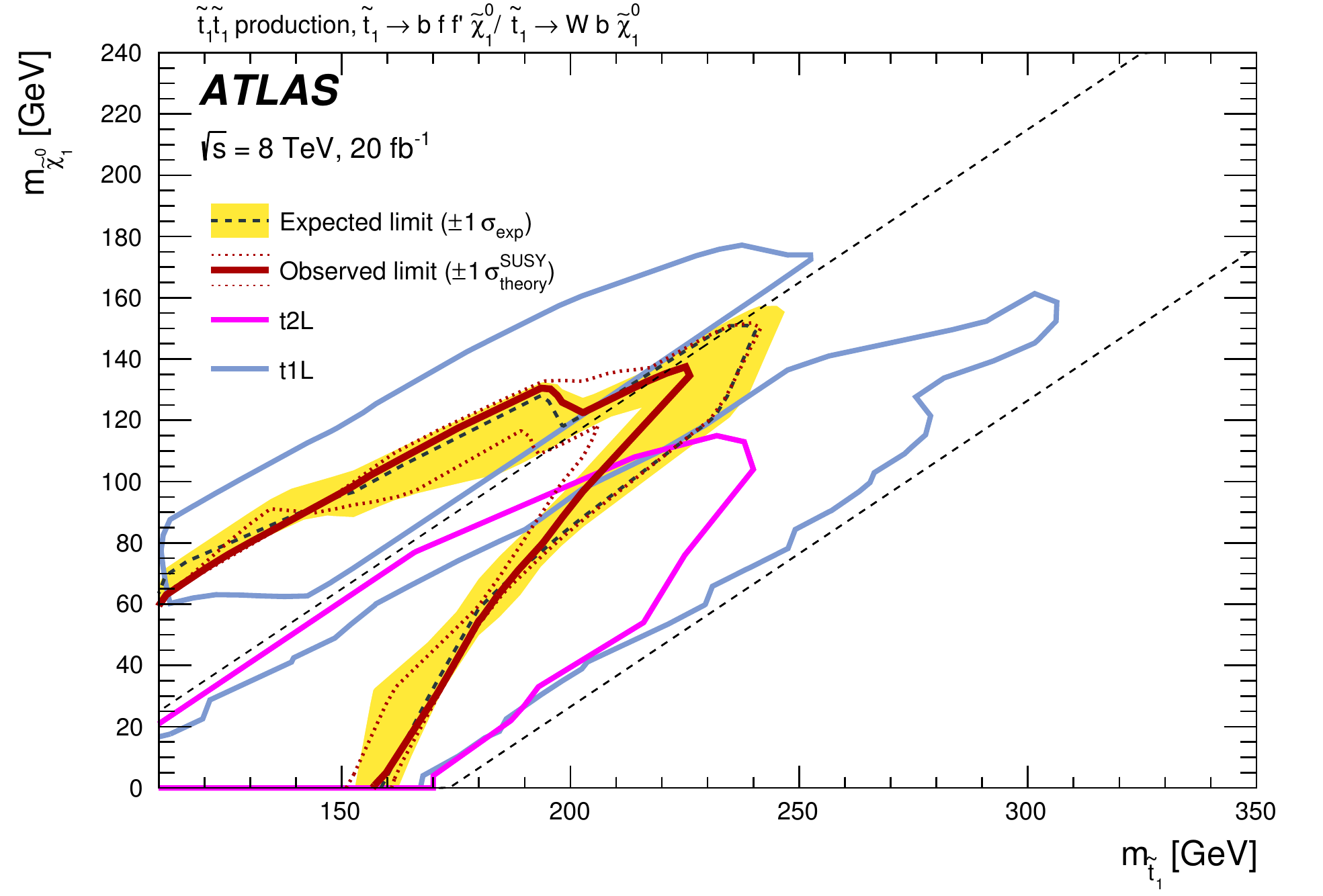}
\vspace{3cm}
}	
\end{center}
\caption{ Exclusion limits at $95\%$ CL in the scenario where both pair-produced stop decay exclusively via (a) $\tone \rightarrow b\chipm$ followed by $\chipm \rightarrow W \neut$, with $\dmTC=10\ \GeV$, and (b) three-body or four-body decay (depending on the neutralino and stop mass). The black dashed line indicates the expected limit, and the yellow
      band indicates the $\pm 1\sigma$ uncertainties, which include all
      uncertainties except the theoretical uncertainties in the
      signal. The red solid line indicates the observed limit, and
      the red dotted lines indicate the sensitivity to
      $\pm 1\sigma$ variations of the signal theoretical
      uncertainties.  For (b), the observed limits achieved by the \stopOneLep\ and \stopTwoLep\ analyses are also shown, and the straight dashed lines correspond to $\DMstopN = m_W + m_b$ and $\DMstopN = \mtop$. }
\end{figure}

Table~\ref{tab:WWresults} compares the predicted and observed numbers of events in each of the signal regions. No excess above the SM prediction is observed, hence the results are first used to derive model-independent 95\% CL exclusion limits on the minimum number of events beyond the Standard Model in the signal region assuming no signal contamination in the control regions, and then to extract limits on $\sigma_{\mathrm{vis}} = \sigma \times \epsilon \times \mathcal{A}$, where $\sigma$ is the cross section for non-SM processes, $\epsilon$ is the selection efficiency and $\mathcal{A}$ is the selection acceptance. These limits are also reported in Table~\ref{tab:WWresults}. Finally, 95\% CL exclusion limits are derived in specific supersymmetric models of direct pair production of stops. The first exclusion limit (Figure~\ref{fig:WW_stop_10}) is derived in a model where the stop is assumed to decay as $\tone \rightarrow b \chipm$ with a branching ratio of 100\%, followed by the decay of the chargino into the neutralino, assumed to be the stable LSP, through $\chipm \rightarrow W^{(*)} \neut$. The chargino mass is assumed to satisfy the relation $m_{\chipm} = m_{\tone} - 10\ \GeV$, and the limit is derived in the $m_{\tone}$--$m_{\neut}$ plane. Stop masses up to about 250 GeV are excluded, almost independently of the neutralino mass. The second limit is derived in a model where the $\tone$ decays through its three-body or four-body decay (depending on its mass and on that of the neutralino) into $\tone \rightarrow b \ell\nu \neut$ with a branching ratio of 100\%, under the assumption that the decay happens through an off-shell top quark and an on- or off-shell $W$ boson. The limit is shown in Figure~\ref{fig:WW_threebody} and fills a gap between the exclusions of the \stopTwoLep\ and \stopOneLep\ analyses.

\subsubsection{Final states containing two top quarks and a Higgs boson (\ttH)}
\label{sec:ttH}

If the lightest stop has a mass such that $\DMstopN \sim \mtop$, the sensitivity of the searches for the production of a $\tone$ pair is greatly reduced. One of the approaches  followed is to search for direct pair production of $\ttwo$ instead. This is the strategy used, for example, by the \stopTwo\ analysis, whose signal regions were optimised to detect the decay of a pair-produced $\ttwo$ followed by the decay $\ttwo \rightarrow Z \tone$. 

Inspired by the search for a SM Higgs boson produced in association with a top quark pair, a search was developed and optimised for the decay $\ttwo \rightarrow \higgs \tone$, where the Higgs boson is assumed to have SM properties, and the $\tone$ is assumed to decay as $\tone \rightarrow t \neut$ with a BR of 100\%. The final state is hence characterised by a large jet multiplicity, by the presence of many  \bjets\ from the top quark and Higgs boson decays and by \etmiss associated with the presence of neutrinos from semileptonic decays of the top quark and of neutralinos.

The selection of electrons, muons, jets and \bjets\ follows the principles outlined in Appendix~\ref{sec:objrec}. The specific choices made for the \pt and pseudorapidity thresholds and working points of the final-state objects, as well as the trigger selection, are the same as those in Ref.~\cite{Aad:2015gra}. The selection requires the presence of exactly one electron or muon with $\pt > 25$ GeV, $\etmiss > 50$ GeV, at least six jets with $\pt > 25$ GeV and $|\eta| < 2.5$, of which at least two are required to be \btagged. The working point chosen for the \btagging is such that the efficiency to tag \bjets\ (evaluated on a MC sample of $\ttbar$ production) is about 70\%.

The modelling of the production of \ttbar\ pairs in association with heavy flavour ($\ttbar$+HF) is of key relevance in this analysis. A detailed categorisation of $\ttbar$+HF is made for the purpose of comparisons with different generators and of the propagation of systematic uncertainties on the different heavy-flavour components. The categorisation is also used to reweight the different flavour components of the $\ttbar$+jets background to obtain a better modelling. These categorisation and reweighting procedures are discussed in detail in Ref.~\cite{Aad:2015gra}. In particular, the $\ttbar+\bbbar$ component, which is simulated with \powheg, is reweighted to a full NLO calculation \cite{Cascioli:2013era} performed in \sherpa 1.4.1+\verb+OpenLoops+\cite{Gleisberg:2008ta,Cascioli:2011va}. The reweighting is done at generator level using a number of kinematic variables such as the top quark \pt, \ttbar\ system \pt, $\Delta R$ and \pt of the dijet system not coming from the top-quark decay.  A different reweighting is applied to the $\ttbar + \ccbar$ and $\ttbar$+ light-jets components, which is based on the ratio of the differential cross sections at $\sqrt{s} = 7$ TeV obtained in data and simulation as a function of the top quark \pt and \ttbar\ system \pt~\cite{Aad:2014zka}. 

The selected events are categorised into different channels, depending on the number of \btagged\ jets (two, three or at least four). The channel with at least four \bjets\ has the largest signal-to-background ratio. The channels with two and three \btagged\ jets are used to calibrate the $\ttbar+$jets background prediction and constrain the associated systematic uncertainties, which, in the channel with at least four \btagged\ jets, are dominated by the $b$-tagging, jet energy scale, and $\ttbar+$jets heavy-flavour content uncertainties.

\begin{figure}[htbp]
  \begin{center}
  	\subfloat[]{
    	\includegraphics[width=0.35\textwidth]{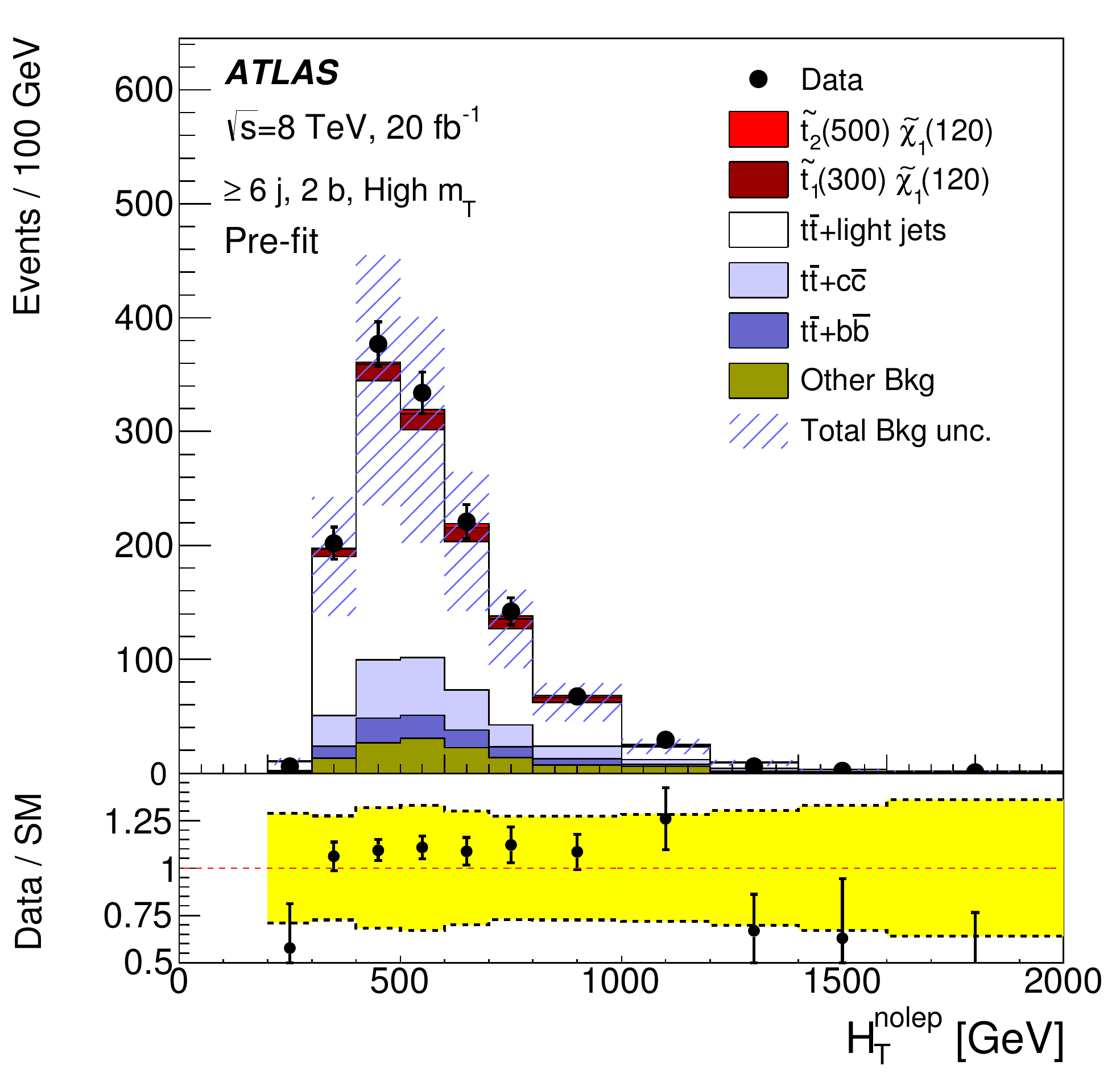}
	}
	\subfloat[]{
    	\includegraphics[width=0.35\textwidth]{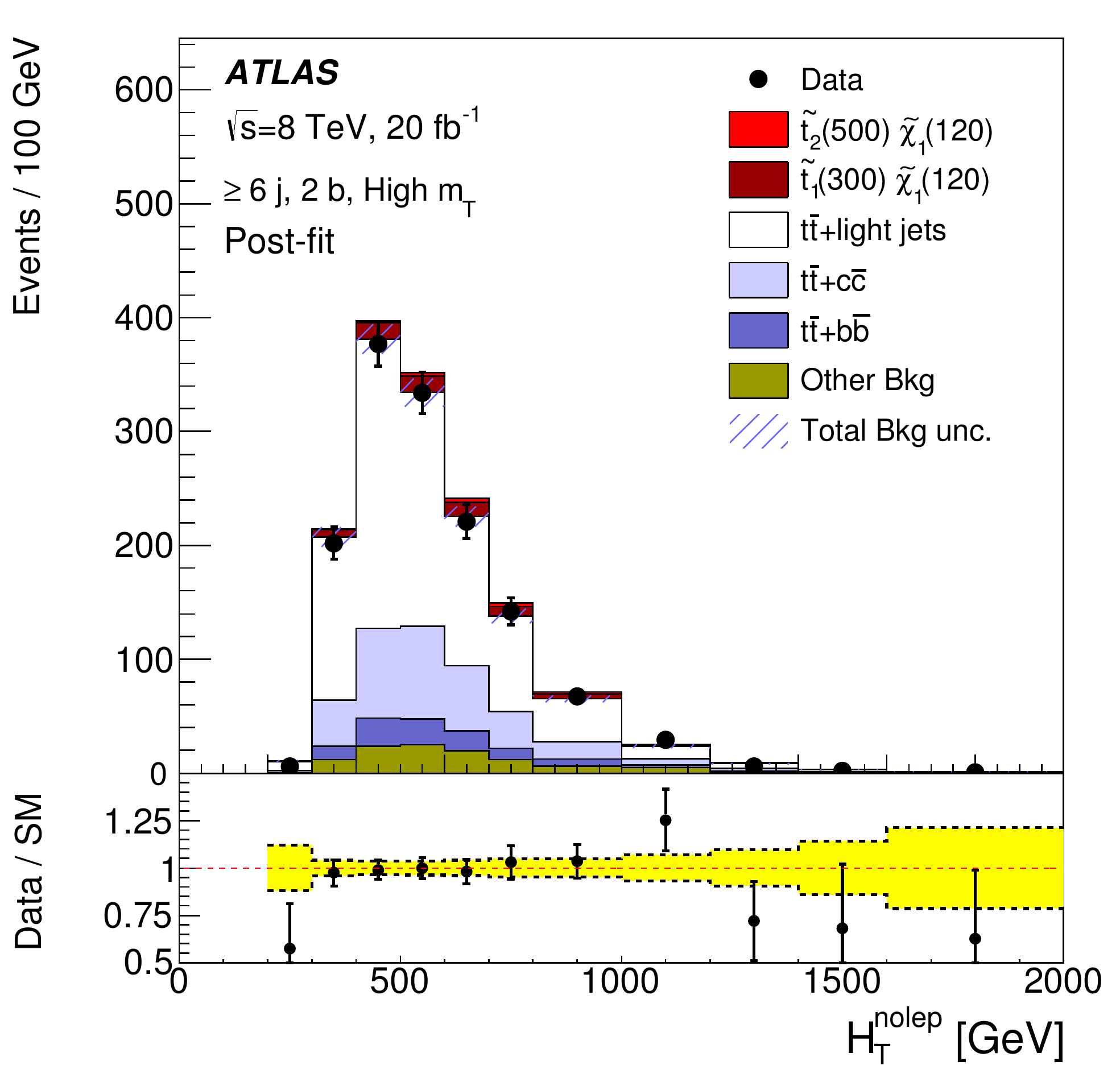}
	}\\
	\subfloat[]{
    	\includegraphics[width=0.35\textwidth]{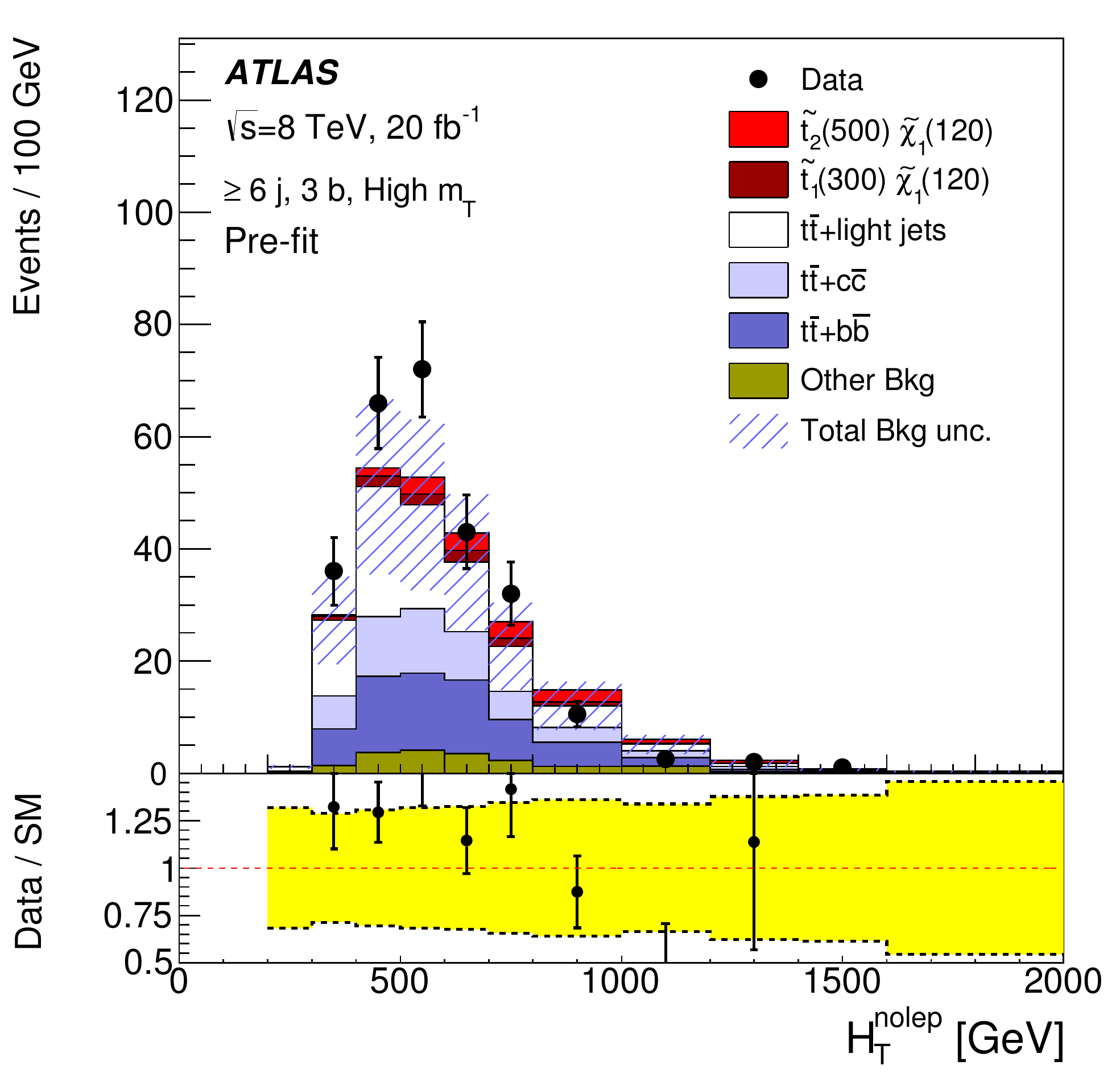}
	}
	\subfloat[]{
    	\includegraphics[width=0.35\textwidth]{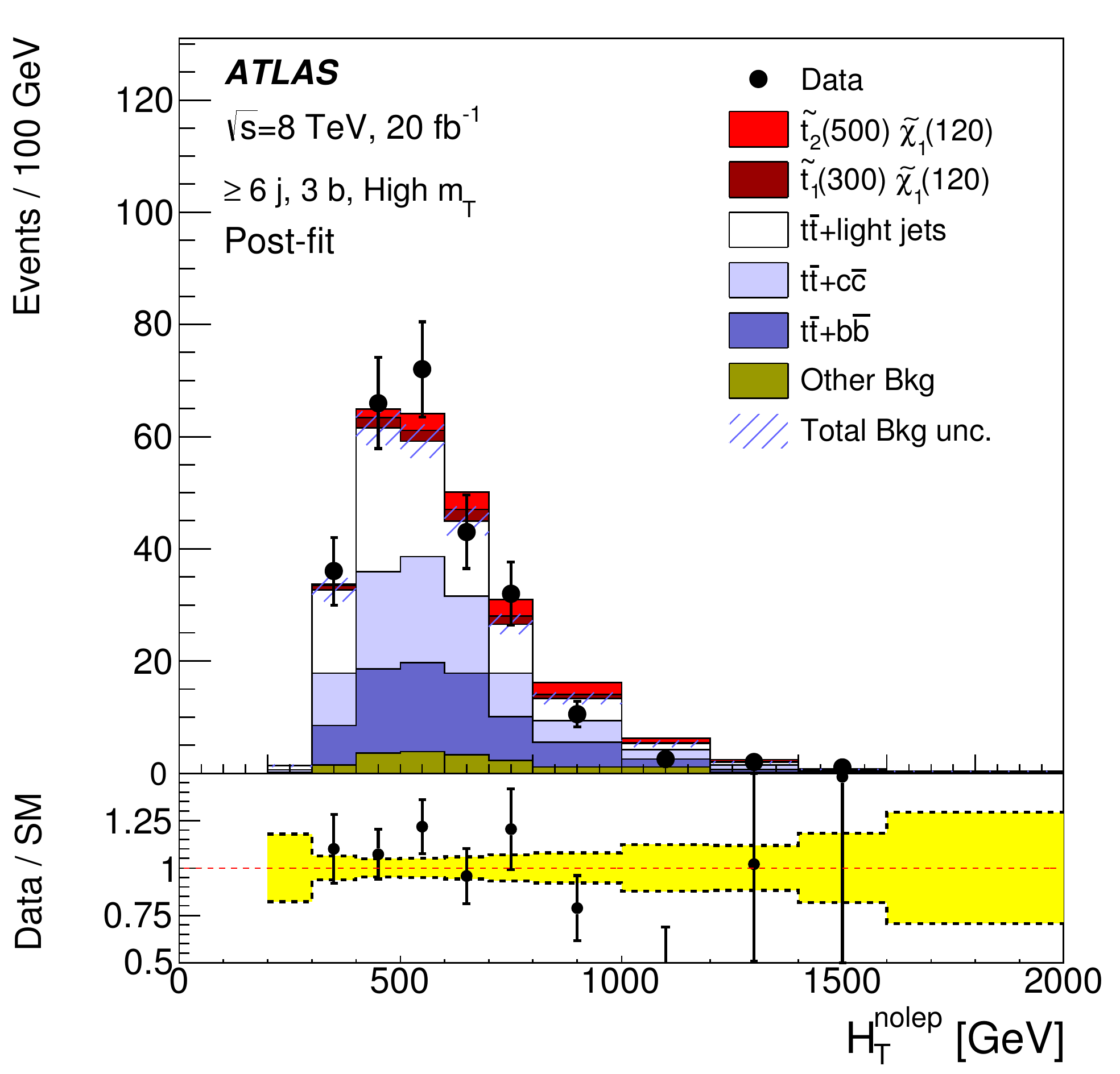}
	}\\
  	\subfloat[]{
    	\includegraphics[width=0.35\textwidth]{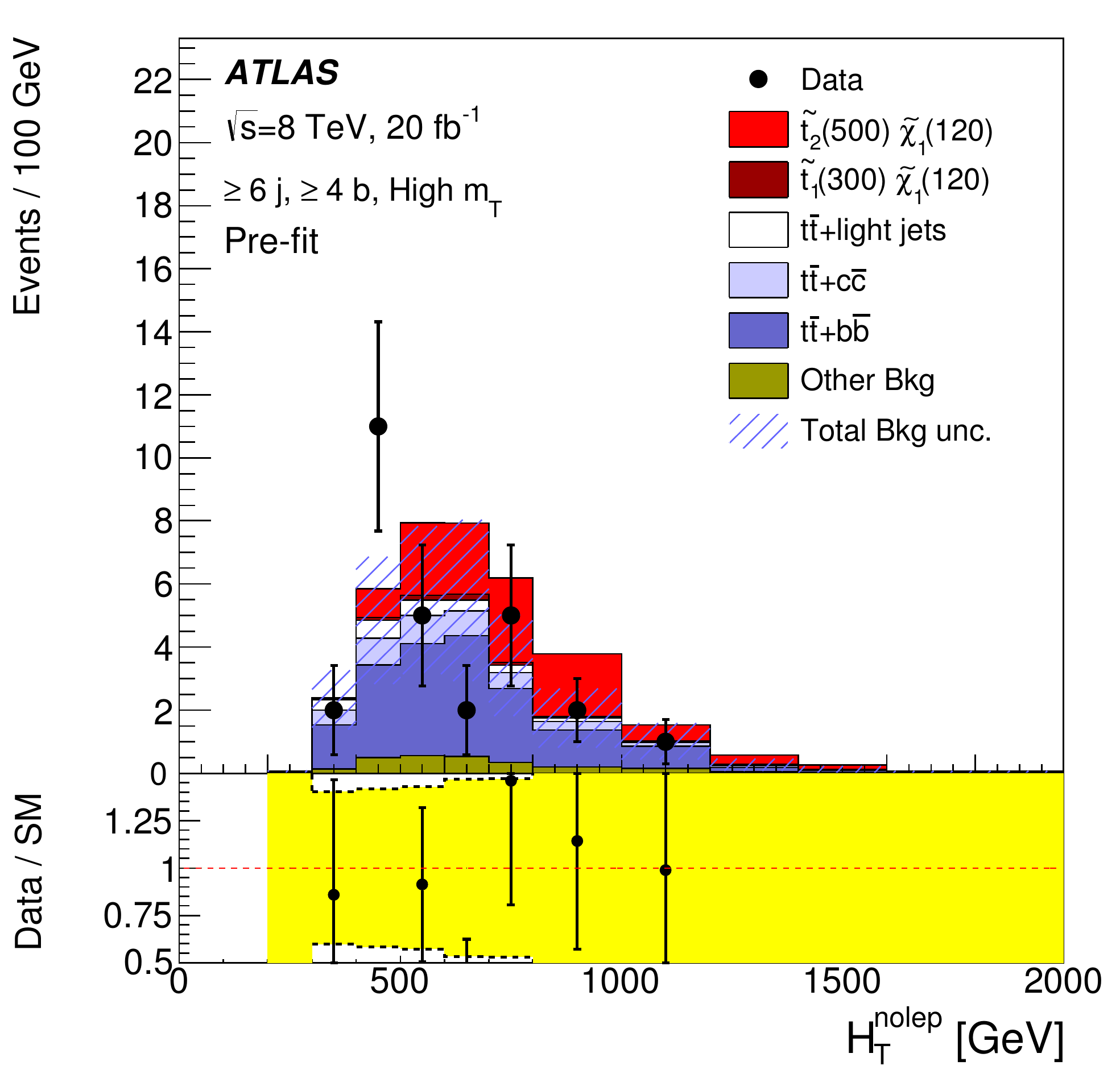}
	}
	\subfloat[]{
    	\includegraphics[width=0.35\textwidth]{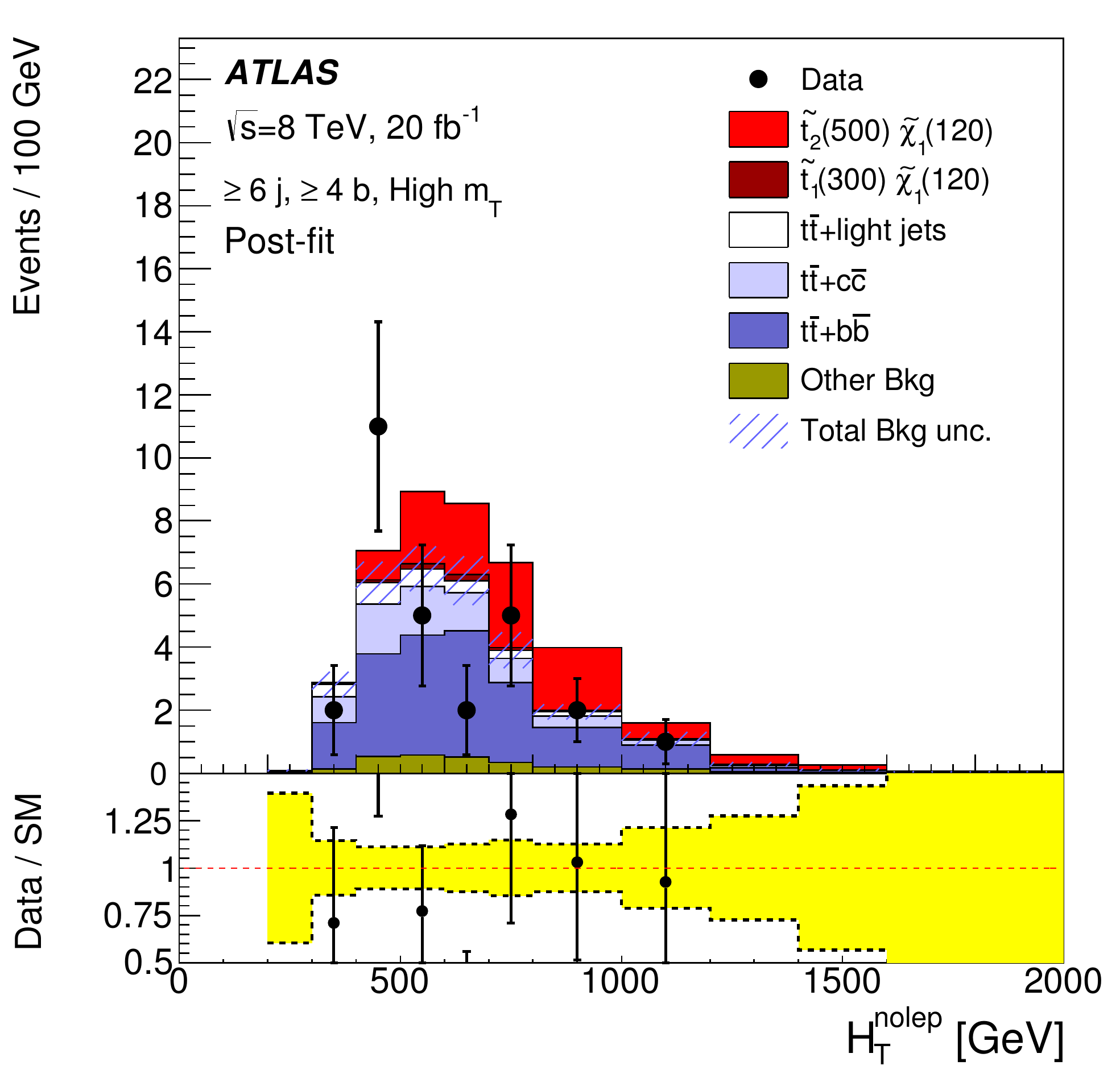}
	}
    \caption{Comparison between data and prediction for the distribution of \htnolep, , defined as the scalar sum of the missing transverse momentum and the transverse momenta of all selected jets, in the high-\mt\ channels considered: (top) two \btagged\ jets, (middle) three \btagged\ jets and (bottom) four \btagged\ jets, (left) before and (right) after the combined fit to data under the background-only hypothesis. The expected signal contributions from $\tone$ and $\ttwo$ pair production, assuming $m_{\ttwo} = 500$ GeV, $m_{\tone} = 300$ GeV, $m_{\neut} = 120$ GeV and a branching ratio of 100\% for $\ttwo \rightarrow \higgs \tone$  are also shown added to the stack (red histograms, in dark red the contribution from direct $\tone$ pair production). 
The bottom panel displays the ratio of the data to the total background prediction. The hashed area represents the statistical and systematics uncertainty on the background.}
    \label{fig:ttH_htnolep}
  \end{center}
\end{figure}

For a given \btag\ multiplicity, events are further categorised depending on the value of the transverse mass $\mt$ of the lepton and the missing transverse momentum. A ``low-\mt'' (``high-\mt'') region is defined by the requirement $\mt < 120$ GeV ($\mt > 120$ GeV). 

The final discriminating variable used is \htnolep, defined as the scalar sum of \etmiss\ and the transverse momenta of all selected jets. The signal is searched for by performing a binned likelihood fit to the \htnolep\ distribution simultaneously in the six channels defined (low/high-\mt\ for three bins in \btagged\ jet multiplicity). The binning used for the \htnolep\ distributions is that used in Figure~\ref{fig:ttH_htnolep}, where the background estimate both before and after the fit is compared to the data  in the high-\mt\ region. The dominant post-fit uncertainties are  those on the absolute normalisation of the $\ttbar+\bbbar$ and $\ttbar+\ccbar$ processes. 

The full list of detector systematic uncertainties considered, discussed in detail in Ref.~\cite{Aad:2015gra}, includes, beside a total uncertainty of 2.8\% on the integrated luminosity, systematic uncertainties on the identification efficiency and energy scale uncertainty of the leptons, reconstruction efficiency and energy scale and resolution uncertainties for jets, $b$-tagging efficiency and mis-tag rate uncertainties. Further modelling uncertainties are considered, which include, beside  production cross-section uncertainties for $W/Z$+jets, single top and \ttbar, dedicated uncertainties on the NLO calculation of the $\ttbar+\bbbar$ process and on the modelling of the $\ttbar+\ccbar$ component.  

No significant excess above the expected background is observed, hence 95\% CL limits are derived in a model where $\ttwo$ production is assumed, followed by the decay $\ttwo \rightarrow \tone \higgs$ (with a branching ratio of 100\%) and $\tone \rightarrow t \neut$ (again with a branching fraction of 100\%).\footnote{Production of $\tone$ pairs is also included in the simplified models. The acceptance of the selection for such events is very small. Nevertheless, this component is considered as signal in the statistical analysis.} The limit is derived as a function of the $\ttwo$ and $\neut$ masses, under the assumption that $\DMstopN = 180$ GeV, and it is presented in Section~\ref{sec:stop2}.

\subsubsection{Final states containing two $b$-jets, a charged lepton, and missing transverse momentum (\TBM)}
\label{sec:tbmet}

Several phenomenological models, where both \chipm and \neut\ are lighter than the 
stop (or the sbottom), allow for the $\tone \rightarrow t \neut$, $\tone \rightarrow b\chipm$ and $\bone \rightarrow b \neut$, $\bone \rightarrow t \chipm$ decay channels to be open with competing branching ratios. Naturalness arguments require the higgsino mass parameter $\mu$ to be smaller than a few hundred GeV, while they impose virtually no constraint on the bino and wino mass parameters $M_1$ and $M_2$. If $\mu \ll M_1, M_2$, then the lightest chargino and neutralino masses are both of the order of $\mu$ and hence \dmCN\ is small. Therefore, pair production of stops can lead to $\tone\tone \rightarrow t\neut b\chipm \rightarrow t b \neut \neut ff'$, where $f$ and $f'$ represents low-\pt\ fermions emitted through $\chipm \rightarrow ff'  \neut$. Assuming both $f$ and $f'$ are too soft to be detected, the final state is characterised by the presence of a top quark, a bottom quark, and neutralinos escaping the detector. Similarly, $\bone$ pair production can lead to the same final state. Dedicated SRs are defined that target this topology, which is not well covered by the \stopZeroLep\  and \stopOneLep\ signal regions aimed at final states containing $t\bar{t}\etmiss$ and the \sbottom\ signal regions targeting $b\bar{b}\etmiss$ final states. 

Both the leptonic and hadronic decays of the
top quark have been studied, and the leptonic channel was found to give a better sensitivity
to the signal models of interest. The dominant SM background processes in the signal regions
are semileptonic \ttbar and single top production. The SM background is evaluated using a
combination of Monte Carlo and partially data-driven techniques.

Events are selected online by a trigger requiring the presence of one electron or muon. The online selection thresholds are such that the plateau efficiency is reached for lepton transverse momenta of 25 GeV.

The identification criteria for electrons, muons, 
jets and \etmiss follow the principles outlined in Appendix~\ref{sec:objrec}. In particular, electrons and muons are required to be isolated: the scalar $\pt$ sum of tracks in a cone $\Delta R = 0.2$ around the electron (muon) is required to be smaller than 10\% of the electron transverse momentum (1.2 GeV). The electron or muon track is excluded from the sum. The \btagging\ algorithm is used at an operating point with 70\% efficiency in simulated top-quark pair production events.
Signal regions are defined as detailed in Table~\ref{tab:tbmetSR}, requiring one and only one electron or muon, 
two \btagged jets and a large \etmiss. Three of the SRs, labelled \TBM--SRin have no additional jet veto applied, while one of them (\TBM-SREx1)
has a veto requirement on the number of jets ($N_{\mathrm{xjets}}$) 
with \pt$>$50~GeV  in addition to the two leading \btagged jets. The final SR optimisation is performed
by using selections on the momenta of the objects, the \mt\ and the \meff\ variables. In addition, the following kinematic 
variables are used in the event selection:

\begin{itemize}
\item[-] $\dphibmin$: The minimum azimuthal distance between
the closest $b$-tagged jet and the $\met$.
This variable is used to remove multijet backgrounds with a cut of $\dphibmin >0.4$.

\item[-] \meff: The scalar sum of the $\pt$ of the two $b$-jets (with $\pt > 25$ GeV and $|\eta| < 2.8$ ) plus at most one light jet (with $\pt > 25$ GeV and $|\eta| < 2.5$)  and the $\met$. 
The number of  light jets, $n$, included in this sum depends on the signal region under study, although $n=1$ was mostly used.

\item[-] $\metsig$: The ratio of the $\met$ to the square root of $H_{{\rm T}}$, which is the scalar sum of the $\pt$ of the two $b$-jets plus one light jet with $\pt > 25$ GeV and $|\eta| < 2.8$.

\item[-] \mt: The transverse mass of the lepton and the missing transverse momentum vector.

\item[-] \mbb: The invariant mass of the two \btagged jets.

\item[-] \mlb: The invariant mass of a \btagged jet and the charged lepton. This variable is bounded from above at $\mtop$ in $\ttbar$ production events. Since two jets are \btagged the variables 
\mlb(1) and \mlb(2) are defined to indicate the invariant mass constructed with the leading and subleading \bjet respectively.
The variable \mlbmin is also defined to indicate the minimum between \mlb(1) and \mlb(2).

\item[-] \amtTwo:
   The asymmetric stransverse mass \cite{Barr:2003rg} is a kinematic variable which can be used to separate processes in which two decays giving missing
transverse momentum occur. It is defined as follows:
  \begin{equation}
    \amtTwo^{2} ( \chi ) = \underset{ {\bf \not{q}}_{\rm T}^{(1)}   +   {\bf {\not q}}_{\rm T}^{(2)} =  {\bf \not{p}}_{\rm T}    }{min} \left [ max \left
 \{  \mt^{2} ( {\bf p}_{\rm T}(v_{1}) , \xspace {\centernot{\bf{q}}_{\rm T}}^{(1)} ; \chi) , \mt^{2} ( {\bf p}_{\rm T}(v_{2}) , {\bf\centernot{q}}_{\rm
T}^{(2)} ; \chi)  \right \}  \right ]
  \end{equation}
  \noindent where ${\bf p}_{\rm T}(v_{i})$ are reconstructed transverse momentum vectors, 
  ${\bf\centernot{q}}_{\rm T}^{(i)}$ represent the missing transverse momenta from the two decays, with a total missing transverse momentum, ${\bf
\centernot{p}}_{\rm T} $, and $\chi$ is a free parameter representing the unknown neutralino mass, which is assumed to be zero in the calculation.
  The \amtTwo\ variable is calculated with different choices for $ {\bf p}_{\rm T}(v_{1})$ and ${\bf p}_{\rm T}(v_{2})$, depending on the value of \mlb$(n)$ ($n = 1,2$), the invariant mass of the $n^{\mathrm{th}}$ \btagged jet $b_n$ and the lepton: 
  \begin{itemize}
  \item If $\mlb(i) < 170$ GeV and $\mlb(j) > 170$ GeV, then \amtTwo\ is calculated with $ v_{1} = b_{i} + \ell $ and $ v_{2} = b_{j}$;
  \item If $\mlb(1) < 170$ GeV and $\mlb(2) < 170$ GeV, then \amtTwo\ is 
evaluated using the two possible combinations for $v_1$ and $v_2$, and the minimum is used;
\item It both $\mlb(1)>170$~GeV and $\mlb(2)>170$~GeV the event is rejected. 
  \end{itemize}

The case of both \mlb(1) and \mlb(2) exceeding 170 GeV is irrelevant: only events with the minimum value of \mlb smaller than 170 GeV populate the control, validation and signal regions.

\end{itemize}

\begin{table}[!h]
\begin{center}
\caption{\label{tab:tbmetSR} Summary of  signal regions used by the \TBM\ analysis.}
\begin{tabular}{l c c c c}
\hline
\hline
SR        & \TBM-SRIn1 & \TBM-SRIn2 & \TBM-SRIn3  & \TBM-SREx1 \\ 
\hline
\bjets & \multicolumn{4}{c}{2 \bjets; \pt$>$25~GeV} \\
1 lepton &  \multicolumn{4}{c}{\pt$>$25~GeV} \\
         &  \multicolumn{4}{c}{$|\eta|<2.5\ (2.47)$ for $\mu\ (e)$} \\
\etmiss (GeV) & $>$200&  $>$120&  $>$220&  $>$160 \\
\mt\ (GeV) &  $>$140& $>$140&  $>$180&  $>$120\\
\meff (GeV) & $>$300 &  $>$450&  $>$650 &  $>$300 \\
\amtTwo (GeV) & $>$180&  $>$200&   $>$180& $>$180 \\
\mlbmin (GeV) & \multicolumn{4}{c}{$<$170} \\
\dphibmin & \multicolumn{4}{c}{$>$0.4} \\
\metsig  ($\GeV^{1/2}$)& $>8$ & $>12$  & $>5$ & $>10$  \\
$N_{\mathrm{xjets}}$ & -- & -- & -- & $<$2  \\
\hline
\hline
\end{tabular}
\end{center}
\end{table}

The optimisation is carried out using both a pMSSM signal model and simplified models where $\Delta m(\chipm, \neut)=5$ or 10~GeV. In the case of the pMSSM model, additional non-\btagged jets are expected in the final state via the production of other SUSY particles, hence the optimisation
points to SRs with no requirement on the $N_{\mathrm{xjets}}$ variable (\TBM-SRIn). In the case of the simplified models, additional 
jets come only from initial- or final-state radiation, and as a consequence a strict selection on $N_{\mathrm{xjets}}$ is applied as in the selection \TBM-SREx1.

The main SM backgrounds are top-pair production, $W$ production in association with heavy-flavour jets and single-top 
production. The MC cross section is used to normalise the single-top background and all the other minor SM backgrounds, such as 
$Z$+jets, diboson production, \ttW and \ttZ. 
The normalisation factors of the \ttbar\ and \Wjets\ backgrounds are determined by a combined profile-likelihood fit. 
Specific control regions, whose event yield is expected to be dominated by each of these production processes, are defined and included in the fit to constrain the normalisation parameters. 
The \ttbar\ control regions (CRT) are defined by inverting the selection on $\amtTwo$, requiring $\amtTwo<$160 (180)~GeV for the inclusive
(exclusive) SRs. The purity of the \ttbar\ process in the CRTs is in excess of 95\%. 
The \Wjets\ control regions (CRW) are defined by requiring $\mt<$120~GeV. For the control regions corresponding to the \TBM-SRIn, events with one \btagged\ jet are included in the CRW. Top quark pair production dominates the CRWs, with a \Wjets\ purity of 30\% or better. The normalisation factors $\mu_W$ and $\mu_{\ttbar}$ are presented
in Table~\ref{tab:tbmetSF}. The background model is then validated using  validation regions, where little signal contamination is expected.

\begin{table}[!htbp]
\begin{center}
\caption{Background scale factors for the \ttbar and $W$ samples, as obtained by the background fit. The errors include both the statistical and systematics uncertainties.\label{tab:tbmetSF}}
\renewcommand\arraystretch{1.2}
\begin{tabular}{ccccc}
\hline
\hline
Norm. Factor & SRinA & SRinB & SRinC & SRexA \\
 \hline
$\quad \mu_{t\overline{t}} \quad$ & $\quad 1.06 \pm 0.07 \quad$ & $\quad 1.12 \pm 0.09 \quad$ & $\quad 0.94 \pm 0.21 \quad$ & $\quad 1.06 \pm 0.07$  \\
$\quad \mu_W \quad$ & $\quad 0.92 \pm 0.20 \quad$ & $\quad 0.61 \pm 0.23 \quad$ & $\quad 0.93 \pm 0.27 \quad$ & $1.10 \pm 0.34 $ \\
\hline 
\end{tabular}
\end{center}
\end{table}

The distributions of the variable $\amtTwo$ in the four SRs are shown in Figure~\ref{fig:tbmet_amt2} together with the expected 
distribution from some of the signal models used to optimise the analysis. Table~\ref{tab:tbmetresults} 
compares the predicted and observed numbers of events in each of the signal regions. 
No excess above the SM prediction is observed, hence the results are first used to derive model-independent 
95\% CL exclusion limits on the  number of events beyond the Standard Model in the signal region, and then to extract limits on 
$\sigma_{\mathrm{vis}} = \sigma \times \epsilon \times \mathcal{A}$, where $\sigma$ is the cross section for non-SM 
processes, $\epsilon$ is the selection efficiency and $\mathcal{A}$ is the selection acceptance. All these 
limits are also reported in Table~\ref{tab:tbmetresults}.

 \begin{figure}[!htb]
\begin{center}
\subfloat[]{
\includegraphics[width=0.45\columnwidth]{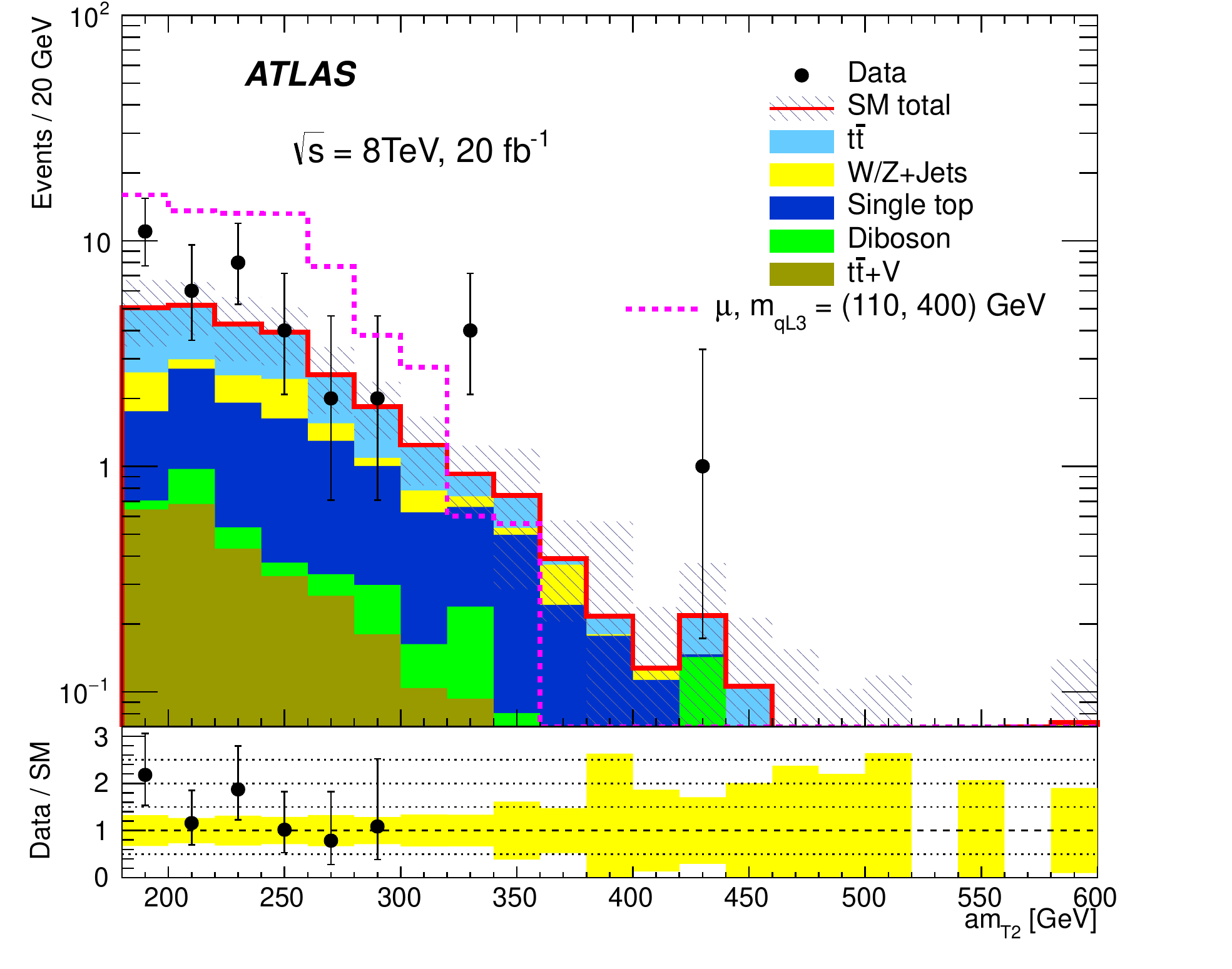}
}
\hfill
\subfloat[]{
        \includegraphics[width=0.45\columnwidth]{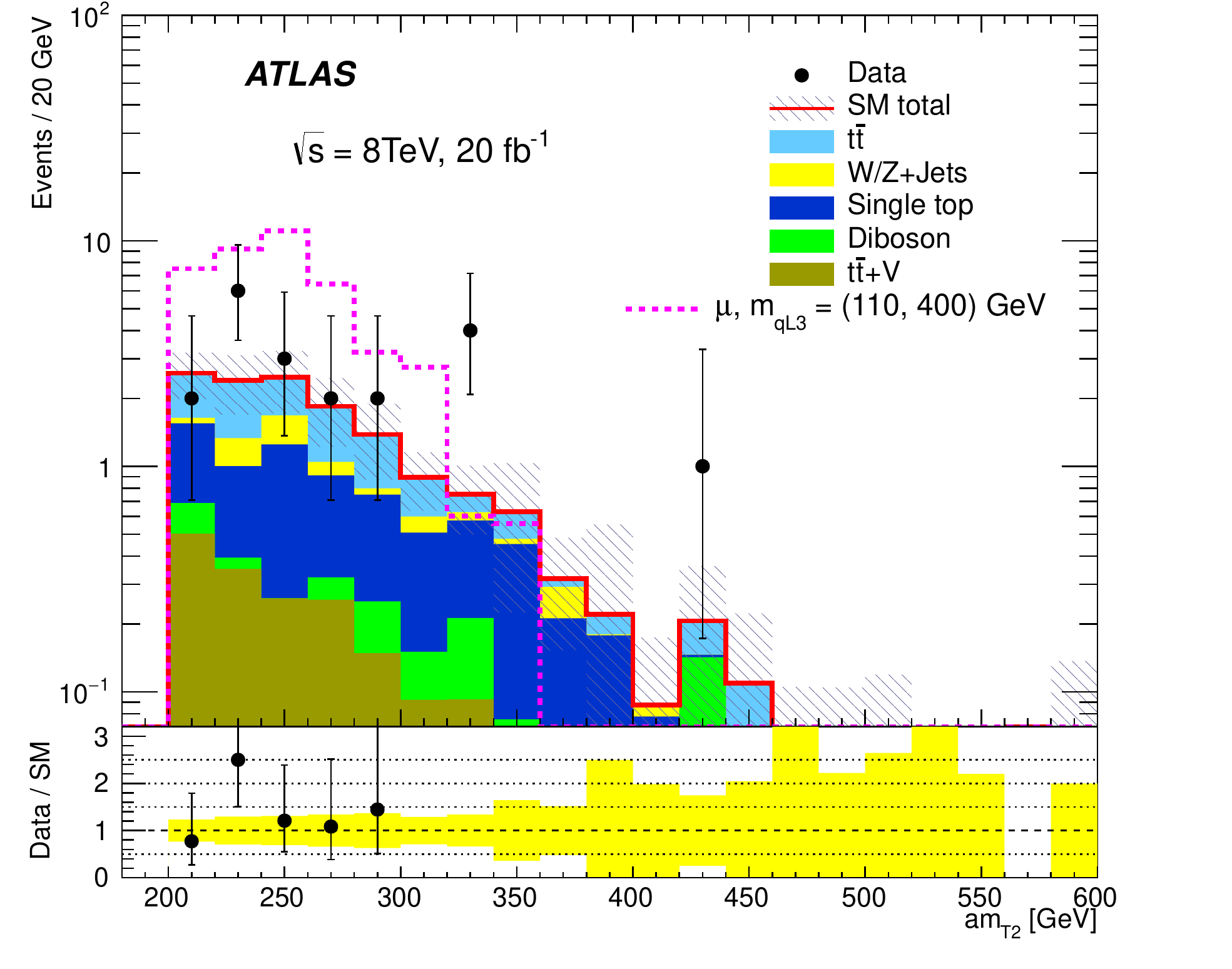}
        }
        \hfill
\subfloat[]{
\includegraphics[width=0.45\columnwidth]{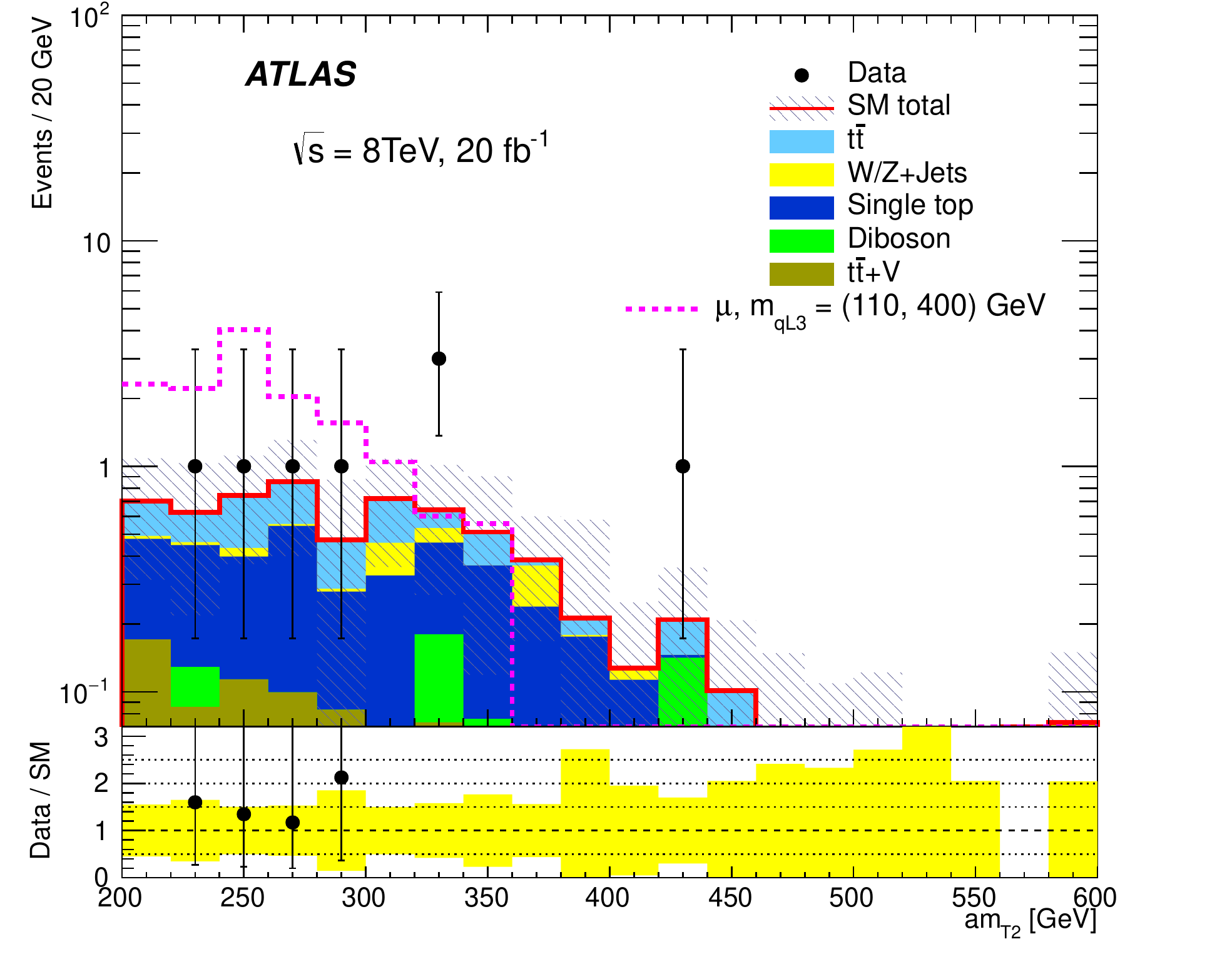}
}
\hfill
\subfloat[]{
\includegraphics[width=0.45\columnwidth]{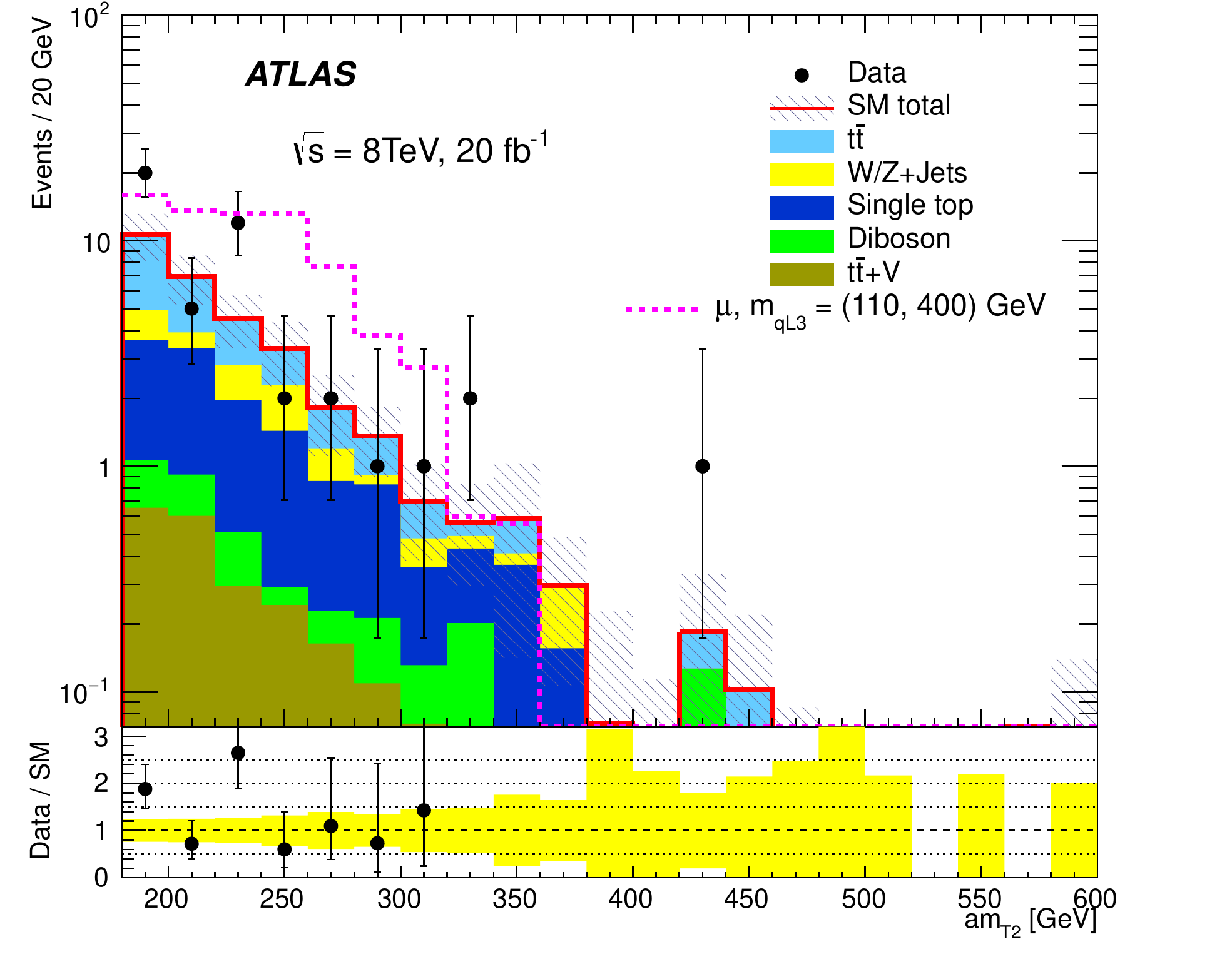}
}
\end{center}
\caption{Distribution of the asymmetric stransverse mass $\amtTwo$ in the (a) SRinA, (b) SRinB (top right), (c) SRinC and 
(d) SRexA defined in the text. The contributions from all SM processes are shown as a histogram stack. 
The contribution from signal points studied by this analysis are also shown.
The lower panels show the ratio between the data and the SM prediction; the  band includes statistical and systematic uncertainties on the SM prediction.
}
\label{fig:tbmet_amt2}
\end{figure}

\begin{table} [htb!]
\begin{center}
  \caption{Observed (Obs) and predicted (Exp) numbers of events in the signal regions of the \TBM\ analysis, 
together with the 95\% CL upper limits on the observed and expected number of signal events ($S_{\rm obs}^{95}$ 
and $S_{\rm exp}^{95}$ respectively), and on the visible cross section ($\langle\epsilon{\rm \sigma}\rangle_{\rm obs}^{95}$).
  \label{tab:tbmetresults}}
\renewcommand\arraystretch{1.2}
\begin{tabular}{lccccc}
\hline
\hline
{\bf Signal channel}                        & Obs & Exp &  $S_{\rm obs}^{95}$  & $S_{\rm exp}^{95}$ &  $\langle\epsilon{\rm \sigma}\rangle_{\rm obs}^{95}$[fb]  \\
\hline
SRinA  &  38 & $27 \pm 7$     &  $28.5$  & ${19.3}^{+7.0}_{-6.1}$ & $1.41$ \\
SRinB  &  20 & $14.1 \pm 2.8$ &  $16.3$  & ${10.7}^{+4.5}_{-2.6}$ & $0.81$   \\
SRinC  &  10 & $7.1 \pm 2.9$  &  $11.9$  & ${9.8}^{+3.3}_{-2.4}$  & $0.58$  \\
SRexA  &  46 & $31\pm 7$      &  $32.1$  & ${20.3}^{+8.0}_{-3.6}$ & $1.58$ \\
\hline
\hline
\end{tabular}
\end{center}
\end{table}

Since the number of events observed agrees with the SM predictions, 95\% CL exclusion limits are derived in specific supersymmetric models of direct pair production of stops. Simplified
models were simulated with the two decays $\tone \rightarrow t \neut$, $\tone \rightarrow b\chipm$ each having a 50\% BR for 
values of \dmCN=5, 20 GeV. Furthermore, by using a weighted combination of these simplified models with models corresponding
to a 100\% BR in either $\tone \rightarrow t \neut$ or $\tone \rightarrow b\chipm$, limits can be obtained for any value of the
stop BR. Figure~\ref{fig:AsymALLBRS} shows the exclusion limits for BR($\tone \rightarrow t \neut$)=25\%, 50\% and 75\% for 
the two values of \dmCN considered.

Finally, 95\% CL exclusion limits are also derived for a natural pMSSM model and are presented in Figure~\ref{fig:tbmetpMSSM}

\begin{figure}[!ht]
\centering
\subfloat[]{
\includegraphics[width=0.45\columnwidth]{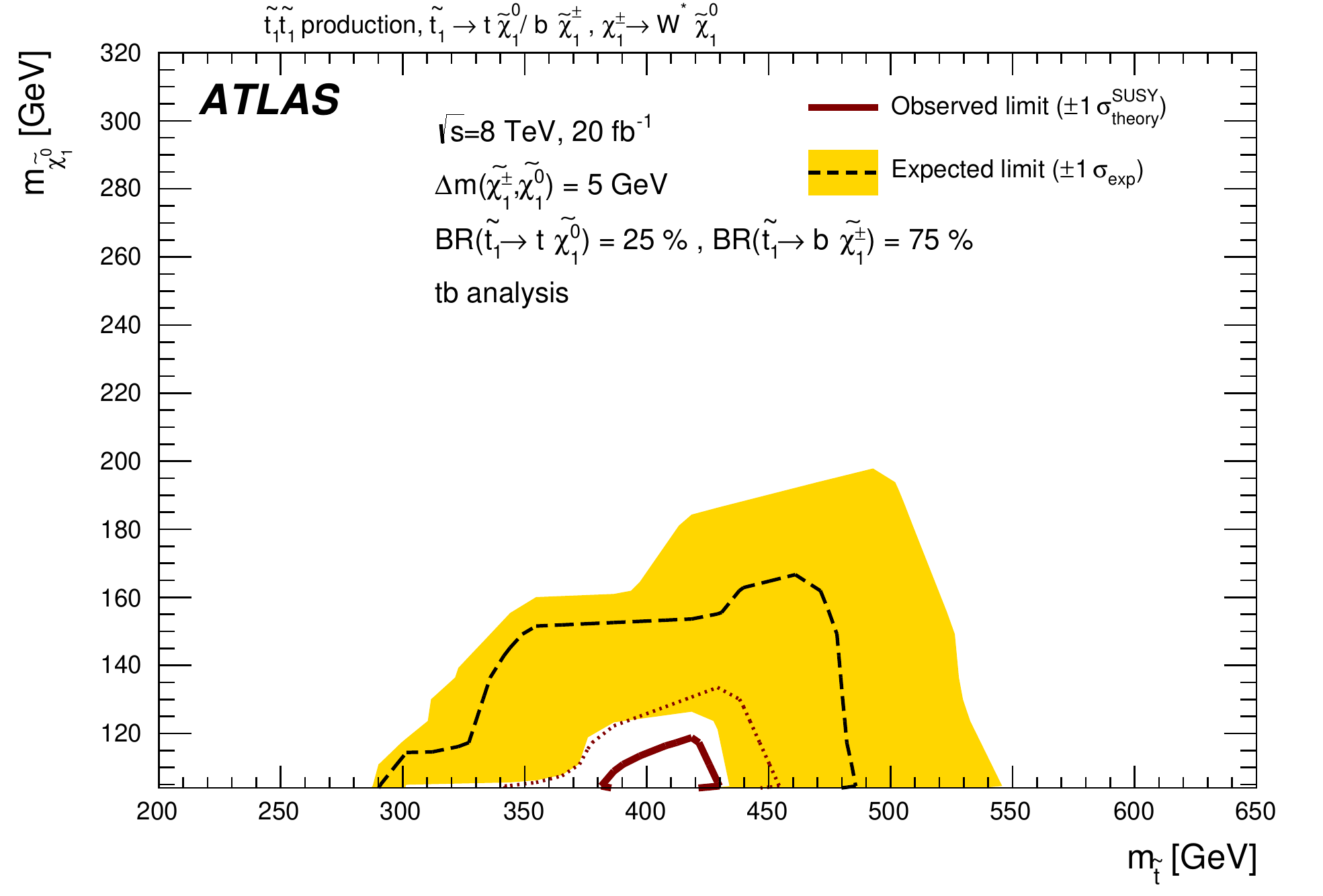}
}
\hfill
\subfloat[]{
\includegraphics[width=0.45\columnwidth]{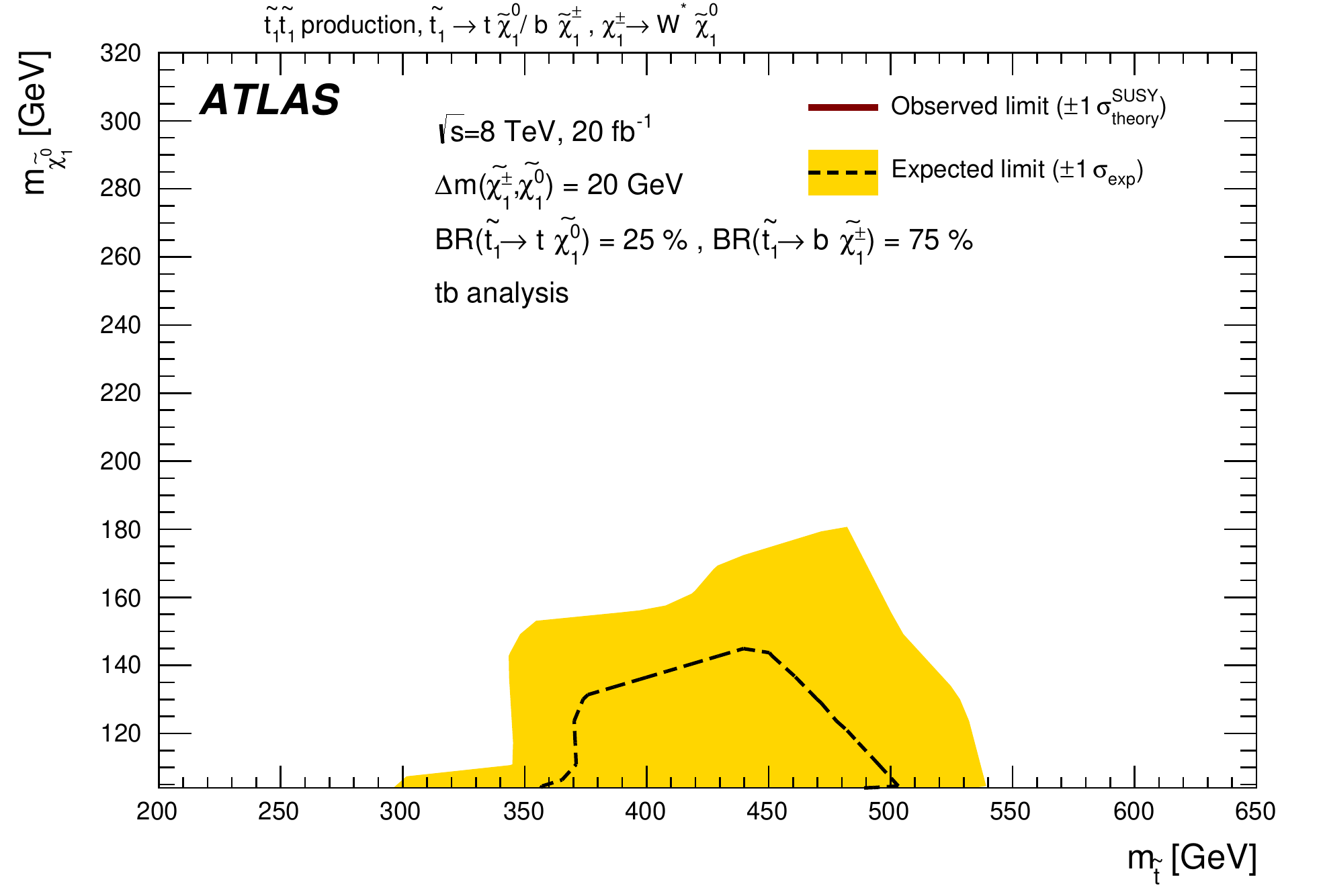}
}
\hfill
\subfloat[]{
\includegraphics[width=0.45\columnwidth]{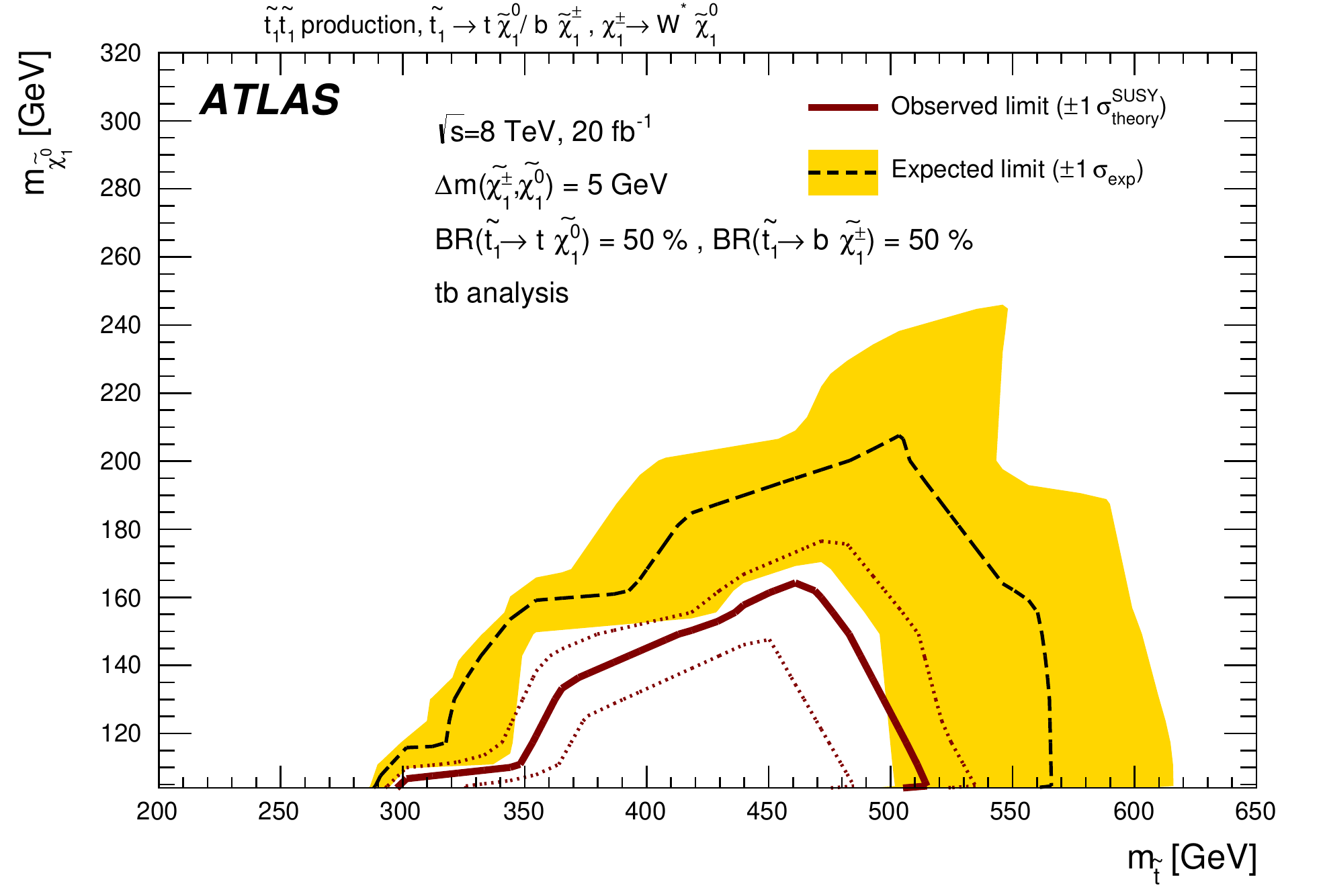}
}
\hfill
\subfloat[]{
\includegraphics[width=0.45\columnwidth]{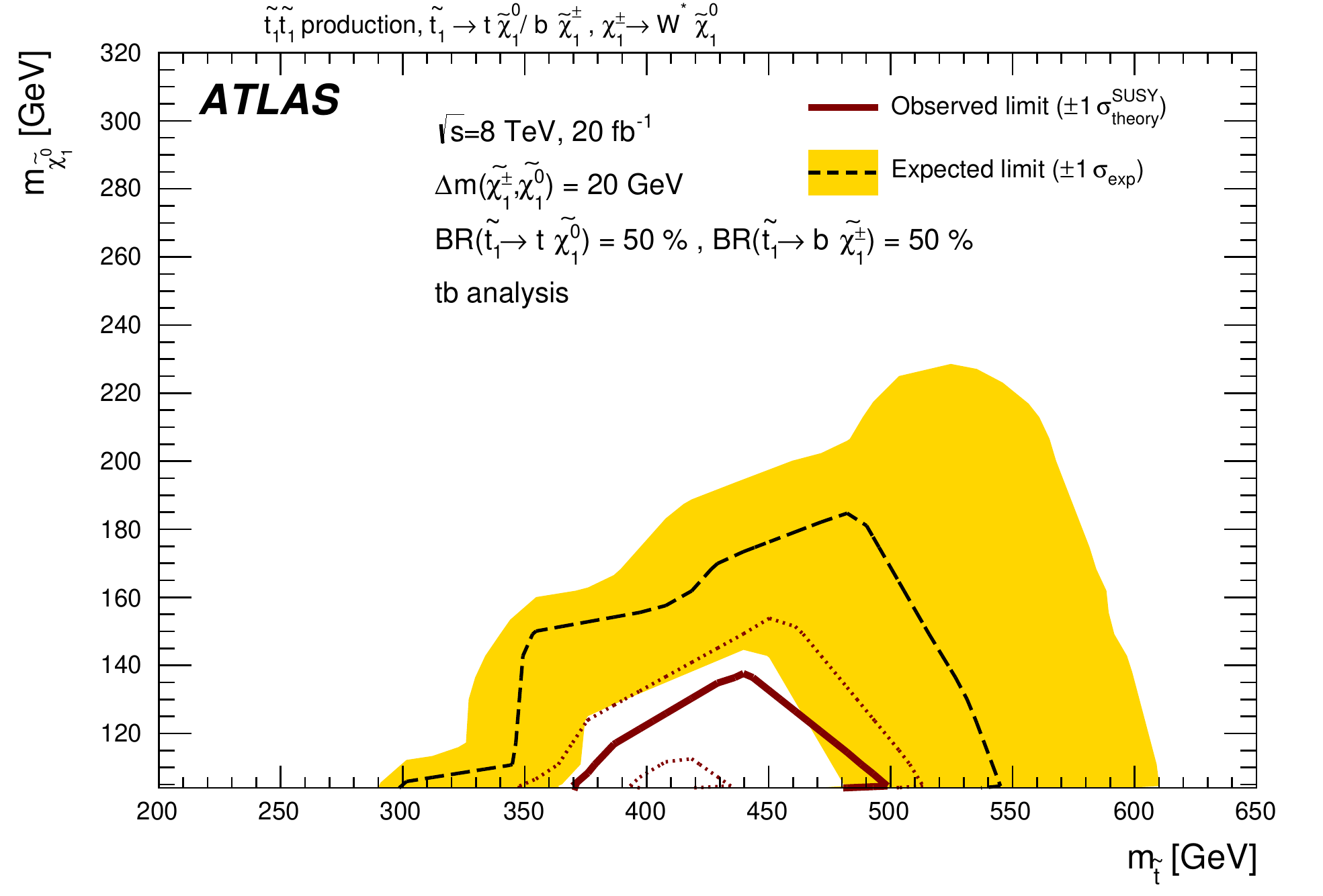}
}
\hfill
\subfloat[]{
\includegraphics[width=0.45\columnwidth]{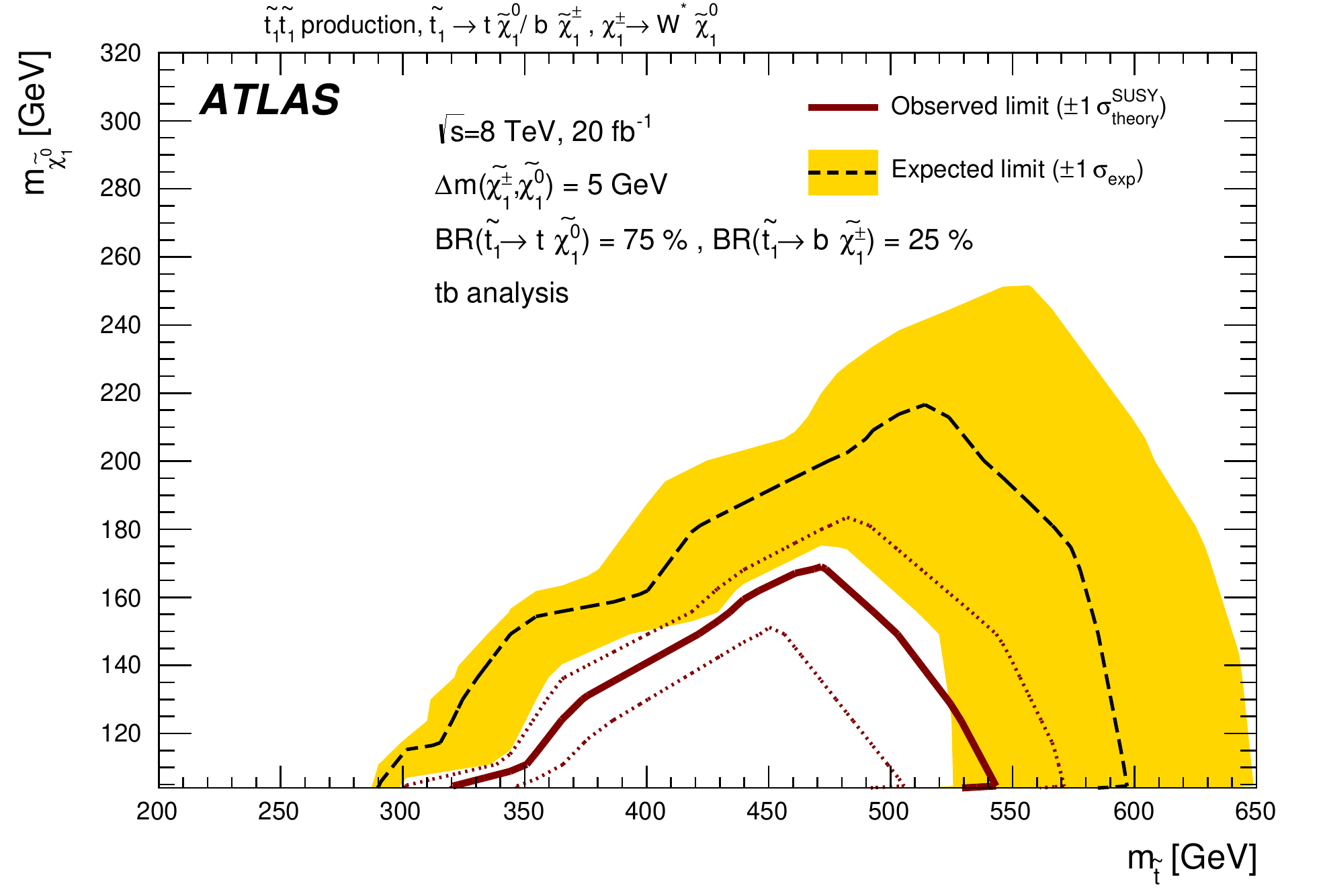}
}
\hfill
\subfloat[]{
\includegraphics[width=0.45\columnwidth]{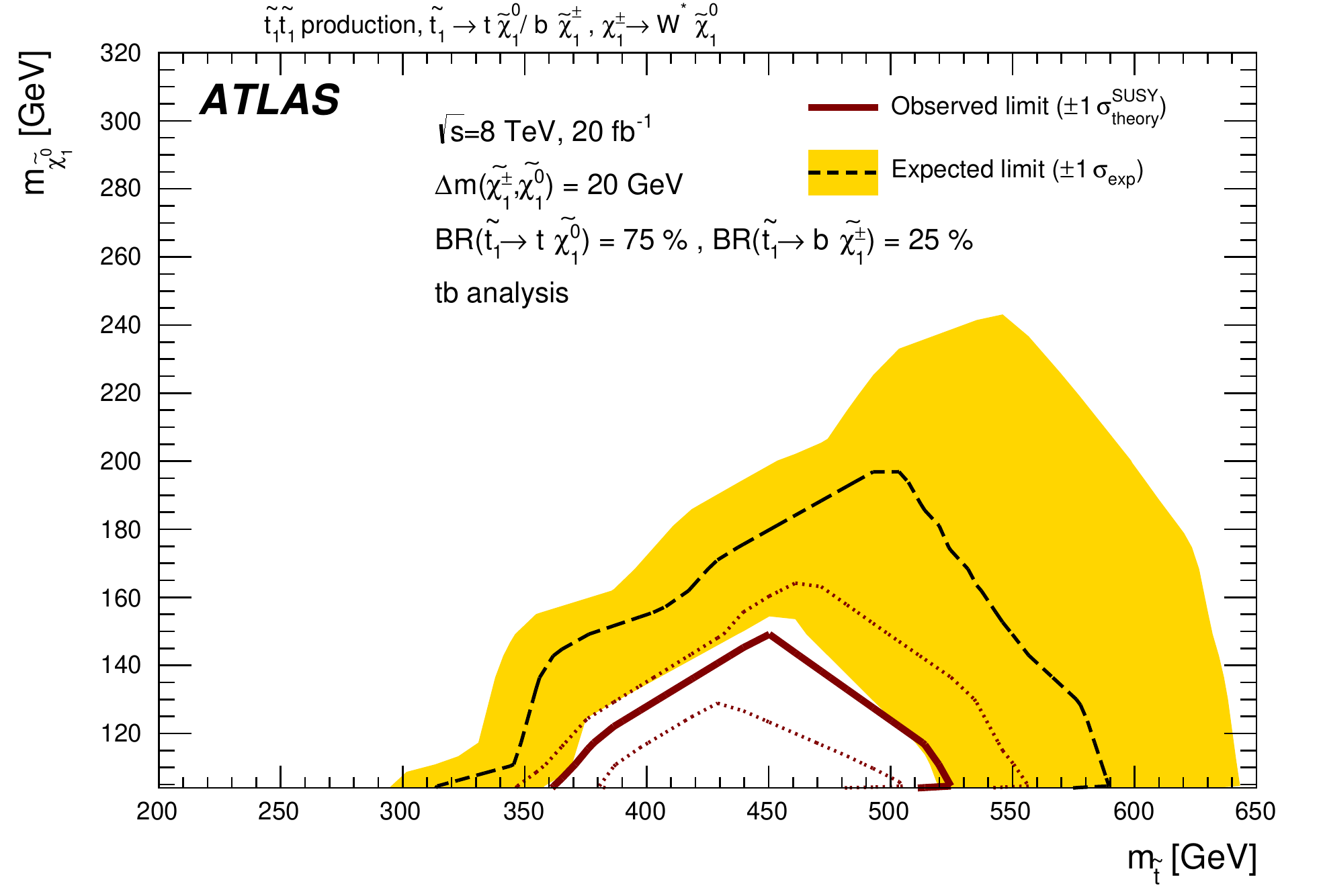}
}
\caption{Exclusion limits at 95\% CL from the \TBM\ signal regions for simplified models with stop decays into both 
$\tone \rightarrow t \neut$ and $\tone \rightarrow b\chipm$ and 
for BR($\tone \rightarrow t \neut$)=25\%, 50\%, 75\% (in descending rows) for the  grids $\dmCN = 5,20$ (left, right columns). The black dashed line indicates the expected limit, and the yellow band indicates the $\pm 1\sigma$ uncertainties, which include all
      uncertainties except the theoretical uncertainties in the
      signal. The red solid line indicates the observed limit, and
      the red dotted lines indicate the sensitivity to
      $\pm 1\sigma$ variations of the signal theoretical
      uncertainties. For each point the SR giving the best expected significance is used.  }
\label{fig:AsymALLBRS}
\end{figure}

\begin{figure}[!ht]
\centering
\includegraphics[width=0.44\columnwidth]{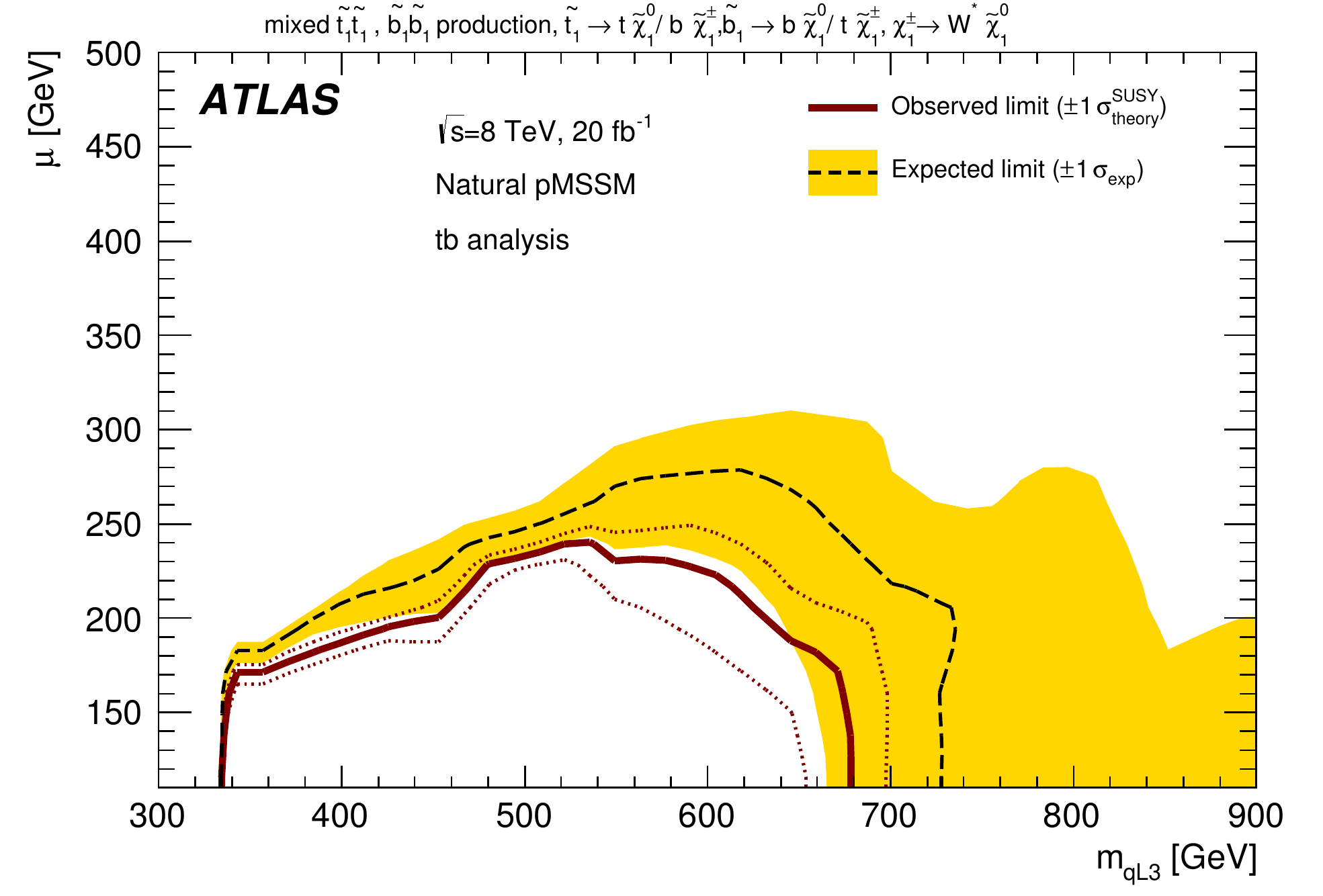} 
\caption{Exclusion limits at 95\% CL from the \TBM\ signal regions for the natural pMSSM model. The black dashed line indicates the expected limit, and the yellow band indicates the $\pm 1\sigma$ uncertainties, which include all
      uncertainties except the theoretical uncertainties in the
      signal. The red solid line indicates the observed limit, and
      the red dotted lines indicate the sensitivity to
      $\pm 1\sigma$ variations of the signal theoretical
      uncertainties. For each point the SR giving the best expected significance is used.}
\label{fig:tbmetpMSSM}
\end{figure}

\section{Further details of the statistical combination of the \stopZeroLep\ and \stopOneLep\ signal regions}
\label{sec:01lepCombine}

This section provides additional details on the combination of the \stopZeroLep\ and \stopOneLep\ signal regions targeting scenarios in which the stop decays into either $\tone \rightarrow t \neut$ or the mixed case where $\tone \rightarrow t \neut$ and  $\tone \rightarrow b \chipm$ are both allowed, as discussed in Section~\ref{sec:01lep_summary}

The statistical combination of the two analyses is performed by running the combined fit simultaneously on the control and signal regions of the two analyses. The detector systematic uncertainties are treated as correlated by using, for each of the uncertainties considered, a single nuisance parameter. The supersymmetric signal parameter strength used is the same for the two analyses, while the normalisation parameters for the background processes are kept independent in each analysis.\footnote{The choice is motivated by the fact that the phase-space regions in which the two analyses determine the normalisation parameters of the \ttbar, \Zjets\ and \Wjets\ (for \stopZeroLep) and \ttbar\ and \Wjets\ (for \stopOneLep) are characterised by different kinematic selections and jet multiplicities.} The nuisance parameters associated with modelling uncertainties of the various processes are also kept independent.

The control regions of the two analyses are not mutually exclusive: the events that belong to both a CR of \stopZeroLep\ and one of \stopOneLep\ are, at most about 2\% of the total number of events of the \stopZeroLep\ CR. The strategy adopted is to remove them from the corresponding \stopZeroLep\ CR for the combination. It has been verified that such removal does not affect the individual results of the \stopZeroLep\ analysis.

For each combination performed, the fit setup is validated by checking that the background normalisation parameters obtained are compatible with those obtained separately by the two analyses, by verifying that no additional constraint on the nuisance parameters is introduced with respect to the individual fits, and by checking that no artificial correlation is introduced between any of the fit parameters.

The 95\% CL limit derived from the combination is shown in Figure~\ref{fig:01lep_combine_100}, where the combined limit is compared to the individual limits obtained by the \stopZeroLep\ and \stopOneLep\ analyses independently.
\begin{figure}[htbp]
  \begin{center}
    \includegraphics[width=0.7\textwidth]{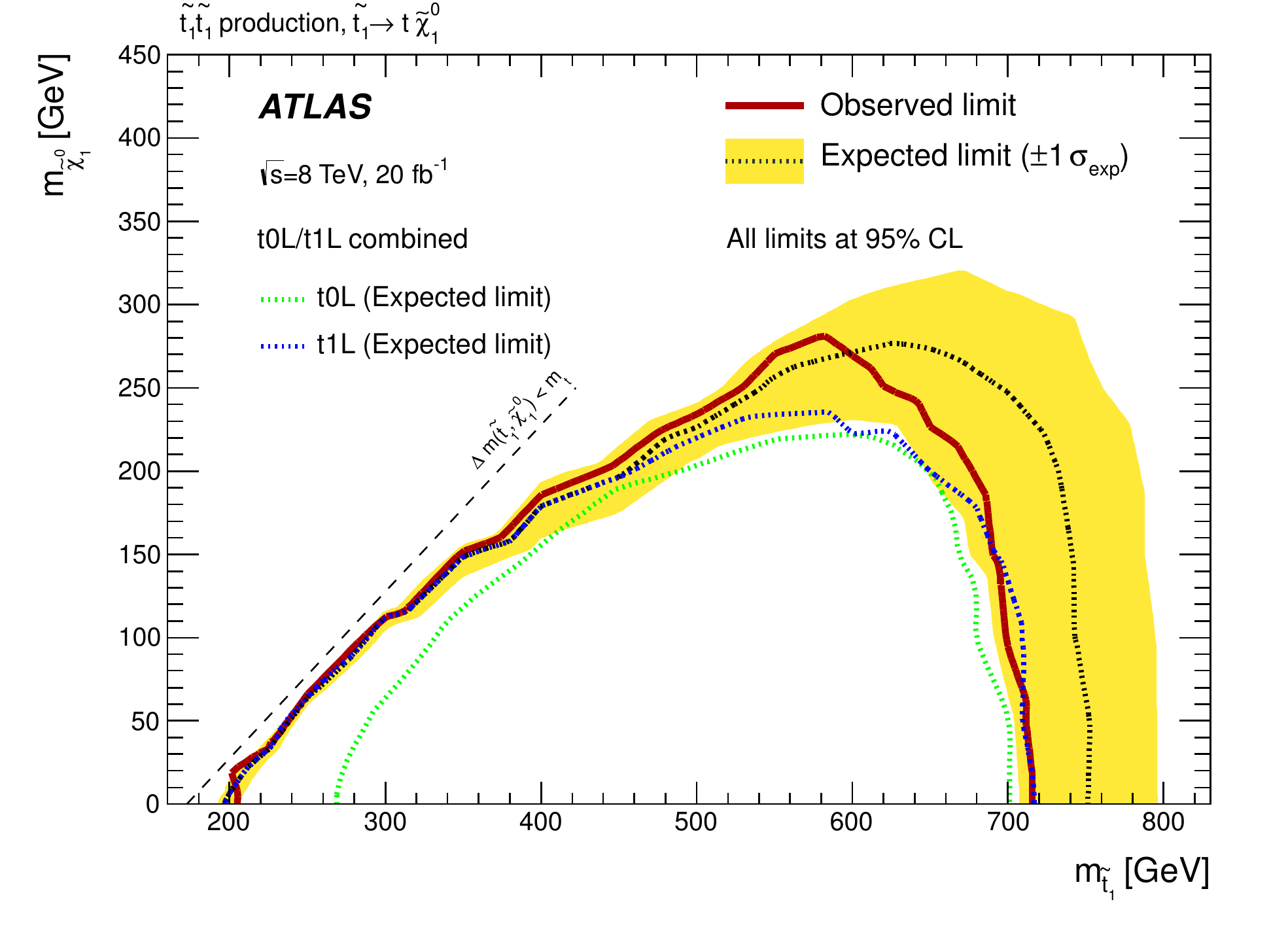}
    \caption{Combined exclusion limits at $95\%$ CL in the scenario where both stops decay exclusively via $\tone \rightarrow t \neut$. The black dashed line indicates the expected limit, and the yellow
      band indicates the $\pm 1\sigma$ uncertainties, which include all
      uncertainties except the theoretical uncertainties in the
      signal. The red solid line indicates the observed limit. For comparison the dotted green and blue lines show the expected
      limits from the standalone \stopZeroLep\ and \stopOneLep\ analyses.}
    \label{fig:01lep_combine_100}
  \end{center}
\end{figure}

\section{Signal generation details}
\label{sec:signal_generation}

Several SUSY models are considered throughout this paper. This section provides the details of how these signal models are generated. For all SUSY models discussed below, the detector response is simulated by passing the generated events  through a detector simulation~\cite{atlassimulation} based on \geant~\cite{geant4} or through a fast simulation using a parametric response to the showers in the electromagnetic and hadronic calorimeters~\cite{ATL-PHYS-PUB-2010-013} and \geant-based simulation elsewhere. All samples are produced with a varying number of simulated minimum-bias interactions overlaid on the hard-scattering event to account for multiple $pp$ interactions in the same or nearby bunch crossings (pileup). The simulation is reweighted to match the number of minimum bias interactions in data, which varies between approximately 10 and 30 interactions in each bunch crossing. Corrections are applied to the simulated samples to account for differences between data and simulation for the trigger and reconstruction efficiencies, momentum scale and resolution of the final-state objects, including the efficiency of identifying jets originating from the fragmentation of \bquarks, together with the probability for mis-tagging light-flavour and charm quarks.

\paragraph{Simplified models:}

The signal samples for the scenario where both stops decay to a top quark and a neutralino are generated using \herwigpp~2.5.2~\cite{Bahr:2008pv} interfaced to \pythia~6.426~\cite{Sjostrand:2006za}. The neutralino is fixed to be a pure bino, enhancing the decay of the $\tright$ component of $\tone$ to a right-handed top quark. Signal samples where the two stops decay as $\tone \rightarrow b \chipm$ are generated with \madgraph~5.1.4.8~\cite{Alwall:2011uj}. For models where the $W$ boson is on-shell, the $\tone$ decay is treated by \madgraph, while if the $W$ is off-shell, \pythia is used to decay the $\tone$. In these samples, the $\tone$ is assumed to be mostly a $\tleft$, and the chargino is assumed to decay through $\chipm \rightarrow W^{(*)} b$ with a branching ratio of 100\%. Several assumptions about the chargino masses are considered as described in the body of the paper.

Models in which the stop is assumed to decay either as $\tone \rightarrow t \neut$ or $\tone \rightarrow b \chipm$ with different branching ratios are obtained by appropriately weighting three samples: one where both stops decay through $\tone \rightarrow t \neut$, a second one where both stops decay through $\tone \rightarrow b \chipm$, and a third one, where one of the two pair-produced stops decays as $\tone \rightarrow t \neut$ and the other one decays as $\tone \rightarrow b \chipm$. This last sample is generated with \madgraph, the $\tone$ is assumed to be a maximal mixing of $\tleft$ and $\tright$. The mass of the chargino in this sample satisfies the gauge-universality relation $m_{\chipm} = 2 m_{\neut}$.

The three-body stop decay samples are generated with \herwigpp, which performs the matrix element calculation of the three-body decay. The four-body stop decay samples are generated with \madgraph.

For all samples considered, the mass of the bottom quark is fixed to 4.8 GeV and its width is assumed to be zero.

The samples where both stops decay as $\tone \rightarrow c \neut$ are generated with \madgraph, with one additional parton from the matrix element. Similarly to the case of the limit derived as a function of the stop branching ratio into $\tone \rightarrow t \neut$, the samples with both stops decaying as  $\tone \rightarrow c \neut$ and those where both stops decay through the four-body decay are appropriately weighted and combined with a third sample where one stop decays as  $\tone \rightarrow c \neut$ and the other decays through the four-body decay to produce a sample of arbitrary branching ratio into $\tone \rightarrow c \neut$ (assuming that $\tone \rightarrow c \neut$ and the four-body decay are the only possible stop decays). Such mixed samples are also generated with \madgraph.

Sbottom pair production samples are also all produced with \madgraph interfaced to \pythia, and no more than one additional parton is added to the matrix element. 
The PDF set used for all signal samples is CTEQ6L1~\cite{Pumplin:2002vw}.

\paragraph{pMSSM models:} In all cases, the particle spectra are generated with SOFTSUSY 3.3.3~\cite{Allanach:2001kg}, while sparticles decays are calculated with SUSY-HIT~\cite{susyhit} (SDECAY 1.3b and HDECAY 3.4). The simulated signal  events are generated using \herwigpp~2.6.3 ~\cite{Bahr:2008pv} with the CTEQ6L1 PDF set.

\bibliographystyle{atlasBibStyleWoTitle}
\bibliography{stopsbottom_summary,analyses}

\newpage 
\begin{flushleft}
{\Large The ATLAS Collaboration}

\bigskip

G.~Aad$^{\rm 85}$,
B.~Abbott$^{\rm 113}$,
J.~Abdallah$^{\rm 151}$,
O.~Abdinov$^{\rm 11}$,
R.~Aben$^{\rm 107}$,
M.~Abolins$^{\rm 90}$,
O.S.~AbouZeid$^{\rm 158}$,
H.~Abramowicz$^{\rm 153}$,
H.~Abreu$^{\rm 152}$,
R.~Abreu$^{\rm 116}$
% ...
\bigskip
\\
% ...
$^{11}$ Institute of Physics, Azerbaijan Academy of Sciences, Baku, Azerbaijan\\
% ...
$^{85}$ CPPM, Aix-Marseille Universit{\'e} and CNRS/IN2P3, Marseille, France\\
% ...
$^{90}$ Department of Physics and Astronomy, Michigan State University, East Lansing MI, United States of America\\
% ...
$^{107}$ Nikhef National Institute for Subatomic Physics and University of Amsterdam, Amsterdam, Netherlands\\
% ...
$^{113}$ Homer L. Dodge Department of Physics and Astronomy, University of Oklahoma, Norman OK, United States of America\\
% ...
$^{116}$ Center for High Energy Physics, University of Oregon, Eugene OR, United States of America\\
% ...
$^{151}$ Institute of Physics, Academia Sinica, Taipei, Taiwan\\
$^{152}$ Department of Physics, Technion: Israel Institute of Technology, Haifa, Israel\\
$^{153}$ Raymond and Beverly Sackler School of Physics and Astronomy, Tel Aviv University, Tel Aviv, Israel\\
% ...
$^{158}$ Department of Physics, University of Toronto, Toronto ON, Canada\\
% ...
\end{flushleft}


\end{document}